\documentclass[12pt]{iopart}
\usepackage{graphicx}
\usepackage{iopams}
\usepackage{harvard}
\usepackage{textcomp}
\usepackage{multirow}
\usepackage{epstopdf}

\graphicspath{{./}}

\newcommand{\apj}{ApJ}
\newcommand{\apjs}{ApJS}
\newcommand{\apjl}{ApJL}
\newcommand{\mnras}{MNRAS}
\newcommand{\prd}{PhRD}
\newcommand{\nat}{Nature}
\newcommand{\aap}{A\&A}
\newcommand{\araa}{ARA\&A}

\newcommand{\actaa}{Acta Astron.}

\begin{document}

\title[$\nu$-$\bar{\nu}$ annihilation above merger remnants: implications of a long-lived MNS]{Neutrino pair annihilation above merger remnants: implications of a long-lived massive neutron star}

\author{A. Perego, H. Yasin, A. Arcones}
 
\address{Institut f\"ur Kernphysik, Technische Universit\"at Darmstadt, Schlossgartenstra{\ss}e 2, 64289 Darmstadt, Germany. \\
         GSI Helmholtzzentrum f\"ur Schwerionenforschung GmbH, Planckstra{\ss}e 1, 64291 Darmstadt, Germany.}
\ead{albino.perego@physik.tu-darmstadt.de}
\vspace{10pt}
\begin{indented}
\item[]December 2016
\end{indented}

\begin{abstract}
Binary neutron star mergers are plausible progenitor candidates for short gamma-ray 
bursts (GRBs); however, a detailed explanation of their central engine is still lacking. 
The annihilation of neutrino pairs has been proposed as one of the possible powering mechanisms.
We present calculations of the energy and momentum deposition operated by neutrino pair 
annihilation above merger remnants. Starting from the results of a detailed, three-dimensional simulation
of the aftermath of a binary neutron star merger, we compute the deposition rates over a time scale comparable 
to the life time of the disk ($t \approx 0.4$ s), assuming a long-lived massive neutron star (MNS).
We model neutrino emission using a spectral leakage scheme and compute the neutrino annihilation rates
using a ray-tracing algorithm.
We find that the presence of the MNS increases the energy deposition rate by a factor $\sim 2$,
mainly due to the annihilation of radiation coming from the MNS with radiation coming from the disk.
We compute the impact of relativistic effects
and discover that, despite they can significantly change the local rate intensity,
the volume-integrated results are only marginally decreased.
The cumulative deposited energy, extrapolated to 1 sec, is $\approx 2.2 \times 10^{49} \, {\rm erg}$.
A comparison with the inferred short GRB energetics reveals that in most cases this energy is  
not large enough, even assuming small jet opening angles and a long-lived MNS.
Significantly more intense neutrino luminosities (a factor ~5-10 larger) are required to explain most of
the observed short GRB. We conclude that it is unlikely that neutrino pair annihilation can explain
the central engine of short GRBs alone.
\end{abstract}

%
\vspace{2pc}

\section{Introduction}

Binary neutron star (BNS) mergers are the catastrophic fate of relativistic binary systems formed by
a pair of neutron stars (NSs). The intense emission of gravitational waves (GW) that characterizes the final
phases of the inspiral and its expected frequencies of $\sim 1$-$10^3$ Hz make merging BNSs one
of the primary targets of ground based GW detectors, such as advanced LIGO, advanced VIRGO, and, in the 
nearby future, KAGRA \citeaffixed{Aasi.etal:2015,Acernese.etal:2015,Aso.etal:2013}{e.g.,}. 
In addition, these events are expected to inject matter and energy into the interstellar medium, 
and to produce a rich variety of paramount astrophysical phenomena, including the nucleosynthesis of the heaviest elements \citeaffixed{Lattimer.etal:1977,Symbalisty.Schramm:1982,Freiburghaus.etal:1999,Korobkin.etal:2012,Goriely.etal:2011,Bauswein.etal:2013,Hotokezaka.etal:2013,Wanajo.etal:2014,Martin.etal:2015,Radice.etal:2016,Wu.etal:2016}{e.g.,}, 
and the emission of characteristic electromagnetic transients from the freshly synthesized radioactive nuclei
\citeaffixed{Li&Paczynski:1998,Metzger.etal:2010a,Roberts.etal:2011,Tanaka&Hotokezaka:2013,Bauswein.etal:2013,Metzger.Fernandez:2014,Grossman.etal:2014}{e.g.,},
and radio flares from the expanding ejecta \citeaffixed{Nakar.Piran:2011,Margalit.Piran:2015}{e.g.,}.

A BNS merger is expected to leave behind a compact object surrounded by a hot and dense accretion disk. Inside this remnant,
gravitational energy is efficiently converted to internal energy and emitted in the form of neutrinos.
The nature of the central object depends primarily on intrinsic attributes of the merging system (such as the NSs masses
and their ratio) and on the still uncertain properties of hot nuclear matter above nuclear saturation density.
The direct collapse of the merging NSs into a black hole (BH) is foreseen only for very massive colliding objects \citeaffixed{Bauswein.etal:2013a}{e.g.,}, 
while in most cases the production of a (possibly metastable) differentially rotating massive neutron star (MNS) is expected.
If the MNS mass exceeds the maximum allowed NS mass, the MNS will collapse to a BH on a characteristic time scale.
The duration of this phase depends sensitively on several parameters and poorly understood processes happening inside the MNS.
Among them, we recall the amount of angular momentum and its distribution, the effects of thermal support
and the consequences of neutrino cooling, the role and the properties of magnetic fields. 

The large amount of energy available in these events, together with the short time scale and the reduced volume that characterize
the merger and its aftermath, led to the hypothesis that BNS mergers represent (together with the merger of BH-NS binaries) 
a plausible progenitor system for short gamma-ray bursts (GRBs) at cosmological distances 
\citeaffixed{Paczynski:1986,Narayan.etal:1992,Woosley:1993,Ruffert.janka:1999a,Rosswog.etal:2003a}{e.g.,}. 
Observational evidences collected over the last years seem to support this hypothesis \citeaffixed{Berger:2014,Fong.etal:2015,Ghirlanda.etal:2016}{e.g.,}. 
However, a satisfactory and detailed explanation of how the merger of a compact binary system involving at least one NS
powers a relativistic jet is still missing and the central engine of short GRBs still remains veiled.
Several different mechanisms have been suggested. Among them, the conversion of energy
by magnetohydrodynamical effects in BH-torus systems \cite{Blandford.Znajek:1977,Paschalidis.etal:2015,Dionysopoulou.etal:2015}, 
the extraction of magnetic energy from highly magnetized, long-lived MNS \cite{Metzger.etal:2011}, and the annihilation of 
neutrino-antineutrino pairs in low density regions above the remnant \cite{Eichler.etal:1989}.
The conditions needed for the formation and collimation of a relativistic jet above
a merger remnant, as well as the interactions with the accretion disk, non-relativistic winds and dynamic ejecta, 
have been recently investigated with increasing details and accuracy \cite{Aloy.etal:2005,Murguia.etal:2014,Nagakura.etal:2014b,Duffell.etal:2015,Just.etal:2016,Murguia.etal:2016}. 

Neutrino emission from merger remnants has been the topic of several dedicated studies 
\cite{Ruffert.etal:1997,DiMatteo.etal:2002,Rosswog.Liebendoerfer:2003,Setiawan.etal:2004,Setiawan.etal:2006,Chen.Beloborodov:2007,Dessart.etal:2009,Caballero.etal:2012,Janiuk.etal:2013,Perego.etal:2014b,Just.etal:2015a,Foucart.etal:2015a}.
The annihilation of neutrino-antineutrino pairs above BH-torus systems has been extensively studied,
also considering the effects of general relativity on the neutrino propagation \cite{Jaroszynski:1993,Popham.etal:1999,Asano.Fukuyama:2000,Miller.etal:2003,Kneller.etal:2006,Birkl.etal:2007,Zalamea.Beloborodov:2011,Just.etal:2016}. 
Annihilation rates above MNS-disk system have been computed by 
\citeasnoun{Dessart.etal:2009} and \citeasnoun{Richers.etal:2015}. In the former case, several snapshots 
from the first 100~ms of an axisymmetric merger aftermath simulation were post-processed 
using both a $S_{n}$ transport scheme and a simpler gray leakage scheme to model neutrino radiation. 
In the latter, four snapshots taken from long-term, axisymmetric simulations of merger aftermaths, 
and covering the 0-3 sec interval were post-processed with a Monte Carlo radiative transfer code and, 
again, with a simpler gray leakage scheme.

In this work, we compute the energy and momentum deposition rates above BNS merger remnant over 
a time scale of 400~ms, comparable with the life time of the disk. We post-process several merger configurations
taken from a long-term, three dimensional simulation of the aftermath of a BNS merger.
We assume a long-lived MNS and specifically investigate the impact of the neutrino emission from the MNS. We also explore 
the impact of relativistic effects in the neutrino propagation on the annihilation rates, in the presence of a MNS.
Finally, we discuss our results in connection with the energy inferred from short GRB observations, to address the question
whether neutrino pair annihilation can be responsible for the production of relativistic jets above BNS mergers. 
The paper is structured as follows: in Section \ref{sec: simulation}, we present the merger aftermath simulation 
from which we take the remnant configuration and the neutrino emission properties. Our method to compute 
the annihilation rates, both in a Newtonian and in a relativistic framework, is outlined in Section \ref{sec: method}.
Section \ref{sec: results} is devoted to the presentation of the annihilation calculations while
in Section \ref{sec: roleMNS} we analyze the impact of a long lived-MNS on the annihilation rates.
Our results are compared with short GRB observations in Section \ref{sec: discussion}.
We summarize and conclude in Section \ref{sec: conclusions}.

\section{Model of binary NS merger remnant}
\label{sec: simulation}

\begin{figure}
 \centering
 \includegraphics[width= 0.7\linewidth]{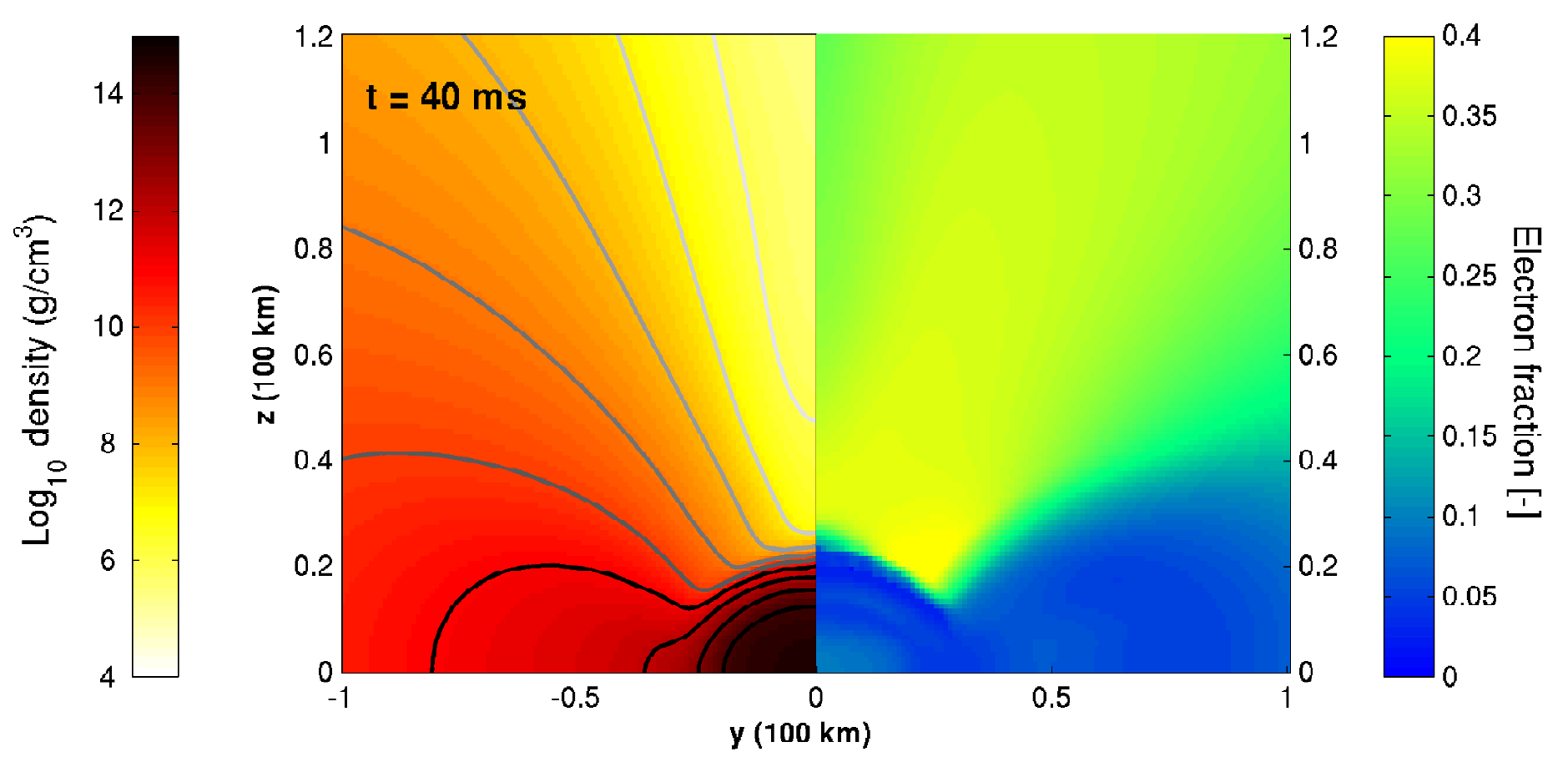}
 \caption{Vertical slice ($x=0$ plane) of the three dimensional domain for the matter density (left) and the 
 electron fraction (right) at 40~ms of our simulation.
 The solid lines on the density profile represent isodensity contours, ranging from 
 $10^{14} \, {\rm g \, cm^{-3}}$ (black line) to $10^{6} \, {\rm g \, cm^{-3}}$ (light gray line).
 The densest part of the remnant corresponds to MNS, while density contours of $10^{11} \, {\rm g \, cm^{-3}}$ and 
 $10^{10} \, {\rm g \, cm^{-3}}$ track the outer boundary of the disk. The expanding matter located above
 the remnant and characterized by low density and reduced neutron richness corresponds to the developing
 neutrino-driven wind.}
 \label{fig: density and Ye profiles}
\end{figure}

Our annihilation rate calculations are based on the numerical results of a three dimensional, Newtonian simulation of the aftermath 
of a binary neutron star merger under the influence of neutrino cooling and heating \cite{Perego.etal:2014b}.
The initial conditions for our model were taken from a high-resolution SPH simulation of the merger of
two equal mass (1.4 $M_{\odot}$) neutron stars \cite{Price2006}. We mapped the remnant configuration at
$\sim 20$~ms after the merger inside our equally spaced Cartesian grid.
At this time, the remnant was composed by an almost axisymmetric, stationary rotating MNS with a mass 
of $M_{\rm NS} \approx 2.6 \, {\rm M}_{\odot}$, surrounded by 
a thick accretion disk with a mass of $M_{\rm disk} \approx 0.18 \, {\rm M}_{\odot}$.
We employed a rather stiff nuclear equation of state (EOS), HS(TM1) \cite{Hempel2012}, for consistency with the initial conditions.
The radius of the MNS, defined as the radial coordinate where the matter density drops below 
$5 \times 10^{12}\, {\rm g \, cm^{-3}}$, was located in a range between 17 km and 
25 km along the vertical and the equatorial directions, respectively.
Neutrinos of all flavors were emitted from the remnant with a total luminosity larger than 
$10^{53} \, {\rm erg \, s^{-1} }$.  
The re-absorption of a fraction of the emitted neutrinos inside the disk caused the formation
of a neutrino-driven wind on a time scale of a few tens of milliseconds. 
At the same time, 
matter expanded from the disk along 
the equatorial direction. This was the beginning of the so-called disk evaporation 
\cite{Fernandez.Metzger:2013,Just.etal:2015a}.
The wind developed mainly from the disk and the funnel above the central neutron star has
a lower density than the material extending in the equatorial direction.
We evolved the system for $\sim 400$~ms.
Since our uniform spatial resolution ($\delta x = 1 \, {\rm km}$) was not sufficient to properly 
resolve the MNS, we assumed
the region inside $\mathcal{S}$ as stationary, where $\mathcal{S}$ is an oblate spheroid
of $z$ as symmetry axis, center in the domain origin, semi-axis $a=30\,{\rm km}$ 
along $x$ and $y$, and $b=23\,{\rm km}$ along $z$, and matter density above 
$5 \times 10^{11} \, {\rm g \, cm^{-3}}$.
Profiles of the matter density and electron fraction inside the MNS and in the innermost part of the disk at 40~ms
inside our simulation are shown in Figure \ref{fig: density and Ye profiles}, together with the 
incipient neutrino-driven wind.

\begin{figure}
 \centering
 \includegraphics[width=\linewidth]{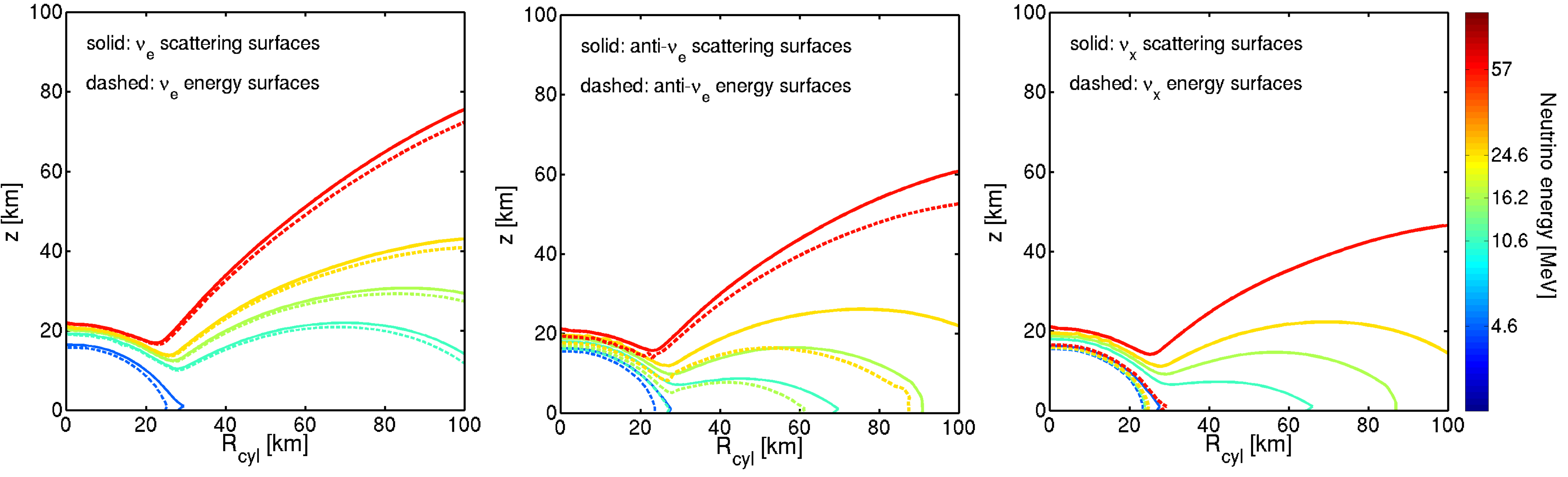}
 \caption{Neutrino scattering (solid lines) and energy (dashed lines) surfaces at 40~ms in the simulation.
 The three panels correspond to $\nu_e$ (left), $\bar{\nu}_e$ (center) and $\nu_x$ (right), while lines with
 different colors and increasing thickness represent increasing neutrino energies.}
 \label{fig: nu_surfaces}
\end{figure}

Neutrino physics is modeled using an energy dependent (spectral) leakage scheme 
\cite{Perego.etal:2014b,Perego.etal:2016}. All neutrino quantities are binned logarithmically in energy,
with twelve energy bins between $2 \, {\rm MeV}$ and $200 \, {\rm MeV}$.
The neutrino reactions included in the calculations are the emission, absorption, and scattering off free nucleons. Neutrino pair emission from electron-positron annihilation and nucleon-nucleon bremsstrahlung are also included in optically thick conditions, as well as absorption from their inverse reactions.
We model three independent neutrino species, $\nu_e$, $\bar{\nu}_e$, and $\nu_x$, the latter being 
a collective species  for $\mu$ and $\tau$ (anti)neutrinos.
Spectral optical depths, $\tau_{\nu}$, are computed by minimizing the line integral of  
inverse mean free paths along several possible straight propagation directions. 
This allows the distinction between optically thin ($\tau_{\nu}\lesssim 1$) and optically thick 
($\tau_{\nu}\gg 1$) regions.
We further distinguish between scattering, $\tau_{\nu,{\rm sc}}$, and energy optical depths, $\tau_{\nu,{\rm en}}$ 
\cite{Raffelt:2001}.
We denote the neutrino surfaces as the surfaces where $\tau_{\nu}=2/3$. In the case of 
$\tau_{\nu,{\rm sc}}$ they correspond to the last scattering surfaces. In the case of 
$\tau_{\nu,{\rm en}}$ they locate the transition from a thermally coupled to a non-thermally coupled regime.
The position of the scattering and energy surfaces inside the remnant at 40~ms in our simulation are represented in
Figure \ref{fig: nu_surfaces}, for all neutrino species and for five representative energies.
Neutrinos are considered emitted isotropically in the half space denoted by 
$\mathbf{n}_{\tau} = - \nabla \tau_{\nu,{\rm sc}}/\left| \nabla \tau_{\nu,{\rm sc}} \right|$,
both from the neutrino scattering surfaces (for radiation diffusing from optically thick conditions) 
and from their production site (for radiation emitted in optically thin conditions) \cite[appendix]{Perego.etal:2014b}.

\subsection{Neutrino luminosities}

\begin{figure}
 \includegraphics[width=0.49 \linewidth]{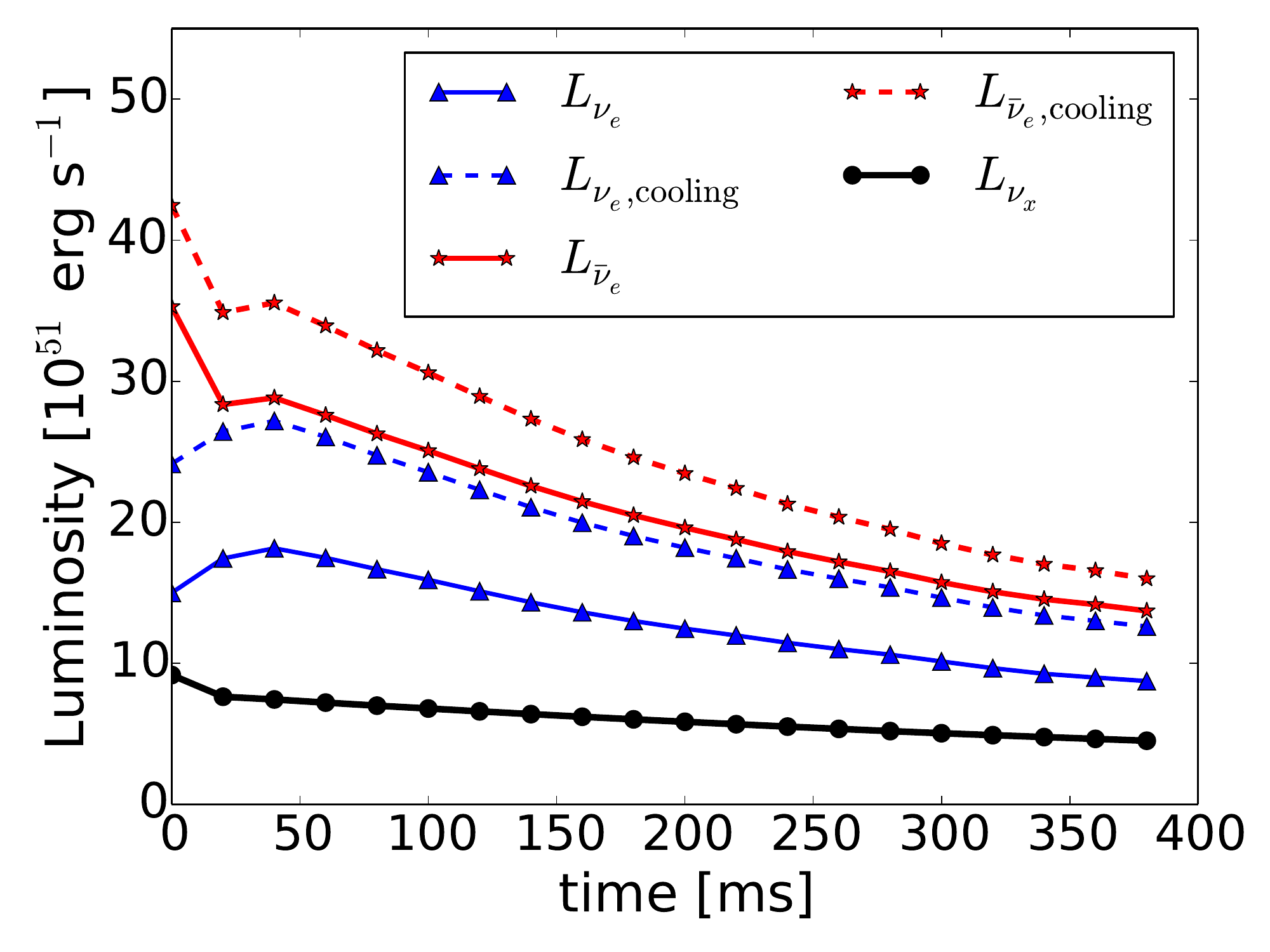}
 \includegraphics[width=0.49 \linewidth]{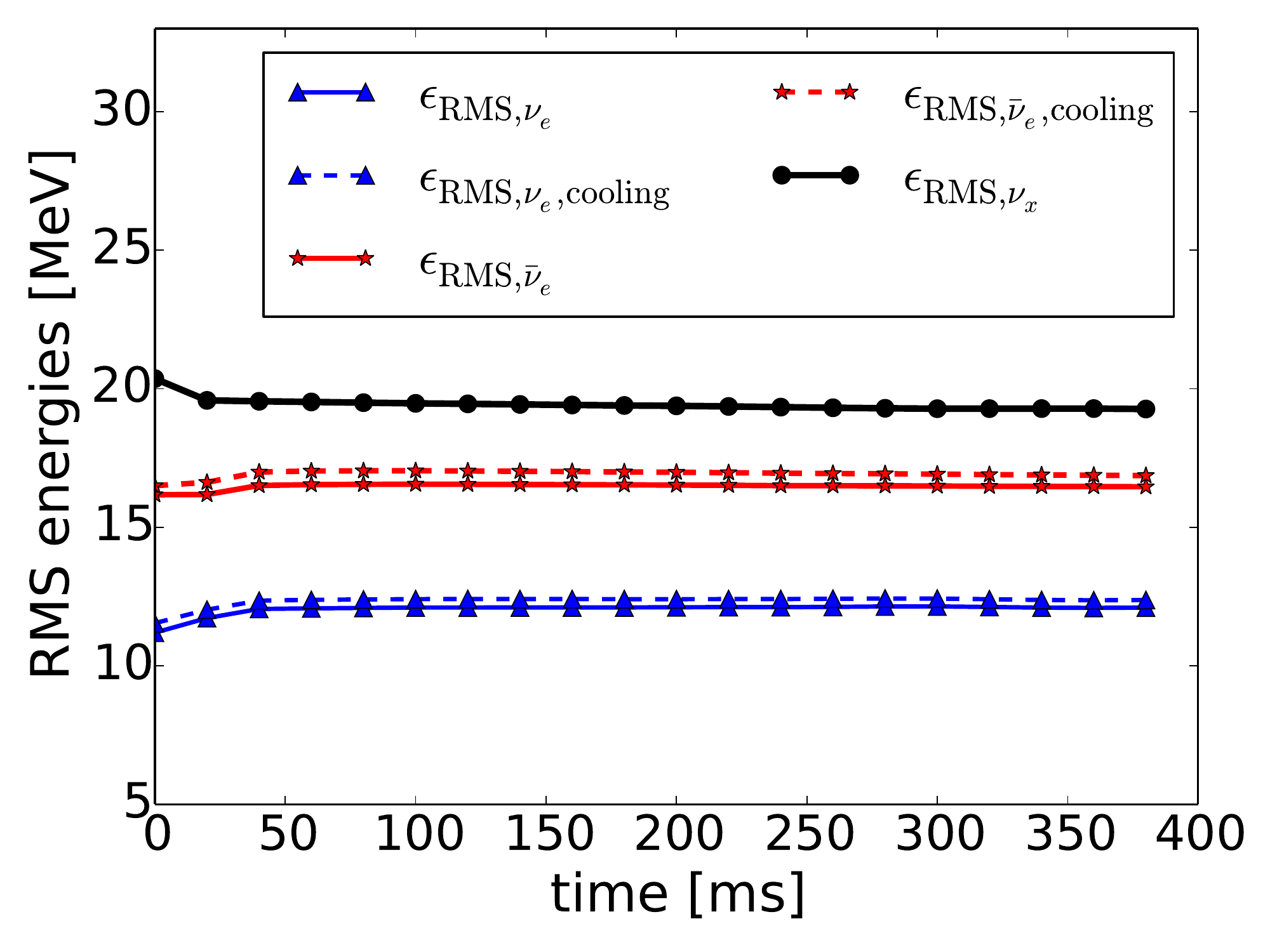}
 \caption{Evolution of the neutrino total luminosities (left) and root mean squared energies (RMS, right).
 Blue lines  (triangles) refer to $\nu_e$, red lines (stars) to $\bar{\nu}_e$,
 black lines (circles) to $\nu_x$. All quantities are measured at infinity.
 Solid and dashed lines refer to net and cooling luminosities (or RMS energies), respectively. In the former case
 neutrino absorption processes outside the last scattering surfaces are included; in the latter, not.}
 \label{fig: neutrino luminosities}
\end{figure}
 
We review the most relevant properties of the neutrino emission obtained from our 
hydrodynamical simulation, since they represent a key ingredient for the calculations of
the annihilation rates.
In Figure \ref{fig: neutrino luminosities}, we present the evolution of the neutrino luminosities 
($L_{\nu}$ left panel) and root mean squared (RMS) energies ($\epsilon_{{\rm RMS},\nu} \equiv \sqrt{\langle \epsilon^2_\nu \rangle}$, 
right panel) for all the three independent neutrino species, over the entire
simulation time.
Both quantities are measured at infinity, taking into account that a fraction 
of the emitted neutrinos (especially $\nu_e$'s and, less significantly,
$\bar{\nu}_e$'s) get absorbed on their way out.
To show the effect of neutrino absorption outside the neutrino surfaces, we plot also the 
luminosities and RMS energies obtained including the neutrinos emitted at the last scattering
surfaces and outside, but assuming the neutrino absorptivity to be zero for $\tau_{\nu,{\rm sc}}< 2/3$.
We call them ``cooling'' luminosities and RMS energies (dashed lines).

Neutrino luminosities are powered by the cooling of the remnant 
and by the process of accretion inside the disk. After an initial phase ($t<40$~ms) when the disk
dissipates internal perturbations and transients, and the total luminosity slightly increases,
all the luminosities decrease with time.
The RMS neutrino energies reflect the different depths and thermodynamical conditions at which 
neutrinos thermally decouple from matter. 
While the values obtained for $\nu_e$ and $\bar{\nu}_e$ are consistent with the ones reported
in the literature, the values obtained for $\nu_x$ are smaller than expected (by $\sim$20\%). As reported in
\citeasnoun{Perego.etal:2014b}, this discrepancy is due to the lack of spatial resolution in our
simulation deep inside the MNS, where the thermal decoupling of $\nu_x$'s occurs.

Due to the stationary treatment of the MNS, the luminosities coming from the spheroid $\mathcal{S}$,
$L_{\nu,\mathcal{S},{\rm sim}}$, are practically constant during the simulation. 
To mimic the decrease produced by the cooling process, we assume the evolution of the luminosities 
coming from the innermost part of the computational domain as an exponential decline:
\begin{equation}
 L_{\nu,\mathcal{S}}(t) =  L_{\nu,\mathcal{S},{\rm sim}} \exp{\left(- t/t_{\rm cool} \right)},
\end{equation}
in analogy to the luminosity produced by a cooling proto-NS \cite{Fischer.etal:2010,Hudepohl.etal:2010}.
The cooling time scale $t_{\rm cool}$ is estimated as 
$t_{\rm cool} \approx \Delta E_{\mathcal{S}}/L_{\nu,\mathcal{S},{\rm sim}}$, where
$\Delta E_{\mathcal{S}}$ is the thermal component of the internal energy
inside $\mathcal{S}$, calculated from the initial remnant configuration.
From our simulation we obtained $t_{\rm cool} \approx 0.725 \, {\rm s}$.

\section{Calculation of the annihilation rates}
\label{sec: method}

\subsection{Newtonian calculation}
\label{sec: method Newtonian}

We consider the annihilation of neutrino pairs into electron positron pairs, i.e. 
$ \nu_i + \bar{\nu}_i \rightarrow e^+ e^- $ with $i \in \{ e,\mu,\tau \}$.
We calculate the energy and momentum deposition rates due to this process outside 
the most relevant neutrino surfaces,
post-processing the input data obtained in our simulation of the aftermath of a binary NS merger.
Since the power associated with neutrino-antineutrino annihilation is expected to be sub-dominant compared with 
the one of charged-current neutrino absorption reactions, we can safely compute the former in a post-processing
step \cite{Dessart.etal:2009}. For consistency with the underlying simulation, we start computing the rate in a Newtonian framework.
Moreover, we assume the neutrino to be massless and we presently do not include the effects of 
neutrino oscillations in our analysis \citeaffixed{Malkus.etal:2016,Zhu.etal:2016,Frensel.etal:2016}{e.g.,}.
We calculate the local energy deposition rate $q_{\nu_i,\bar{\nu}_i} (\mathbf{x}, t)$ for the neutrino
flavor $i$ according to 
the following expression \cite[and references therein]{Kneller.etal:2006,Dessart.etal:2009}:
\begin{eqnarray}
q_{\nu_i,\bar{\nu}_i}(\mathbf{x}, t) = & \frac{\sigma_0 \ (c_A^2 + c_V^2)_{\nu_i,\bar{\nu}_i}}{6 \ c \ (m_e c^2)^2} 
\int_{\Omega_{\nu_i}} {\rm d} \Omega_{\nu_i} \int_{\Omega_{\bar{\nu}_{i}}} {\rm d}\Omega_{\bar{\nu}_{i}} 
\int_0^{\infty} {\rm d}\epsilon_{\nu_i} 
\int_0^{\infty} {\rm d}\epsilon_{\bar{\nu}_{i}} \nonumber \\
& (\epsilon_{\nu_i} + \epsilon_{\bar{\nu}_i}) \ I_{\nu_i}
\ I_{\bar{\nu}_i}
\ (1 - \cos \Phi)^2,
\label{eq: dessart}
\end{eqnarray}
We extend the previous formula to compute the deposited momentum $\mathbf{p}_{\nu_i,\bar{\nu}_i}$:
\begin{eqnarray}
\mathbf{p}_{\nu_i,\bar{\nu}_i}(\mathbf{x}, t) = & 
\frac{\sigma_0 \ (c_A^2 + c_V^2)_{\nu_i,\bar{\nu}_i}}{6 \ c^2 \ (m_e c^2)^2} 
\int_{\Omega_{\nu_i}} {\rm d}\Omega_{\nu_i} 
\int_{\Omega_{\bar{\nu}_{i}}} {\rm d}\Omega_{\bar{\nu}_{i}} 
\int_0^{\infty} {\rm d}\epsilon_{\nu_i} 
\int_0^{\infty} {\rm d}\epsilon_{\bar{\nu}_{i}} \nonumber \\
& (\epsilon_{\nu_i} \mathbf{n}_{\nu_i} + \epsilon_{\bar{\nu}_i}\mathbf{n}_{\bar{\nu}_i} ) \ I_{\nu_i}
\ I_{\bar{\nu}_i}
\ (1 - \cos \Phi)^2,
\label{eq: dessart momentum}
\end{eqnarray}
In Eqs.~(\ref{eq: dessart}) and (\ref{eq: dessart momentum}),
$c$ is the speed of light, $m_e$ the mass of the electron, 
$\sigma_0 = 4 m_e^2 G_F^2/ \pi \hbar^4 \approx 1.71 \times 10^{-44} \, {\rm cm^2}$ 
the typical neutrino cross section, with $G_F$ the Fermi constant, and $\hbar$ the reduced Planck constant.
Furthermore, $I_\nu = I_\nu(\mathbf{x},t, \epsilon_\nu, \mathbf{n}_{\nu})$ denotes 
the radiation intensity at position $\mathbf{x}$ and time $t$, for neutrinos with 
energy $\epsilon_\nu$ along the direction $\mathbf{n}_{\nu}$. 
The infinitesimal solid angle and energy in the momentum space of the propagating radiation is
represented by ${\rm d}\Omega_\nu$ and ${\rm d}\epsilon_\nu$, respectively.
The cosine of the angle between the momenta of the colliding neutrinos is given by 
$\cos{\Phi} = (\mathbf{n}_{\nu} \cdot \mathbf{n}_{\bar{\nu}})$. Equations
(\ref{eq: dessart}) and (\ref{eq: dessart momentum}) apply to all neutrino species, provided the 
appropriate values of the weak coupling constants $c_A$ and 
$c_V=c_A + 2 \sin^2{\theta_W}$, where $\sin^2{\theta_W} = 0.23$. 
In the case of $\nu_e$ and $\bar{\nu}_e$, $c_A=1/2$, while for $\nu_x$ and $\bar{\nu}_x$, $c_A=-1/2$.
Note that in the expressions for $q_{\nu_i,\bar{\nu}_i}$ and $\mathbf{p}_{\nu_i,\bar{\nu}_i}$ 
we have neglected phase space blocking for $e^{\pm}$ in the final state, as well as
a term proportional to $(m_ec^2)^2/(\epsilon_{\nu_i} \epsilon_{\bar{\nu}_i})$ in the integrand.
In fact, the latter is $10^{3}$ times smaller for typical neutrino energies \citeaffixed{Perego.etal:2014b}{e.g.,}.

Since the neutrino intensities required in Eqs.~(\ref{eq: dessart}) and (\ref{eq: dessart momentum}) 
are not directly provided by our simulation, we compute them
based on the local emissivities.
We fix one point $B$ outside the most relevant neutrino surfaces.
We denote the amount of energy ${\rm d}E_B$ transported across a surface element ${\rm d}A_B$
in a time ${\rm d}t$ by ${\rm d}N_B$ neutrinos in an energy interval ${\rm d}\epsilon_B$ around
$\epsilon_B$ propagating inside a solid angle ${\rm d}\Omega_B$ around $\mathbf{n}_B$,
in terms of radiation intensity $I_B \equiv I_{\nu}(\mathbf{x}_B,\epsilon_B,\mathbf{n}_B)$ 
as ${\rm d}E_B = \epsilon_B ~ {\rm d}N_B = I_B \ {\rm d}A_B \ {\rm d}t \ {\rm d}\Omega_B \ {\rm d}\epsilon_B$.
We further express the amount of energy emitted at any point $A$ 
by a volume ${\rm d}V_A$ in a time ${\rm d}t$ in ${\rm d}N_A$ neutrinos
with an energy interval ${\rm d} \epsilon_A$ around $\epsilon_A$ and
into a solid angle ${\rm d}\Omega_A$ around $\mathbf{n}_A$, 
as ${\rm d}E_A = \epsilon_A ~ {\rm d}N_A = \eta_A \ {\rm d}V_A \ {\rm d}t \ {\rm d}\Omega_A \ {\rm d}\epsilon_A$, where 
$\eta_A \equiv \eta_{\nu}(\mathbf{x}_A,\epsilon_A,\mathbf{n}_A)$ is the local emissivity.
Conservation of particles and energy implies:
\begin{equation}
\epsilon_B~{\rm d}N_B  = \epsilon_A~{\rm d}N_A \, \exp (-\Delta \tau_{{\rm en},AB}),
\label{eq: energy conservation between A and B}
\end{equation}
where we took into account that a fraction of the emitted neutrinos get absorbed on their way to $B$.
This amount is estimated using an exponential damping based on the difference of the energy optical
depth $\tau_{\rm en}$ between $A$ and $B$, $\Delta \tau_{{\rm en},AB} = \tau_{{\rm en},A} - \tau_{{\rm en},B}$.
Inserting the definitions of intensity and emissivity into Eq.~(\ref{eq: energy conservation between A and B}), 
and rewriting $dA_B = d_{AB}^2 \ d\Omega_A$,
where $d_{AB}$ denoted the distance between the two points, we obtain:
\begin{equation}
I_B \ d\Omega_B \ d\epsilon_B = \frac{\eta_A \ dV_A \ d\epsilon_A}{d_{AB}^2} \, \exp (-\Delta \tau_{{\rm en},AB}).
\label{eq: intensity_omega_e}
\end{equation}
We model the emission of neutrinos from the neutrino surfaces and from the optically transparent regions
consistently with the treatment used in the post-merger simulation. Specifically,
\begin{equation}
\eta_{\nu}(\mathbf{x}_A,\epsilon_{A},\mathbf{n}_{A}) =
\left\{
 \begin{array}{lr}
  {\rm const.}  & \quad {\rm if} \; \langle \mathbf{n}_{A} \rangle \cdot \mathbf{n}_{A} > 0 \, , \\
  0             & \quad {\rm otherwise},
 \end{array}
 \right.
 \label{eq: isotropic emissivity}
\end{equation}
i.e., we assume that radiation is emitted isotropically in the half space identified by
the unitary vector $\langle \mathbf{n}_A \rangle$ and we take
$\langle \mathbf{n}_A \rangle=\mathbf{n}_{\tau,A}$ (see Section \ref{sec: simulation}).

We discretize the three dimensional space with cylindrical coordinates 
$\{R_i,\phi_j,z_k\}$ where $i$, $j$ and $k$ can vary between 1 and $N_R$, $N_\phi$ and $N_z$, respectively.
For the neutrino energy, we use the same energy bins that we employed in our post-merger simulation, 
$\{ \epsilon_m \}_{m=1,N_e}$, with $N_e=12$.
From the results of our simulation, we extract cylindrically averaged values 
for the spectral optical depths $\left( \tau_{\nu} \right)_{i,k,m} \equiv \tau_{\nu}(R_i,z_k,\epsilon_m)$
and for the (solid angle integrated) neutrino emission rates in each energy bin $\epsilon_m$,
$\left( R_{\nu} \right)_{i,k,m} \equiv R_{\nu}(R_i,z_k,\epsilon_m)$.
Integrating Eq.~(\ref{eq: isotropic emissivity}) over the whole solid angle, 
we can express $\eta_{\nu}$ in terms of $R_{\nu}$:
\begin{equation}
(R_\nu)_{i,k,m} =
\frac{2 \pi \, \left( \eta_{\nu} \right)_{i,k,m} \, {\rm d}\epsilon_m}{\epsilon_m}.
\label{eq: eta and R_nu relationship}
\end{equation} 
If we discretize Eq.~(\ref{eq: dessart}) according to
\begin{eqnarray}
q_{\nu_i,\bar{\nu}_i}(\mathbf{x}_B,t) = 
\frac{\sigma_0 \ (c_A^2 + c_V^2)_{\nu_i,\bar{\nu}_i}}{6 \ c \ (m_e c^2)^2} \, \sum_{A(i,j,k)} \, \sum_{A'(i',j',k')} \,
\sum_{m,m'} \, (\epsilon_m + {\epsilon_{m'}}) \nonumber \\
\left( {\rm d} \Omega_{\nu}       \, {\rm d} \epsilon_{\nu}       \, I_{\nu} \right)_{A(i,j,k,m) \rightarrow B} \,
\left( {\rm d} \Omega_{\bar{\nu}} \, {\rm d} \epsilon_{\bar{\nu}} \, I_{\bar{\nu}} \right)_{A'(i',j',k',m') \rightarrow B} 
\, \ (1 - \cos \Phi_{A,A'})^2 ,
\label{eq: dessart_discretized}
\end{eqnarray}
inserting Eqs.~(\ref{eq: intensity_omega_e}) and (\ref{eq: eta and R_nu relationship}) we obtain
\begin{eqnarray}
& q_{\nu_i,\bar{\nu}_i}(\mathbf{x}_B,t) =
\frac{\sigma_0 \ (c_A^2 + c_V^2)_{\nu_i,\bar{\nu}_i}}{6 \ c \ (m_e c^2)^2} \, \sum_{A(i,j,k)} \, \sum_{A'(i',j',k')} \, \sum_{m,m'} \: 
(\epsilon_m + \epsilon_{m'}) \epsilon_m \epsilon_m'  \nonumber \\
& \Theta(\langle \mathbf{n}_{A,m} \rangle \cdot \mathbf{n}_{AB} ) \: \Theta(\langle \mathbf{n}_{A',m'} \rangle \cdot \mathbf{n}_{A'B}) \;
\frac{ (R_\nu)_{i,j,k,m}     \ {\rm d}V_{i,j,k}}{2 \pi d_{AB}^2} \;
\frac{ (R_{\bar{\nu}})_{i',j',k',m'} \ {\rm d}V_{i',j',k'}}{2 \pi d_{A'B}^2} \nonumber \\
&  \exp{(-( \left( \Delta  \tau_{\nu,{\rm en}} \right)_{AB,m} + 
   \left( \Delta \tau_{\nu,{\rm en}} \right)_{A'B,m'}))} \, (1 - \cos \Phi_{A,A'})^2,
 \label{eq: dessart_discretized 2}  
\end{eqnarray}
where ${\rm d}V_{i,j,k}$ is the volume element in cylindrical coordinates, 
$\mathbf{n}_{A^{(\prime)}B} = (\mathbf{x}_B-\mathbf{x}_{A^{(\prime)})}/ |\mathbf{x}_B-\mathbf{x}_{A^{(\prime)}} | $,
and $\Theta$ is the Heaviside step function.
For the discrete version of Eq.~(\ref{eq: dessart momentum}), it is enough to replace
$(\epsilon_m + \epsilon_m')$ with $(\epsilon_m \mathbf{n}_{AB} + \epsilon_m' \mathbf{n}_{A'B})/c$ in Eq.~(\ref{eq: dessart_discretized 2}). 

Starting from the local expression of the energy deposition rate, Eq.~(\ref{eq: dessart}),
we compute the volume-integrated energy deposition rate, $Q_{\nu_i,\bar{\nu}_i}(t)$, as
\begin{equation}
 Q_{\nu_i,\bar{\nu}_i}(t) = \int_{\tilde{V}} \: q_{\nu_i,\bar{\nu}_i}(t,\mathbf{x}) \, {\rm d}V \, .
 \label{eq: Q non rel}
\end{equation}
Following \citeasnoun{Birkl.etal:2007}, we define $\tilde{V}$ as the volume 
outside the most relevant neutrino surfaces, and in which the radial component of the total
momentum deposited by neutrinos is pointing outwards.
In addition, we designate the annihilation efficiency as \citeaffixed{Eichler.etal:1989,Just.etal:2016}{e.g.,}:
\begin{equation}
\eta_{\nu,\bar{\nu}} = Q_{\nu,\bar{\nu}}/(L_{\nu} + L_{\bar{\nu}}) \, .
\label{eq: annihilation efficiency}
\end{equation}

Since the integrand in Eq.~(\ref{eq: Q non rel}) is proportional to 
$(c_A^2 + c_V^2)_{\nu_i,\bar{\nu_i}} (\epsilon_{\nu_i} + \epsilon_{\bar{\nu}_i}) I_{\nu_i}I_{\bar{\nu}_i}$,
it is natural to assume that \citeaffixed{Goodman.etal:1987,Janka:1991,Setiawan.etal:2006}{e.g.,}:
\begin{equation}
Q_{\nu_i,\bar{\nu}_i} \approx \frac{\sigma_0 (c_A^2 + c_V^2)_{\nu_i,\bar{\nu}_i}}{96 \pi^2 c (m_e c^2)^2}
L_{\nu_i} L_{\bar{\nu}_i} 
\left[
\frac{\langle \epsilon^2_{\nu_i} \rangle}{\langle \epsilon_{\nu_i} \rangle} G_{\nu_i,\bar{\nu}_i} +
\frac{\langle \epsilon^2_{\bar{\nu}_i} \rangle}{\langle \epsilon_{\bar{\nu}_i} \rangle} H_{\nu_i,\bar{\nu}_i}
\right] \, ,
\label{eq: simple parametrization}
\end{equation}
where $G_{\nu_i,\bar{\nu}_i}$ and $H_{\nu_i,\bar{\nu}_i}$ are geometrical factors that contain 
the integral of all the residual spatial and geometrical dependencies. 
In the next sections, we will test the validity of this parametrization, based on detailed numerical
results, using both the luminosities at infinity and the cooling luminosities.

Finally, we define
$E_{\nu_i,\bar{\nu}_i}(t)$ as the deposited cumulative energy:
\begin{equation}
E_{\nu_i,\bar{\nu}_i}(t) = \int_0^t \: Q_{\nu_i,\bar{\nu}_i}(t') \, {\rm d}t' \, ,
 \label{eq: E non rel}
\end{equation}
alongside with a global annihilation efficiency, 
\begin{equation}
\bar{\eta}_{\nu,\bar{\nu}}=E_{\nu_i,\bar{\nu}_i}/\int_0^t \; (L_{\nu}+L_{\bar{\nu}}) \, {\rm d}t' \, . 
\end{equation}

\subsection{Relativistic calculation}
\label{sec: method rel}

The hydrodynamics simulation and the annihilation rate calculations presented in the previous 
sections are performed in a Newtonian framework. However, matter in the disk moves close to Keplerian, 
and typical orbital speeds are given by:
\begin{equation}
 \frac{v_{\phi}(R)}{c} \sim \sqrt{\frac{GM_{\rm NS}}{Rc^2}} \approx 0.309 
 \left( \frac{M_{\rm NS}}{2.6 M_{\odot}} \right)^{1/2}
 \left( \frac{R_{\rm NS}}{20 {\rm km}} \right)^{-1/2}
 \left( \frac{R}{2 \, R_{\rm NS}} \right)^{-1/2} \, ,
\end{equation}
where $M_{\rm NS}$ and $R_{\rm NS}$ are the MSN mass and radius, respectively. 
Moreover, the ratio between the the Schwarzschild radius $R_g$ and the actual MNS radius is
\begin{equation}
 \frac{R_g}{R_{\rm NS}} = 
 \frac{2 G M_{\rm NS}}{R_{\rm NS}c^2} \approx 0.382   
 \left( \frac{M_{\rm NS}}{2.6~M_{\odot}} \right)
 \left( \frac{R_{\rm NS}}{20~{\rm km}} \right)^{-1}.
\end{equation}
Thus, relativistic effects due to the fast motion of matter and to the intense gravitational field
are potentially relevant. A consistent relativistic calculation requires to
start from the results of relativistic hydrodynamical simulations of a neutron star merger, and to 
compute the annihilation rates in a relativistic framework. Due to the Newtonian character of
our input simulations, this is not entirely possible. However, we can compute the impact of special and
general relativistic effects on the neutrino propagation and on the annihilation process using
our Newtonian input data.
We consider the following effects on the propagation from the emission to the interaction
points: the Doppler and beaming effects due to 
the fast motion of the emitting fluid elements, and the redshift 
and light bending effects due to motion of the radiation in the gravitational potential.
A detailed explanation of the adopted spacetime model, as well as of the implementation of the 
relativistic effects is presented in \ref{apx: implementation rel effects}.

Following \citeasnoun{Birkl.etal:2007} and \citeasnoun{Zalamea.Beloborodov:2011}, we compute the volume-integrated energy
deposition rates in the local frames
\begin{equation}
 Q^{{\rm rel}}_{\nu_i,\bar{\nu}_i}(t) = 
 \int_{\tilde{V}} \: q_{\nu_i,\bar{\nu}_i}(t,\mathbf{x}) \, \sqrt{{\rm det}(g_{ab})} \, {\rm d}r \, {\rm d}\theta \, {\rm d} \phi \, ,
 \label{eqn: relativisitic local volume integrated energy rate}
\end{equation}
as well as the energy deposition rates measured by an infinitely distant observer:
\begin{equation}
 Q^{{\rm rel},\infty}_{\nu_i,\bar{\nu}_i}(t) = 
 \int_{\tilde{V}} \: q_{\nu_i,\bar{\nu}_i}(t,\mathbf{x}) \, \sqrt{{\rm det}(g_{\mu \nu})} \, {\rm d}r \, {\rm d}\theta \, {\rm d} \phi \, ,
 \label{eqn: relativisitic at infinity volume integrated energy rate}
\end{equation}
where the redshift due to the gravitational field is also included\footnote{Note that Eqs.~(\ref{eqn: relativisitic local volume integrated energy rate})
and (\ref{eqn: relativisitic at infinity volume integrated energy rate}) differ by the indexes 
of the metric tensor: in the former expression only the spatial
part of $g_{\mu \nu}$ is considered, while the full metric is used in the latter.}.
Each of these two rates can be integrated over time to obtain the cumulative energies, 
$ E^{\rm rel}_{\nu_i,\bar{\nu}_i}(t)$
and $E^{\rm rel,\infty}_{\nu_i,\bar{\nu}_i}(t)$.

\subsection{Numerical setup}

We perform the calculation of the annihilation rates for twenty equally spaced time steps, ranging from 0 
to 380~ms.
Since the input data are cylindrically symmetric, we expect also $q_{\nu_i,\bar{\nu}_i}$ and $\mathbf{p}_{\nu_i,\bar{\nu}_i}$
to be axisymmetric. Thus, we compute the rates only on the $\phi = 11 \pi/45$ plane and we assume them to be valid for
any polar angle $\phi$ .
For the discretization of the $\phi$ angle required by our calculations, we choose $N_{\phi}=45$
($\Delta \phi = 8$\textdegree) and we test the convergence of our results with respect 
to this parameter ($\delta Q_{\nu_i,\bar{\nu_i}} \lesssim 3 \%$ for $N_{\phi}=60$).
To save computing time, we consider the emission of neutrinos only from cells with 
$\rho \geq 2 \times 10^{9}~{\rm g~cm^{-3}}$. Also in this case, this simplification
introduces an error lower, at most, than
a few percents.

\section{Energy and momentum: deposition rates}
\label{sec: results}

\begin{figure*}
 \centering
 \includegraphics[width=0.48 \linewidth]{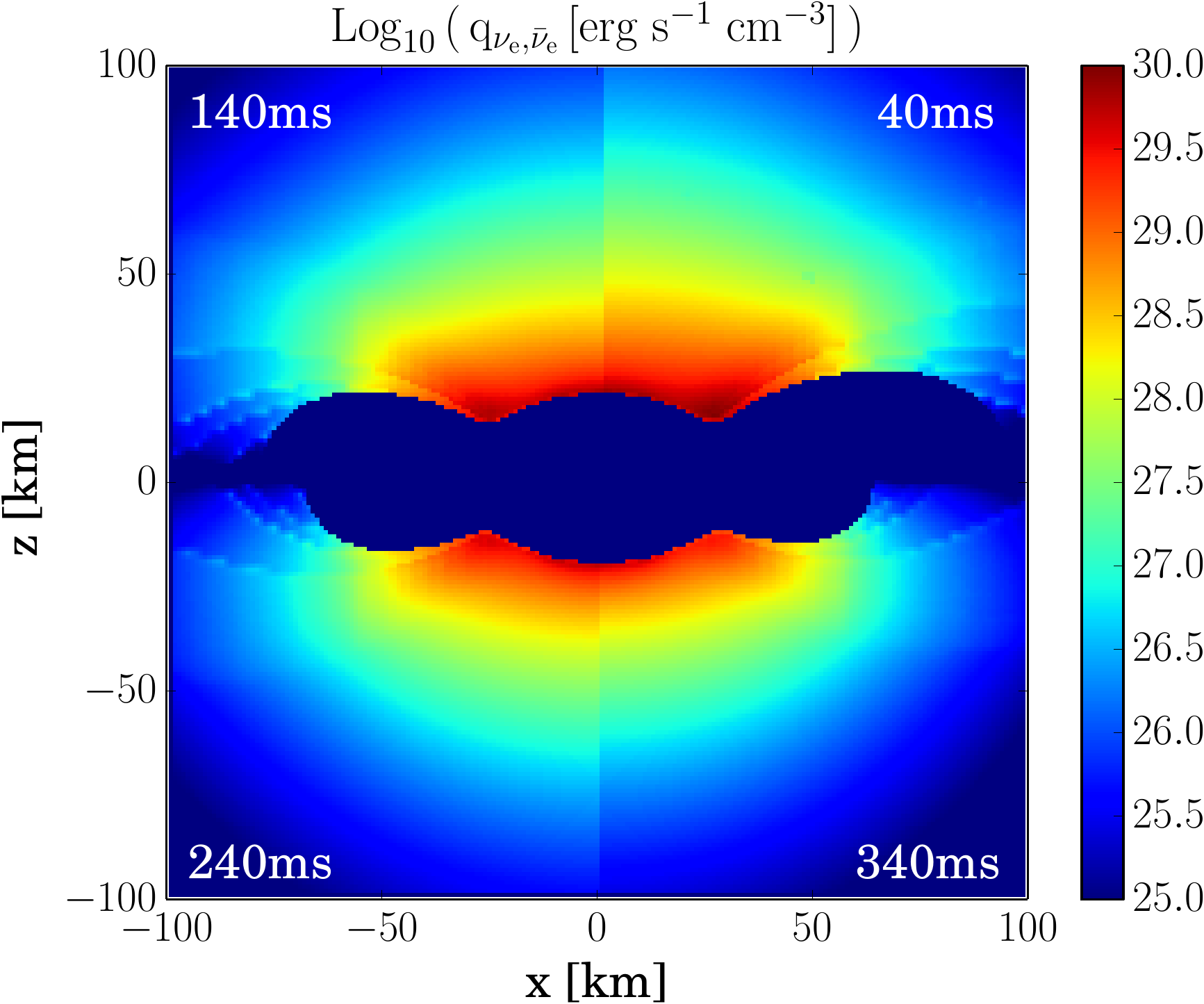}
 \hspace{0.1cm}
 \includegraphics[width=0.48 \linewidth]{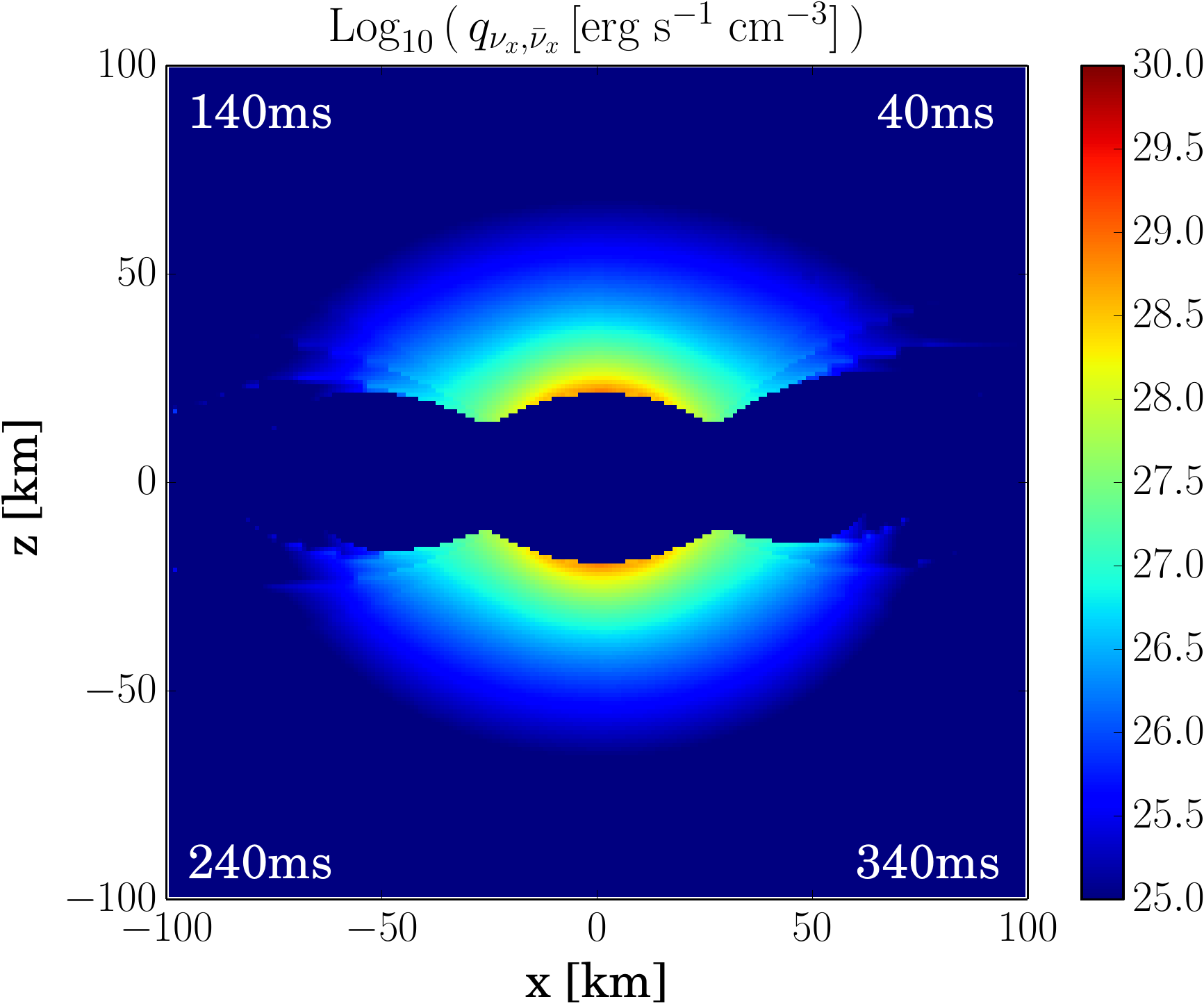}
 \caption{Color coded is the local energy deposition rate $q_{\nu,\bar{\nu}}$, as a function of 
 the cylindrical coordinates, in the $\phi=11 \pi/45$ plane and in the 
 intervals $[0,100]~{\rm km} \times [0,100]~{\rm km}$.
 In the left panel we plot the rate for $\nu_e,\bar{\nu}_e$, in the right one for $\nu_x,\bar{\nu}_x$.
 The four quadrants refer to different times, as indicated by the labels.
 If $\rho > 10^{11}{\rm g \, cm^{-3}}$ we set the rates to the minimum (blue area).}
 \label{fig: anni rate different timesteps}
\end{figure*}

In Figure \ref{fig: anni rate different timesteps}, we show
the local energy deposition rate $q_{\nu_i,\bar{\nu}_i}(\mathbf{x},t)$ at four different times, 
both for electron flavor (left panel) and heavy flavor (right panel) neutrino pairs. These rates
have been obtained using our standard (i.e., non-relativistic) approach. We notice that 
in both cases and at each time the energy deposition rate is more intense close to the remnant 
and decreases for increasing distances, as a consequence of the decrease of the neutrino intensity. 
Moreover, even if the rates decrease with time
due to the decline of the neutrino luminosities, their spatial distribution remains qualitatively similar
during the evolution of the system. However,
the results obtained for the different flavors differ significantly. In the $\nu_e$-$\bar{\nu}_e$ case, the
energy deposition is more intense and distributed over a wider volume, including the regions above 
the MNS and the innermost part of the disk.
In the $\nu_x$-$\bar{\nu}_x$ case, the energy deposition is
concentrated above the MNS and the decrease of
$q_{\nu_x,\bar{\nu}_x}(\mathbf{x},t)$ with time is less pronounced, due to the milder reduction 
of the $\nu_x$ luminosities.

\begin{figure}
 \centering
 \includegraphics[width=0.48 \linewidth]{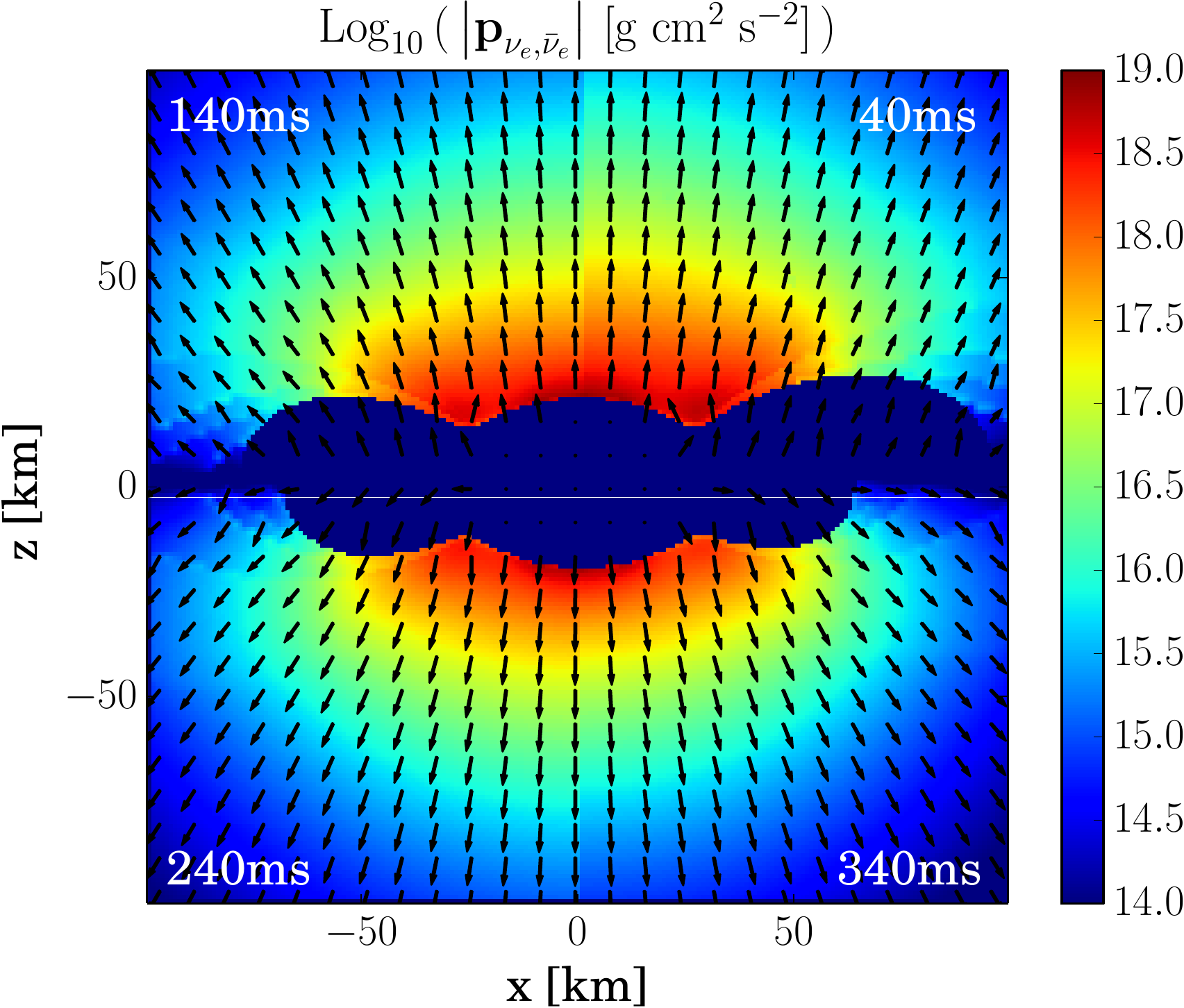}
 \hspace{0.1cm}
 \includegraphics[width=0.48 \linewidth]{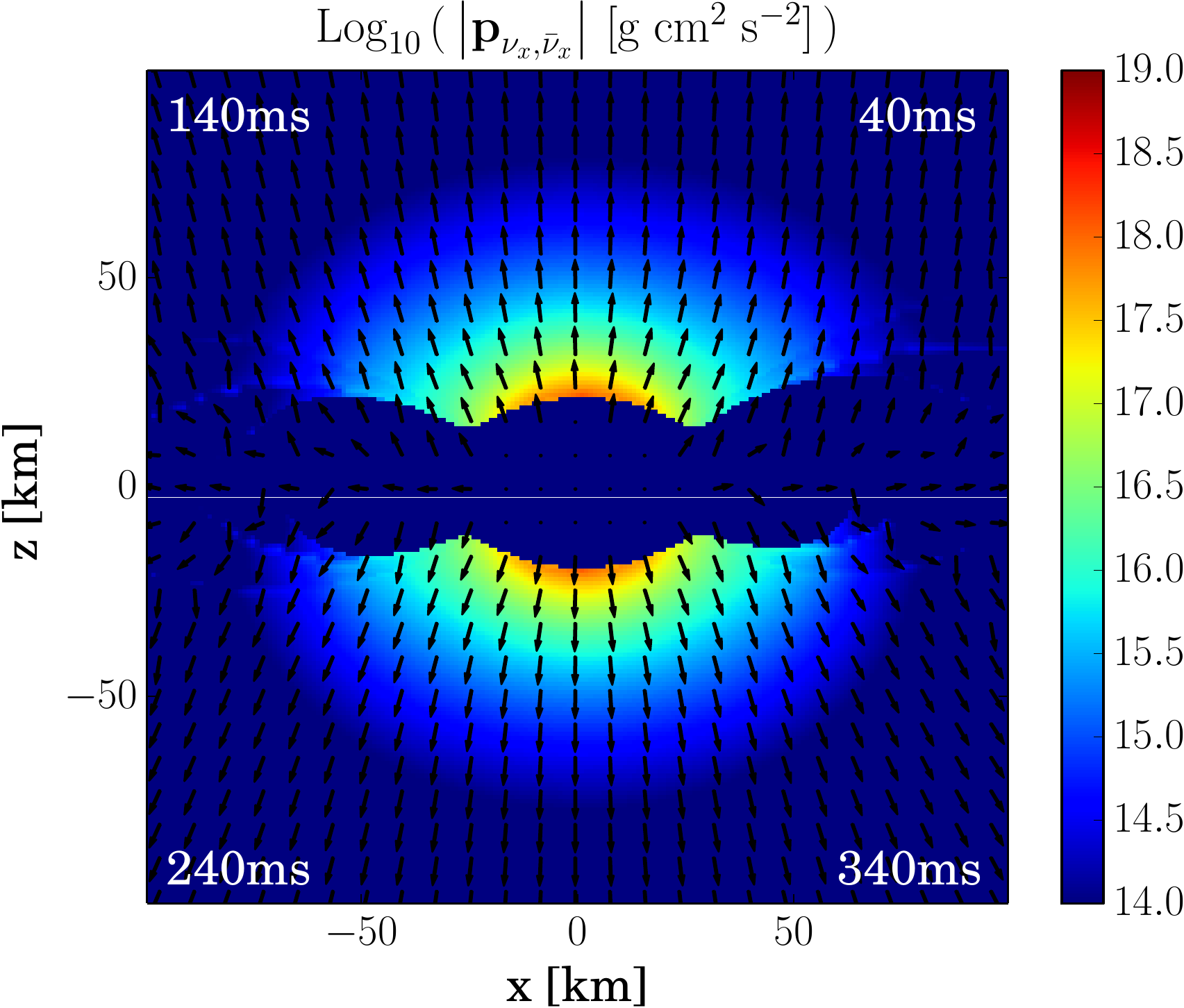}
 \caption{Same as in Figure \ref{fig: anni rate different timesteps}, but for the
 momentum deposition rate $\mathbf{p}_{\nu,\bar{\nu}}$. The color code indicate the modulus 
 of the vector, $|\mathbf{p}_{\nu,\bar{\nu}}|$, while the arrow the direction projected on
 the plane. Due to the symmetry of the neutrino sources, 
 $\left(p_{\nu,\bar{\nu}} \right)_{\phi} \approx 0 $.}
 \label{fig: anni momentum rate different timesteps}
\end{figure}

\begin{figure}
 \centering
 \includegraphics[width=0.48 \linewidth]{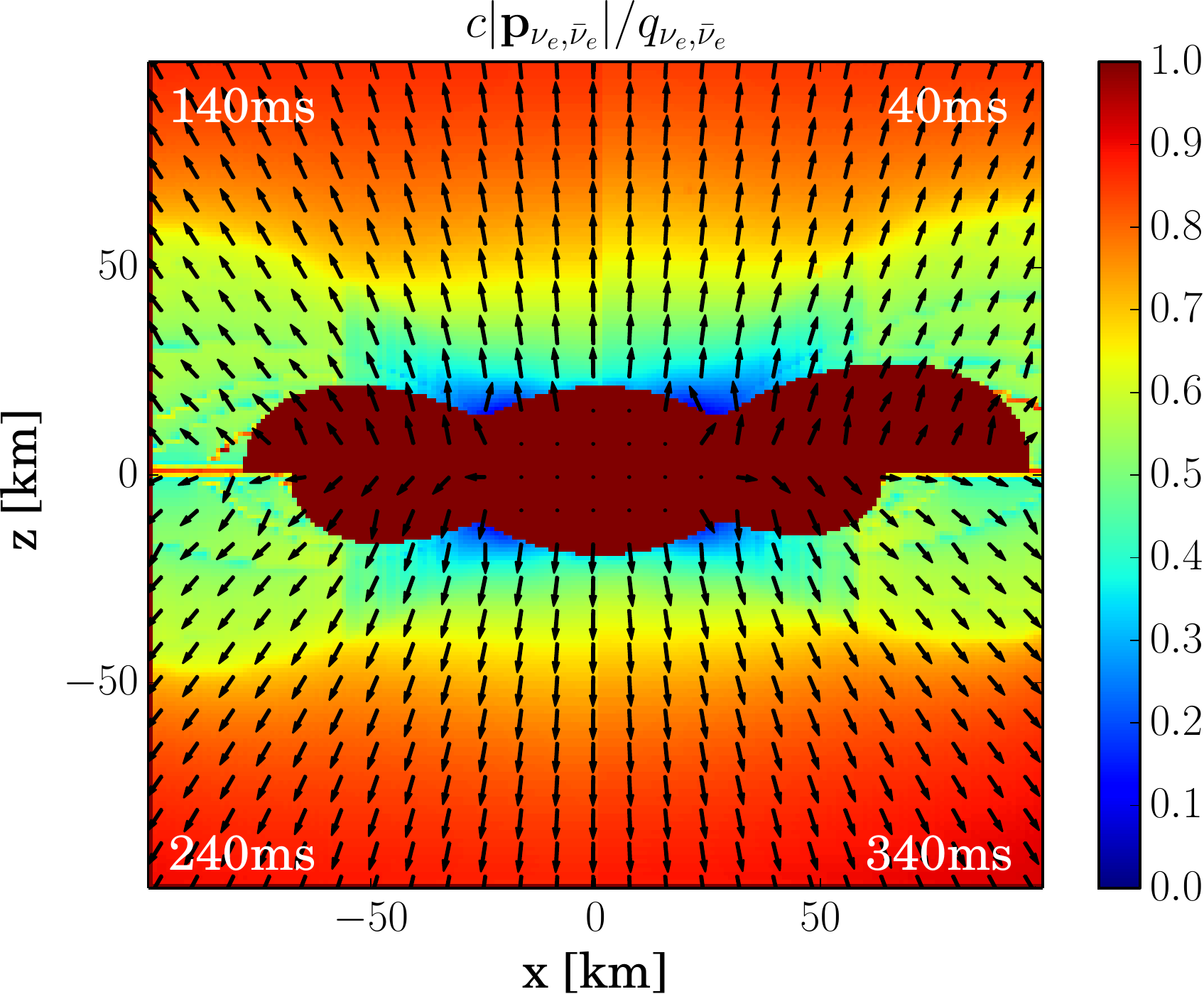}
 \hspace{0.1cm} 
 \includegraphics[width=0.48 \linewidth]{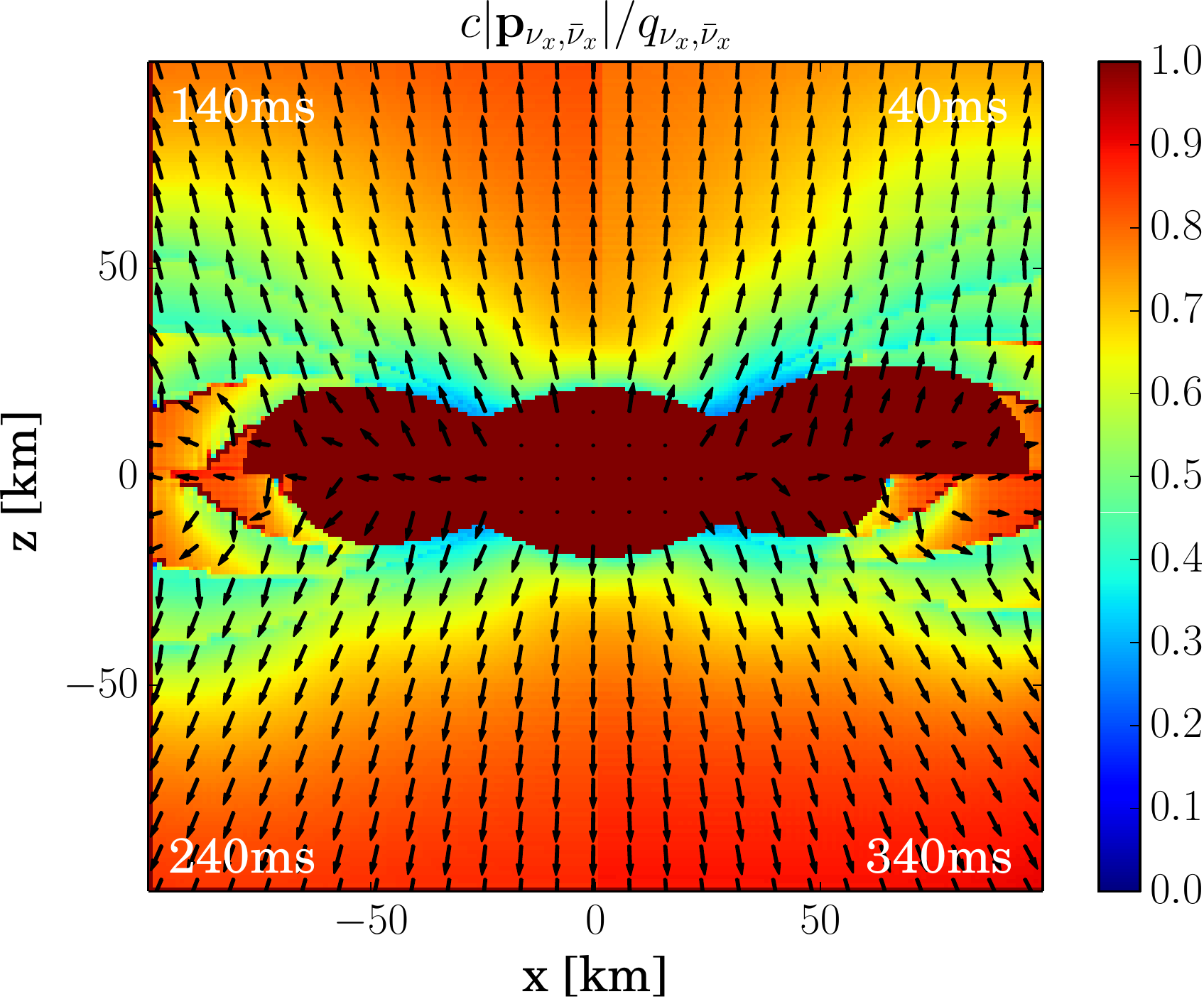}
 \caption{Same as in Figure \ref{fig: anni momentum rate different timesteps}, but for the ratio
 between the modulus of the momentum deposition rate, $ | \mathbf{p}_{\nu,\bar{\nu}} |$, times $c$
 and the energy deposition rate, $ q_{\nu,\bar{\nu}}$. If $\rho > 10^{11}{\rm g \, cm^{-3}}$ we set the ratio 
 to the maximum (red area).}
 \label{fig: anni rate momentum ratio different timesteps}
\end{figure}

The deposition rate for the linear momentum is presented in 
Figure~\ref{fig: anni momentum rate different timesteps}, again for four different times.
The arrows indicate the direction of $\mathbf{p}_{\nu,\bar{\nu}}$ projected on the vertical plane.
Due to the axisymmetry of the neutrino emissivities, we expect the azimuthal component of the deposited
momentum to be negligible. The geometry of the MNS and of the disk influences the emission direction of the
neutrinos and, in turn, the direction of $\mathbf{p}_{\nu,\bar{\nu}}$, according to Eq.~(\ref{eq: dessart momentum}). 
In particular, the MNS produces an almost spherical emission 
inside the funnel above its surface. This combines with the axisymmetric emission from the 
disk. The latter happens predominantly at the transition between the MNS and the disk, 
where neutrinos are emitted towards the funnel and, on average, perpendicularly to 
the neutrino surfaces (see Figure \ref{fig: nu_surfaces}).
At larger radii, the average emission direction from the disk becomes parallel to the $z$ axis
or even pointing outwards. 
Since in the case of $\nu_x$ the neutrino emission comes mainly from the MNS, 
$\mathbf{p}_{\nu_x,\bar{\nu}_x}$ presents a more radial structure.
We notice that at larger times, when the disk has shrunk its radius 
and consumed its mass, $\mathbf{p}_{\nu_e,\bar{\nu}_e}$ also develops into a more radial configuration.

A comparison between the energy and momentum deposition reveals that the intensity of the latter
tracks the one of the former. However, the dependence on the colliding angle $\Phi$ contained in
Eqs.~(\ref{eq: dessart}) and (\ref{eq: dessart momentum}), and the vector sum that characterizes
the momentum deposition imply that $c |\mathbf{p}_{\nu,\bar{\nu}}| < q_{\nu,\bar{\nu}}$.
In particular, the more isotropic and symmetric the local neutrino distribution, the less efficient is the momentum
deposition compared to the energy one. In Figure~\ref{fig: anni rate momentum ratio different timesteps}, we present 
$c |\mathbf{p}_{\nu,\bar{\nu}}| / q_{\nu,\bar{\nu}}$ and confirm that close to the MSN and to 
the disk, where the propagation directions are expected to be more isotropic, the ratio is significantly
smaller than 1, reaching even 0.2, while it increases up to 0.8 at larger distances.

\begin{figure}
 \centering
 \includegraphics[width=0.48 \linewidth]{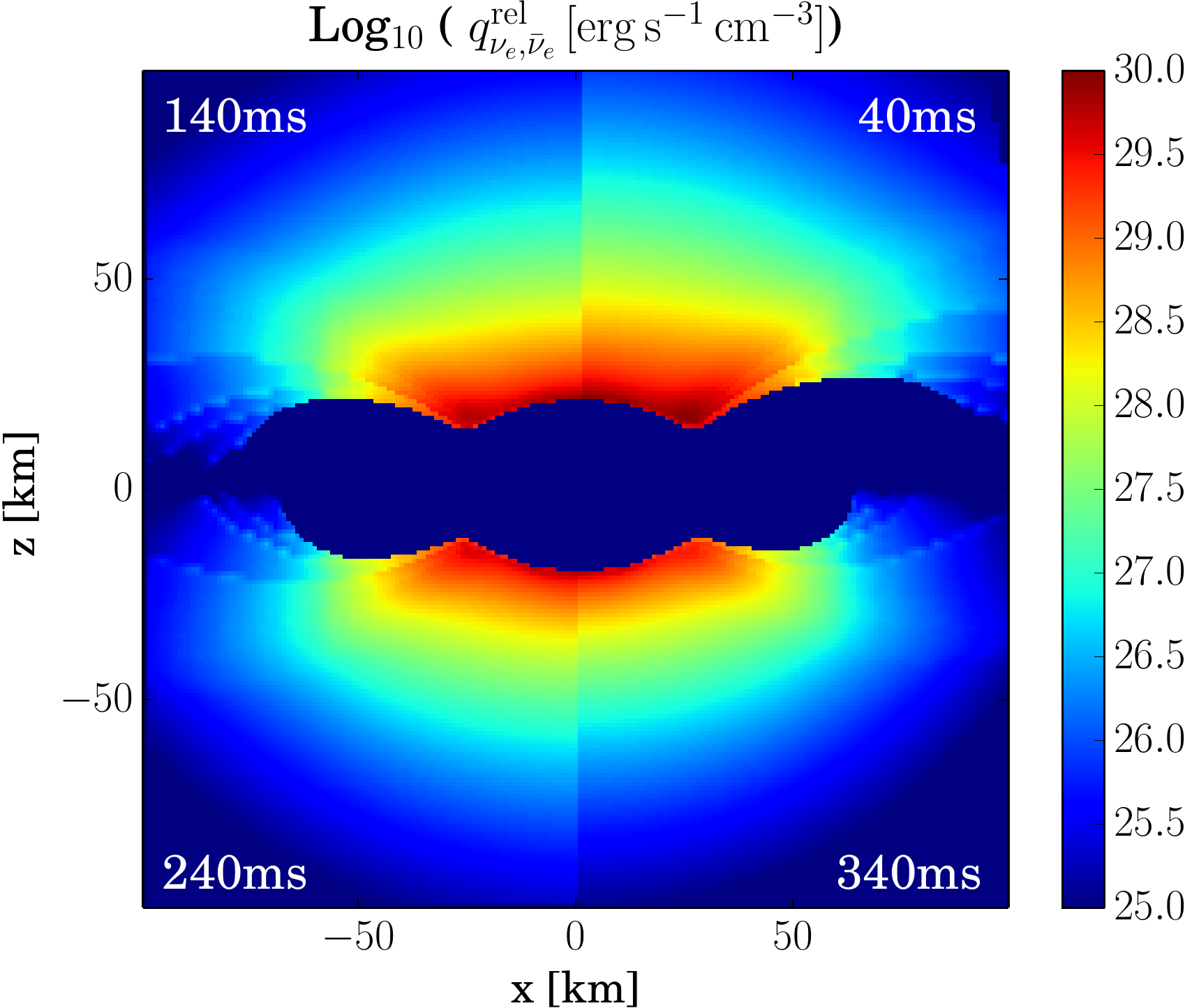}
 \hspace{0.1cm}
 \includegraphics[width=0.48 \linewidth]{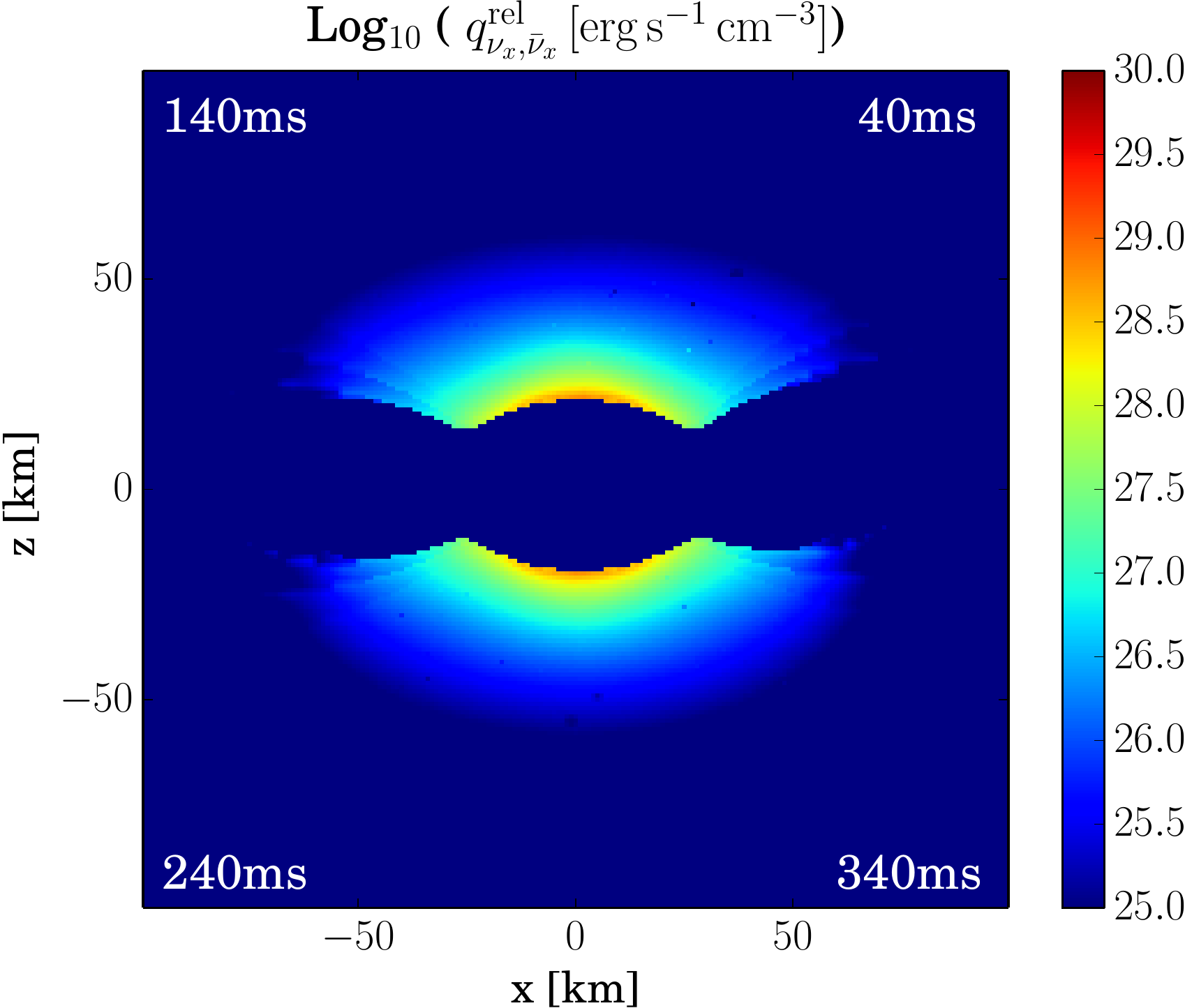}
 \caption{Same as in Figure \ref{fig: anni rate different timesteps}, but including the special and
 general relativistic effects for the neutrino propagation outside the neutrino surfaces.}
 \label{fig: anni rate with GR different timesteps}
\end{figure}

\begin{figure}
 \centering
 \includegraphics[width=0.48 \linewidth]{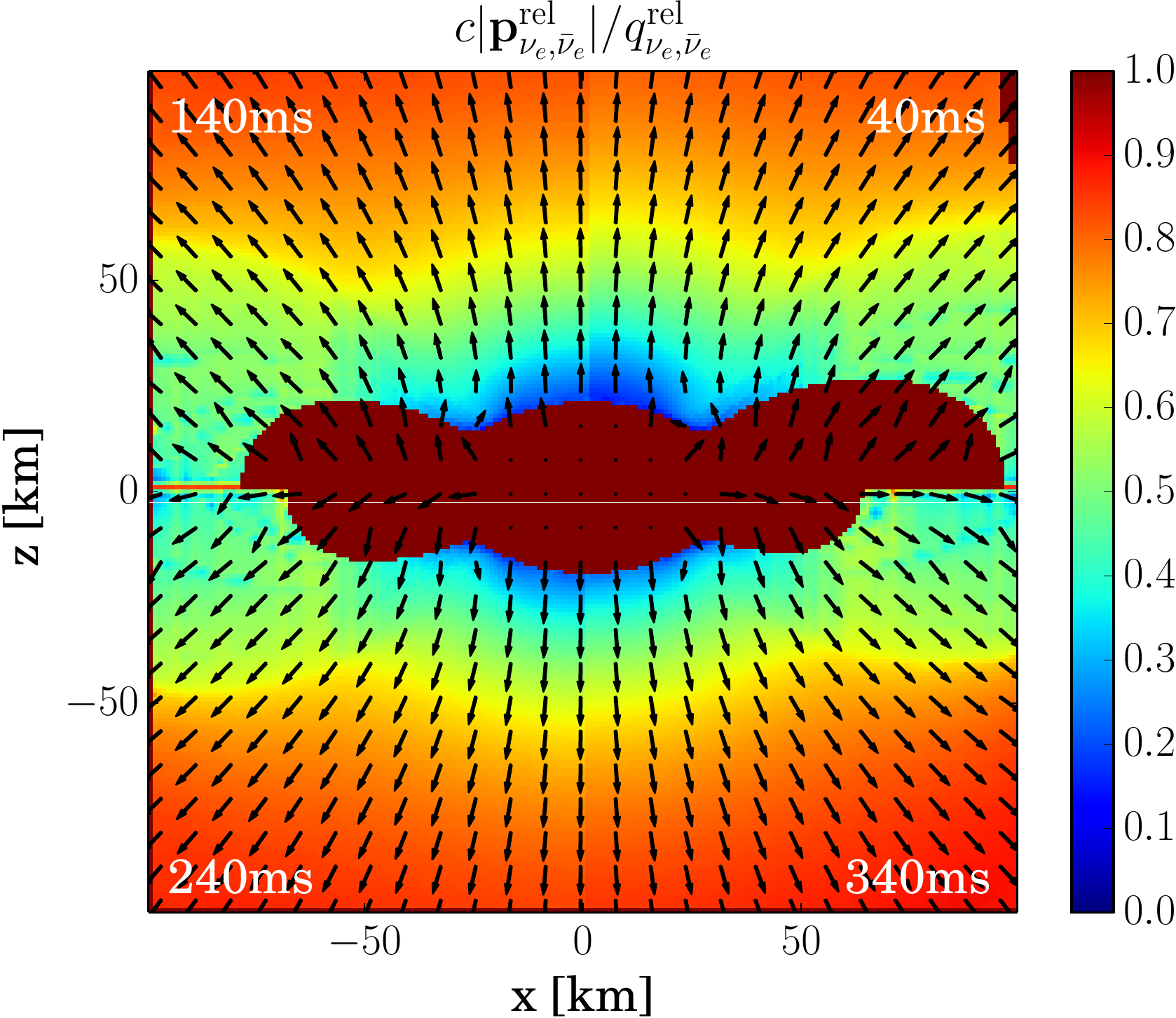}
 \hspace{0.1cm}
 \includegraphics[width=0.48 \linewidth]{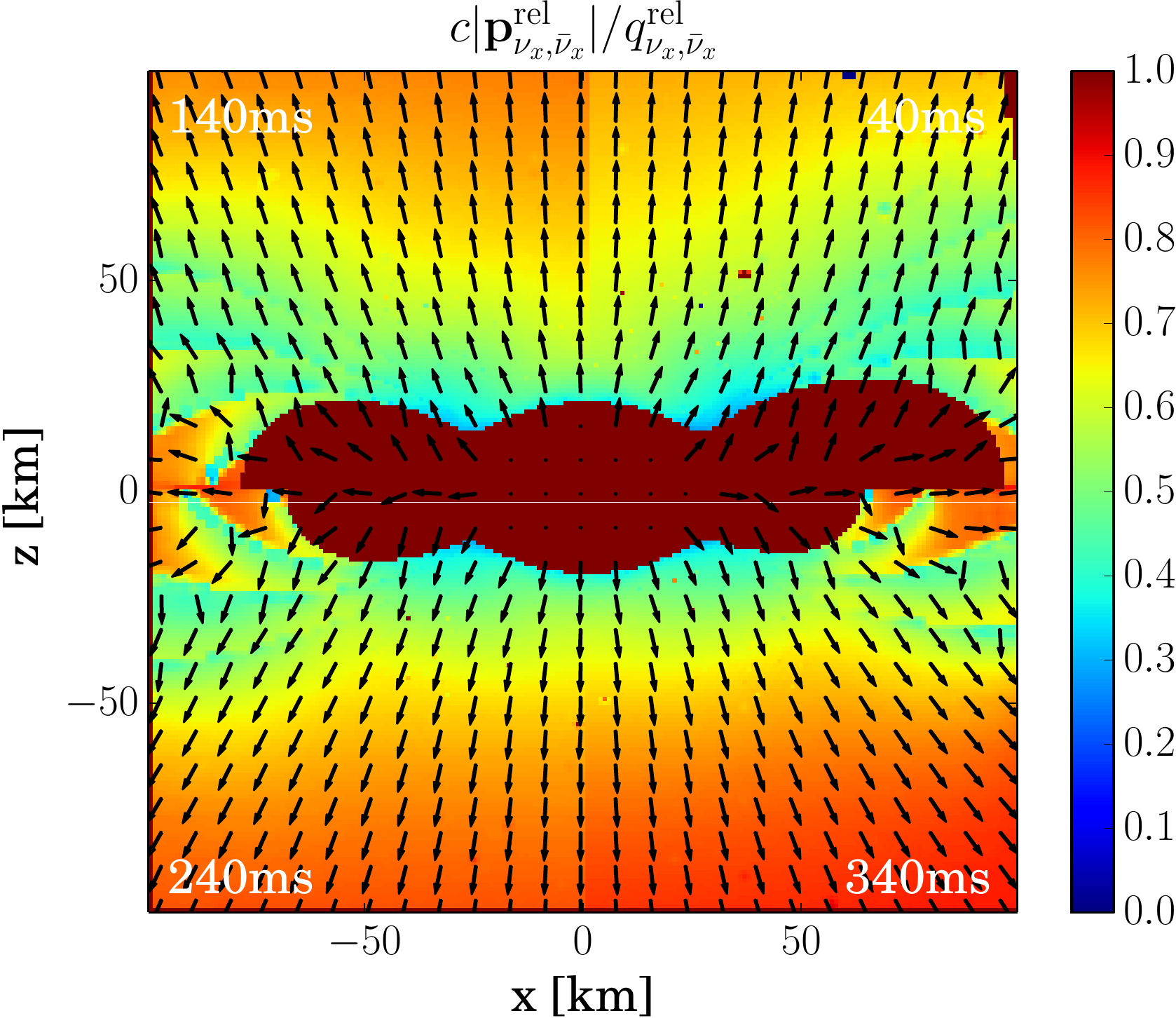}
 \caption{Same as in Figure \ref{fig: anni rate momentum ratio different timesteps}
 but including the special and general relativistic effects for the neutrino propagation 
 outside the neutrino surfaces.}
 \label{fig: anni rate momentum ratio with GR different timesteps}
\end{figure}

We repeat the pair annihilation calculations including special and general relativistic effects, and we present
the results for the local energy deposition rate in 
Figure~\ref{fig: anni rate with GR different timesteps}. The inclusion of the relativistic
effects does not change the qualitative features of the energy deposition rate.
A careful comparison between the energy deposition rates provided by $\nu_e,\bar{\nu}_e$ pairs 
shows that the light bending and the gravitational blueshift increase the annihilation
rates immediately above the MNS (by $\lesssim 20 \%$), while the gravitational redshift, 
together with the beaming of the radiation emitted by matter rotating 
inside the disk, reduces the energy deposition rate in the funnel and above the innermost part of the 
disk. Where the rates are larger, the reduction is usually $\sim 20\%$, but it increases when the distance
from the MNS and the disk increases, as well as at later times. 
In the case of $\nu_x,\bar{\nu}_x$ pairs, the dominant
effect is provided by the gravitational redshift and the result is a reduction 
of the energy deposition rate everywhere in the computational domain.
The gain in annihilation efficiency due to the light bending is attenuated by
the dominant contribution of the MNS to the $\nu_x$ luminosities.

The qualitative features of the momentum deposition rate computed by taking into account the 
relativistic effects in the neutrino propagation are very similar to the ones obtained in
our reference Newtonian calculations. In Figure \ref{fig: anni rate momentum ratio with GR different timesteps},
we present the ratio between the modulus of the momentum and the energy deposition rates. A comparison
with the analogue figure obtained in the Newtonian framework, Figure \ref{fig: anni rate momentum ratio different timesteps}, 
shows that the light bending reduces the efficiency at which a net momentum is deposited more significantly, 
with respect to the energy deposition rate, especially close to the MNS surface and in the funnel above it.

\begin{figure}
 \centering
 \includegraphics[width=0.49 \linewidth]{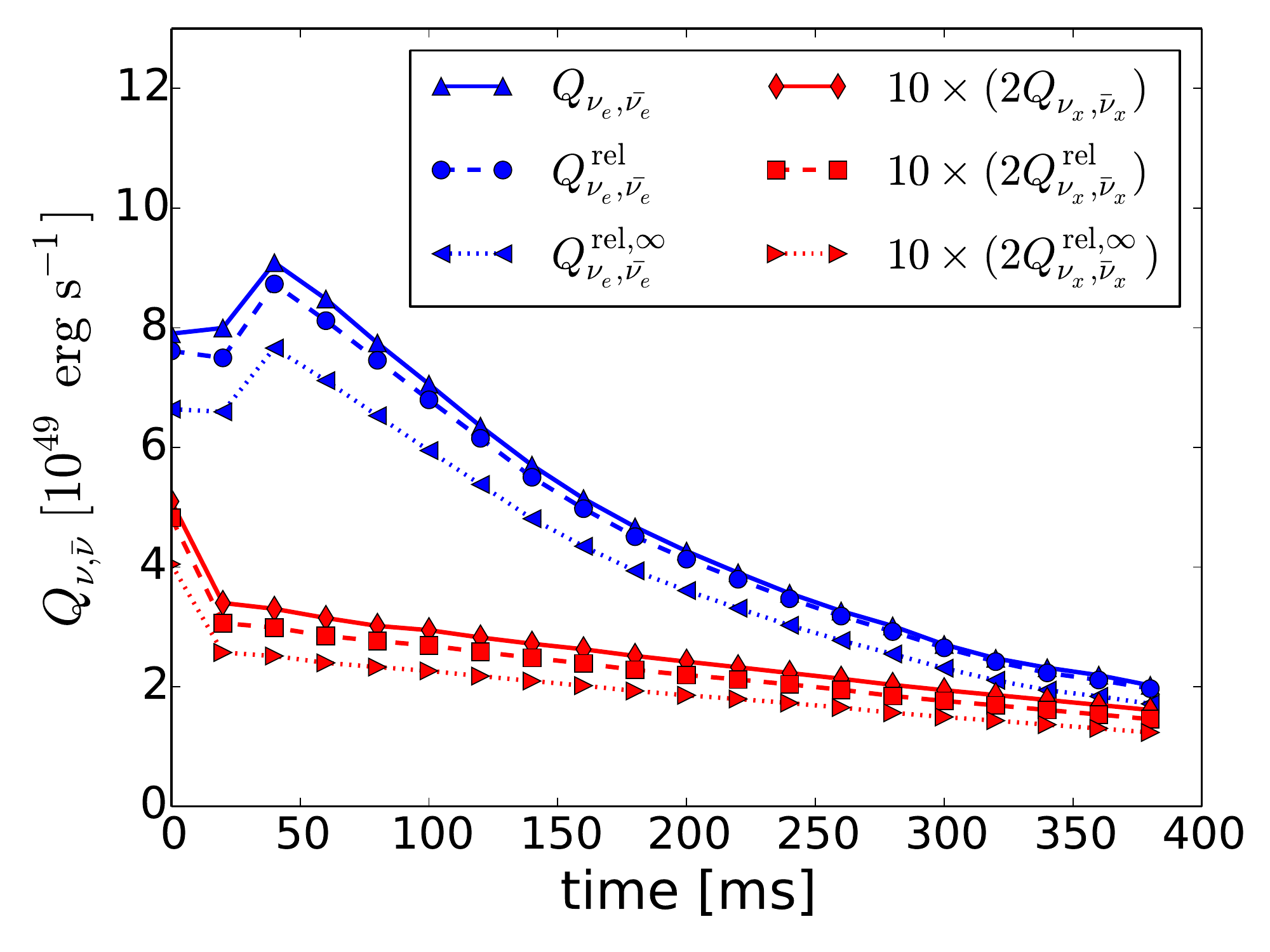}
 \includegraphics[width=0.49 \linewidth]{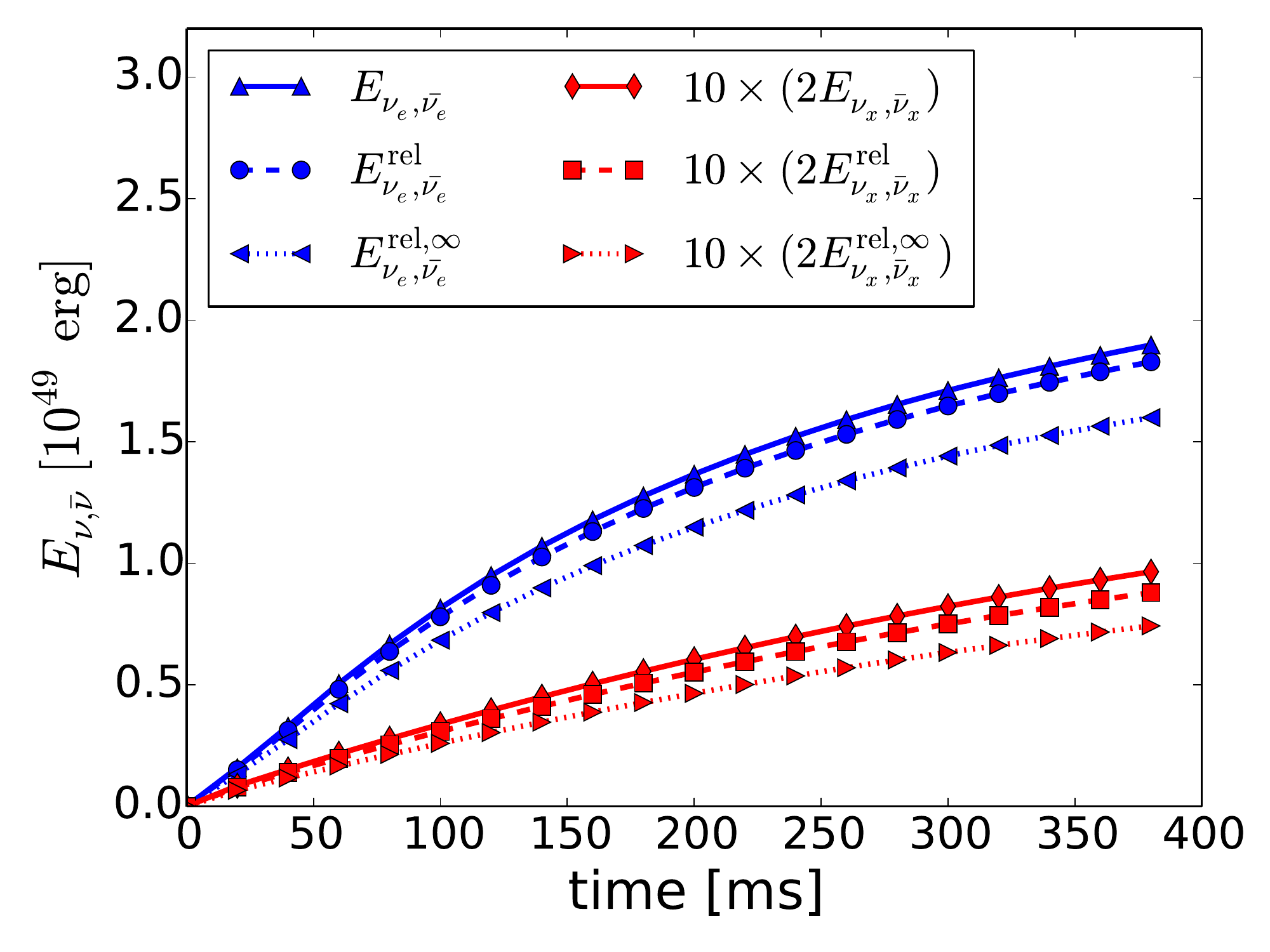}
 \caption{Volume-integrated energy rate (left) and cumulative deposited energy (right) 
 by neutrino pair annihilation, for our reference Newtonian calculations ($Q_{\nu,\bar{\nu}}$, $E_{\nu,\bar{\nu}}$)
 and for the calculations including relativistic effect in the neutrino propagation 
 (local quantities, $Q^{\rm rel}_{\nu,\bar{\nu}}$ and $E^{\rm rel}_{\nu,\bar{\nu}}$, and quantities measure by a far observer
 $Q^{\rm rel,\infty}_{\nu,\bar{\nu}}$ and $E^{\rm rel,\infty}_{\nu,\bar{\nu}}$).
 Blue and red lines refer to electron and heavy flavor neutrinos, respectively. The factor 2 in $Q_{\nu_x,\bar{\nu}_x}$
 and $E_{\nu_x,\bar{\nu}_x}$ takes into account the $\mu$ and $\tau$ flavors, while the factor 10 is only a magnification factor.}
 \label{fig: anni rate integrated}
\end{figure}

In Figure \ref{fig: anni rate integrated}, we present the volume integrated rates, 
$Q_{\nu_i,\bar{\nu}_i}(t)$ (left panel) and 
cumulative deposited energy, $E_{\nu_i,\bar{\nu}_i}(t)$ (right panel), for both the reference Newtonian
calculations, Eqs.~(\ref{eq: Q non rel})-(\ref{eq: E non rel}), and including relativistic corrections, 
(\ref{eqn: relativisitic local volume integrated energy rate}). 
We first notice that, for the volume element in our cylindrical discretization, one has
\begin{equation}
{\rm d}V \propto R~\Delta R~\Delta z \, .
\label{eq: volume element dependences}
\end{equation}
Thus, intense deposition regions located at larger cylindrical radii contribute more significantly than close
to the rotational axis. The integration over the volume allows 
to quantify the difference
between the different neutrino flavors: $Q_{\nu_e,\bar{\nu}_e}(t)$ is $\sim$~60 times larger than
$Q_{\nu_x,\bar{\nu}_x}(t)$. 
Due to this large difference, the contribution of $\nu_x$'s is significantly sub-dominant for the
overall energy deposition process. 
The approximated expression (\ref{eq: simple parametrization}) accounts for this considerable disparity.
In particular, comparing the electron and the heavy flavors,
$(c_A^2 + c_V^2)_{\nu_e,\bar{\nu}_e}/(c_A^2 + c_V^2)_{\nu_x,\bar{\nu}_x} \approx 4.6$ 
and $L_{\nu_e} L_{\bar{\nu}_e} / (L_{\nu_x} L_{\bar{\nu}_x}) \approx 12$.
The inclusion of relativistic effects in the neutrino propagation changes only marginally the
intensity of the energy deposition rate. In particular, the energy blueshift and light bending
happening close to the MNS are globally compensated by the redshift and the light beaming in the funnel
and above the disk. Overall, the integrated energy deposition measured by local observers differs 
only by a few percent in the Newtonian and in the relativistic case. The difference becomes more 
relevant ($\sim 15\%$) only when $Q^{{\rm rel},\infty}_{\nu_i,\bar{\nu}_i}$ and 
$E^{{\rm rel},\infty}_{\nu_i,\bar{\nu}_i}$ are considered: this is due to the inclusion of the 
gravitational redshift required for the deposited energy to travel from the deposition site to
infinity.

The volume-integrated energy deposition rate for $\nu_e$ and $\bar{\nu}_e$ ranges between 
$9 \times 10^{49}{\rm erg \, s^{-1}}$ and $2 \times 10^{49}{\rm erg \, s^{-1}}$, with a clear
decreasing trend with time.
The total amount of energy deposited by
neutrino pair annihilations in the funnel above the merger remnant reaches $1.95 \times 10^{49}$~erg 
at 380~ms in our Newtonian calculation, while it reduces to $1.6 \times 10^{49}$~erg in 
the relativistic calculations when the 
energy at infinity is considered. We notice that, even if the deposition rate has significantly 
decreased at the end of our simulation, this energy has not saturated yet.
This is due to the longer time scales of neutrino diffusion from the MNS, 
$t_{\rm cool} \sim 0.7 \, {\rm s}$, and of disk consumption, 
$t_{\rm disc} \sim 0.5 \, {\rm s}$ \cite{Perego.etal:2014b}.

\section{Role of the MNS}
\label{sec: roleMNS}
In order to investigate the role of the MNS in the process of neutrino
pair annihilation, we decompose the local emissivity $\eta_{\nu}$ into
a contribution from the disk ($\eta_{\nu,{\rm disk}} $) and one from the
neutron star ($\eta_{\nu,{\rm NS}}$). All neutrinos that originate
inside the spheroid $\mathcal{S}$ are tagged as ``neutron star (NS)
neutrinos'', while all the others are identified as ``disk (DS)
neutrinos''. We notice that this distinction refers to the origin
place, but not necessarily to the emission place.  While all DS
neutrinos are emitted from the disk region, a fraction of the NS
neutrinos can diffuse out from the disk before being emitted. This
happens especially for those neutrinos emitted by the MNS along the
equator.

\subsection{Neutrino luminosity and accretion rate}
\label{sec: lum_acc}

In the case of an accretion disk around a BH, the disk luminosity can
be related to the accretion rate at the BH horizon, $L \approx
\eta_{\rm acc} \dot{M}c^2$, where $\eta_{\rm acc}$ is a coefficient
that quantifies the efficiency at which the rest mass of the accreting
matter is converted into emitted radiation (typically, $\eta_{\rm acc}
\lesssim 0.05 $, \citeasnoun{Just.etal:2016}). Here we present
the behavior of the luminosity and accretion rate based on our
simulation and discuss the significant differences due to the presence
of a MNS.

\begin{figure}
 \includegraphics[width=0.49 \linewidth]{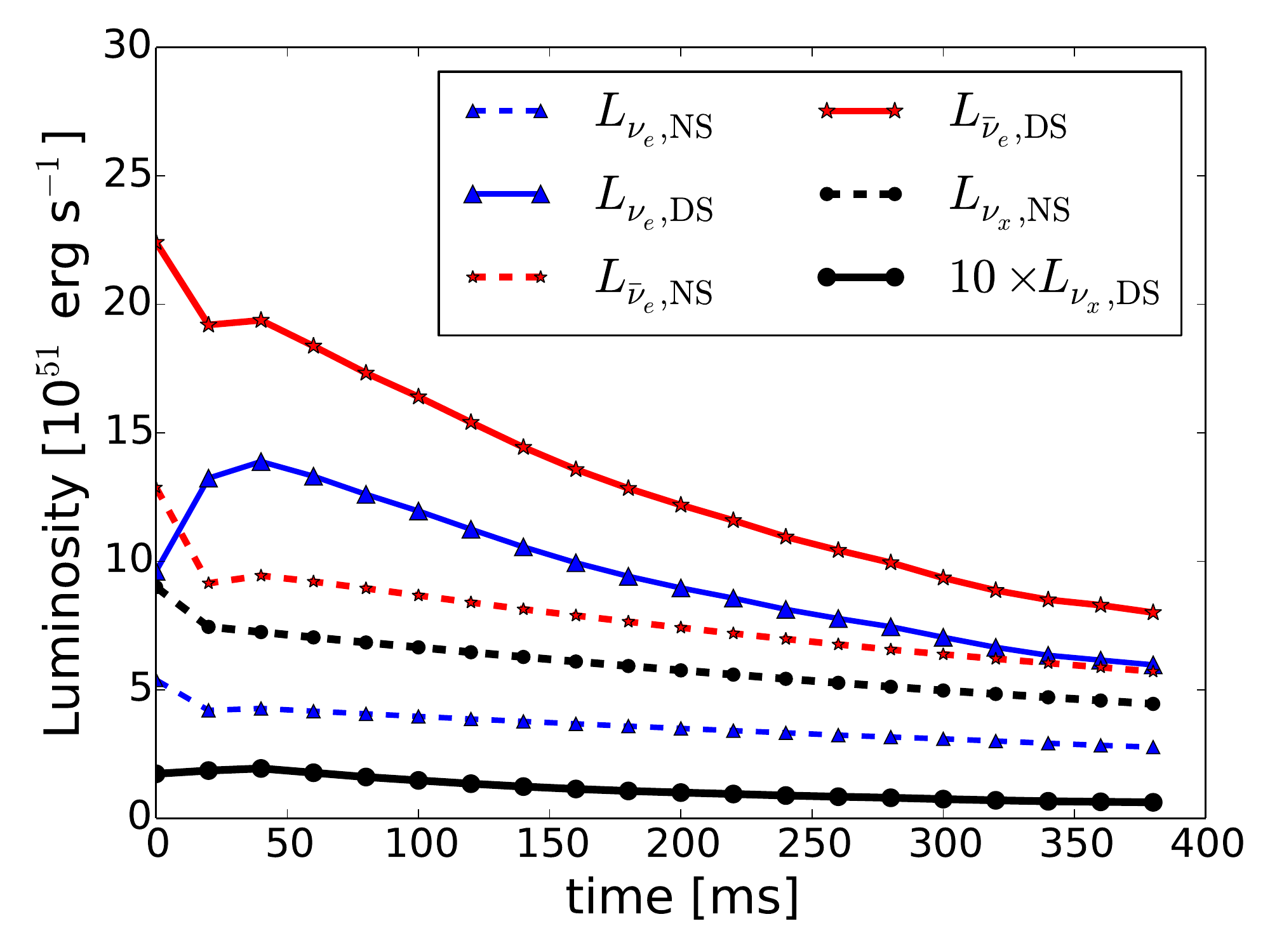}
 \includegraphics[width=0.49 \linewidth]{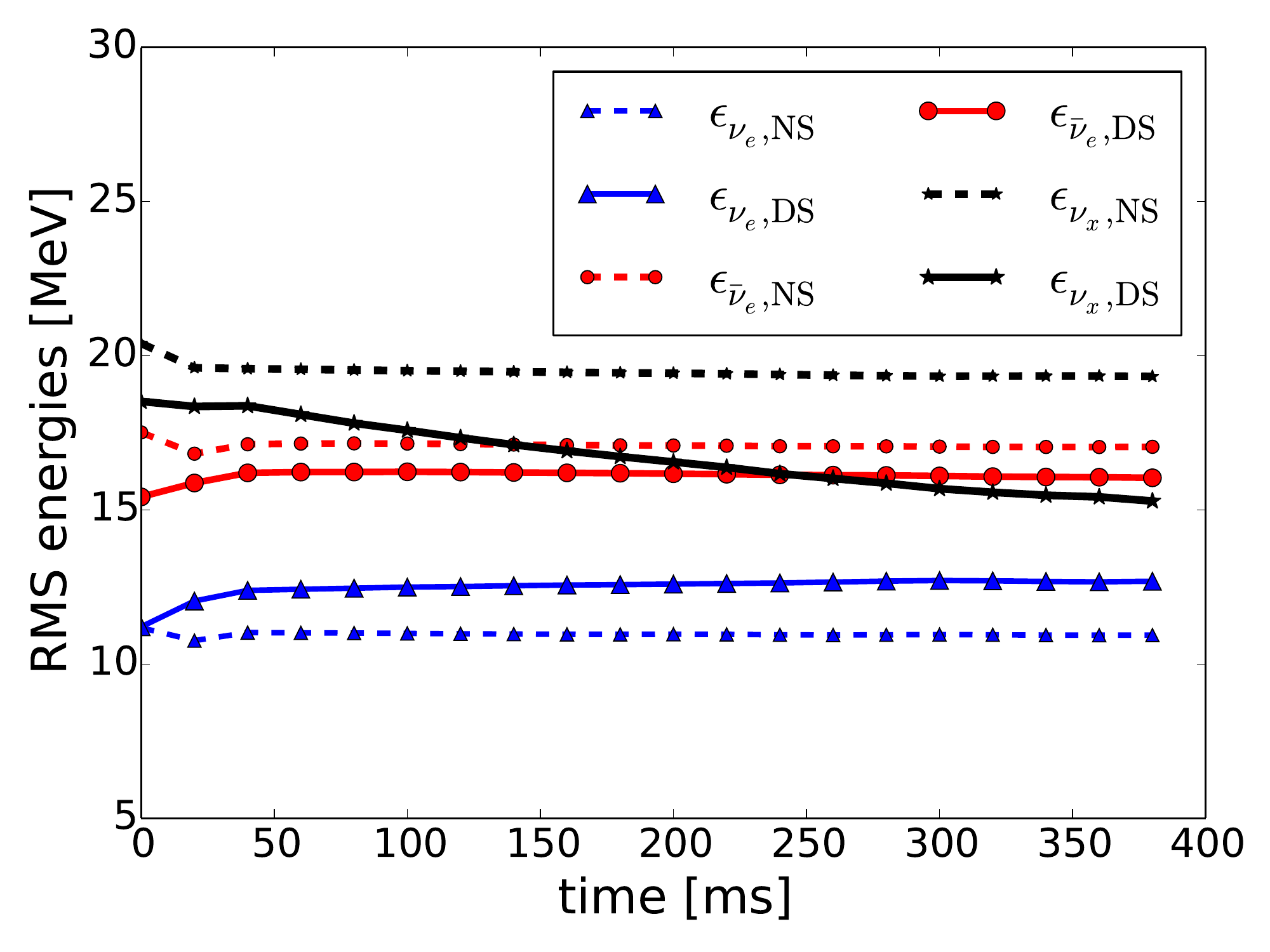}
 \caption{Disk (solid lines) and neutron star (dashed lines) contributions to the total luminosities 
 (left) and RMS energies (right) obtained from our simulation of the aftermath of a binary neutron 
 star merger.
 Blue lines  (triangles) refer to $\nu_e$, red lines (stars) to $\bar{\nu}_e$,
 black lines (circles) to $\nu_x$. All quantities are measured at infinity. For this figure
 we used only the luminosities including the absorption of neutrinos in optically thin conditions.}
 \label{fig: decomposed neutrino luminosities}
\end{figure}

The distinction between neutrinos coming from the MNS and from the disk allows to decompose 
each neutrino luminosity into a MNS and a disk contribution.
In Figure \ref{fig: decomposed neutrino luminosities}, we show for each independent species 
the disk and the MNS components. The MNS contributions originate mainly from the diffusion of neutrinos
from the hot and dense central remnant. Their decrease reflects the progressive cooling of the MNS and of
the matter accreting on its surface. The larger disk components are more related to the accretion process and 
present a more pronounced reduction. 

\begin{figure}
 \includegraphics[width=0.49 \linewidth]{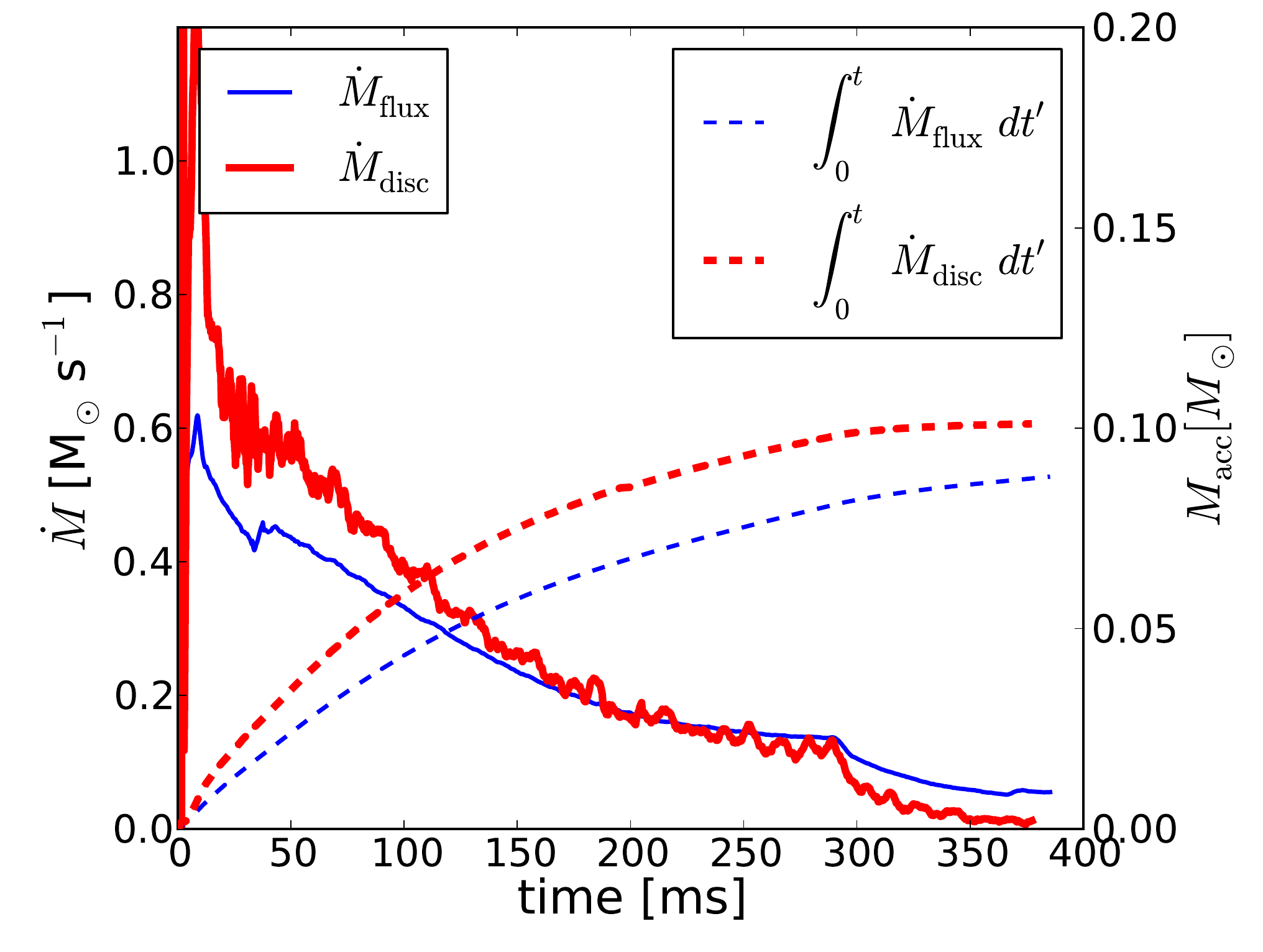}
 \includegraphics[width=0.49 \linewidth]{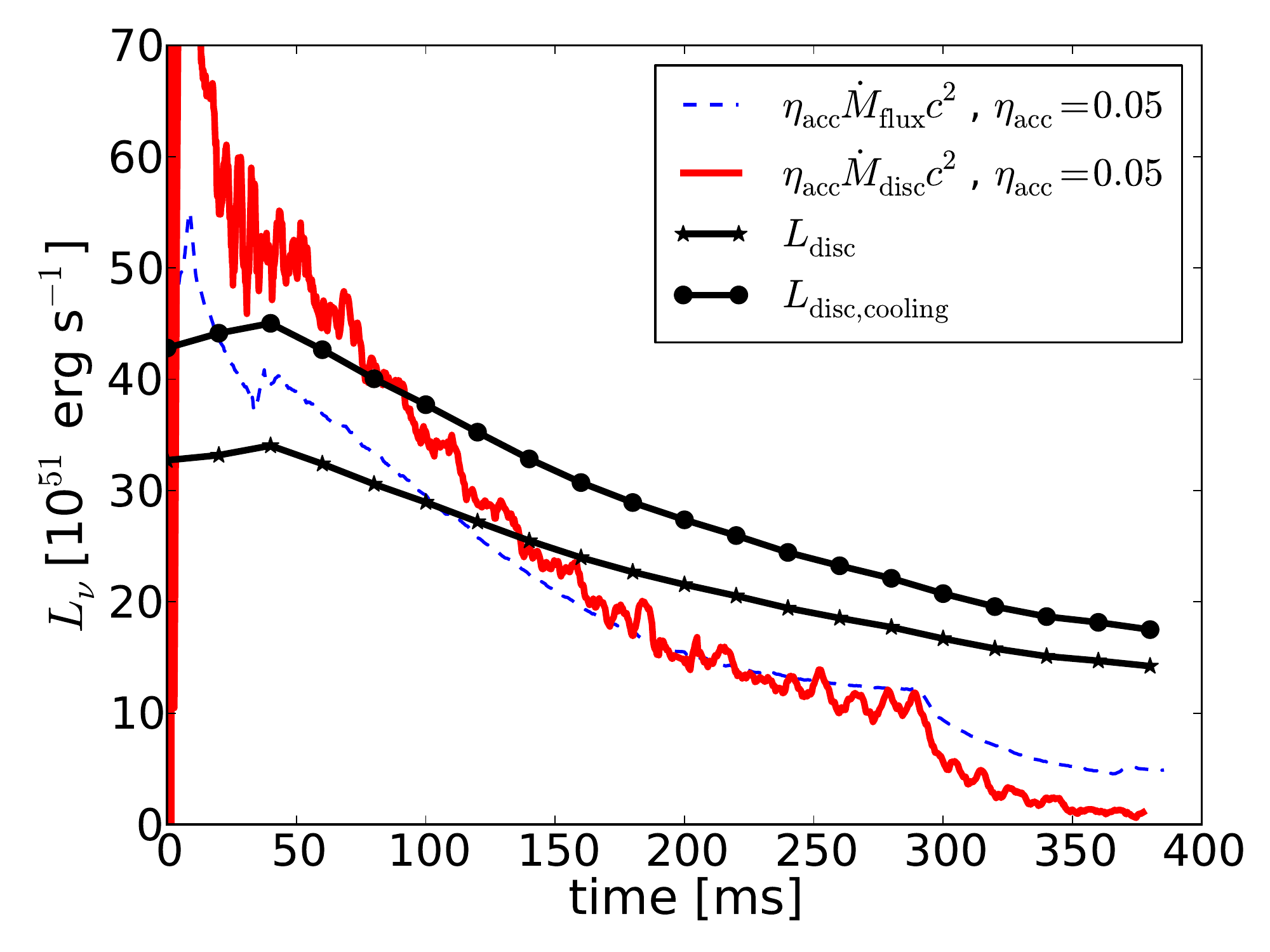}
 \caption{Left: mass accretion rates (solid lines) and cumulative accreted matter (dashed lines) computed
 inside the accretion disk. The red thick lines refer to the accretion rate as time derivative of the mass
 enclosed in the innermost part of the disk, while the blue thin lines as matter flux at a cylindrical radius 
 of $\tilde{R}=40~{\rm km}$. Right: comparison between the temporal evolution of the neutrino luminosities (black lines)
 and of the luminosities derived from the accretion rate, assuming $L = \eta \dot{M}c^2$.}
 \label{fig: accretion rate and luminosity} 
\end{figure}

The evolution of the accretion rate inside the disk, $\dot{M}_{\rm disk}$ is shown in the left panel of Figure \ref{fig: accretion rate and luminosity}. Since there is no clear inner boundary where to measure the flux of the infalling matter, we
compute the accretion rate as the time derivative of the mass contained inside the densest part
of the disk ($\rho > \rho_{\rm min}=5 \times 10^{10} \, {\rm g \, cm^{-3}}$), outside a 
cylindrical radius $R_{\rm in}= 30 \, {\rm km}$:
\begin{equation}
 \dot{M}_{\rm disk} \equiv - \frac{{\rm d} M_{\rm disk}}{{\rm d} t} = 
 - \frac{{\rm d} }{{\rm d} t} \int_{0}^{2 \pi} \: \int_{-\infty}^{\infty} \: \int_{R_{\rm in}}^{\infty} \: 
 R^2 \rho \: \Theta(\rho - \rho_{\rm min}) \, {\rm d}R \, {\rm d}z \, {\rm d}\phi \, ,
\end{equation}
where $\Theta$ is the Heaviside step function. 
The choice of $R_{\rm in}=30 \, {\rm km}$ ensures the exclusion of the MNS.
Since the expansion of matter in the $\nu$-driven wind and in the viscous component happen around
a few times $10^{10} {\rm g \, cm^{-3}} \lesssim \rho_{\rm min}$, our definition of 
$\dot{M}_{\rm disk}$ takes into account only the disk consumption due to the accretion process.
To verify our assumption, we also compute
\begin{equation}
 \dot{M}_{\rm flux} \equiv 
 \int_{\Sigma(\tilde{R}=40 \, {\rm km})} \: \rho \, \mathbf{v} \, \cdot  \, {\rm d}\boldsymbol{\Sigma} \, ,
\end{equation}
where $\Sigma(\tilde{R})$ is the cylindrical surface of radius $\tilde{R}$ and 
${\rm d}\boldsymbol{\Sigma}$ is an infinitesimal surface element pointing in the positive radial
direction. We choose $\tilde{R}=40 \, {\rm km}$ since for R $\gtrsim 40 \, {\rm km}$ the disk 
is characterized by a global infalling bulk motion along the radial direction, while at smaller 
radii, radially outgoing flows are also observed.

In the initial phase ($t < 100\, {\rm ms}$), accretion onto the MNS occurs mainly from
high density regions located within $\tilde{R}$, thus $\dot{M}_{\rm disk} > \dot{M}_{\rm flux}$. 
Once the disk has reached an almost stationary configuration ($t > 100\, {\rm ms}$), the 
two accretion rates behave similarly and $\dot{M}_{\rm disk} \approx \dot{M}_{\rm flux}$.
The integral of $\dot{M}_{\rm disk}$ reveals that the MNS has accreted $\sim$~0.1~$M_{\odot}$ within
400~ms.

In the right panel of Figure~\ref{fig: accretion rate and luminosity}, we compare the total disk luminosity
$L_{\rm DS} = L_{\nu_e,{\rm DS}} + L_{{\bar{\nu}_e},\rm DS} + 4 L_{{\nu_x},\rm DS}$
(using both the total and the cooling luminosities) with the luminosity obtained from the accretion rate, 
$L_{\rm acc} = \eta_{\rm acc} \dot{M}_{\rm disk}c^2$
assuming $\eta_{\rm acc}=0.05$. We notice that in the presence of a MNS $L_{\rm DS}$ is no more
proportional to $\dot{M}$. In particular, the decline of
the disk luminosity is slower than the decrease of the accretion rate. 
The difference with the BH case is due to the presence of the long-lived MNS, which has several consequences:

1) The MNS surface does not act as an inner, open boundary, and the transition between 
the MNS and the disk happens in a high density, high temperature region where the accreted matter
settle on the MNS surface.
The steep density gradient that characterizes this innermost part
of the disk and the MNS surface translates into a large pressure gradient.
While for $R \gtrsim \tilde{R} \approx 40 {\rm km} $ the radial 
flow inside the disk is distinguished by an inward bulk motion, at smaller radii 
it is characterized by the presence of both inward and outward radial flows, resulting 
from the combination of large scale inflow, bounces on the MNS surface and 
fast orbital motion. 

2) Even if the innermost part of the disk is moderately optically thick 
$(\tau_{\nu,{\rm sc}}) \sim 10$, matter inside it can still emit neutrinos.
They diffuse out on a time scale of a few ms and are emitted at the neutrino surfaces.

3) Due to the disk finite size and consumption, the density
profile inside the disk decreases and the disk becomes progressively more transparent to neutrinos. 
In particular, the neutrino diffusion time scale, 
$t_{\nu,{\rm diff}} \sim \alpha_{\rm diff} \tau^2_{\nu} \lambda_{\nu}/c $ where $\alpha_{\rm diff} \sim 3$, 
decreases with time. At the same time, the accretion process partially compensates 
the internal energy emitted by the neutrino radiation. 
If $\epsilon_{\rm int}$
is the specific internal energy of matter 
the ratio $\epsilon_{\rm int}/t_{\nu,{\rm diff}}$
increases with time in the innermost part of the disk. 

In summary, in the presence of a MNS, 
the larger efficiency at which the internal energy can be radiated by neutrinos, together with the longer 
time that matter has to cool, partially compensates the disk consumption and the accretion rate attenuation.

\subsection{MNS and disk contributions to the energy deposition rate}

\begin{figure}
 \includegraphics[width=0.48 \linewidth]{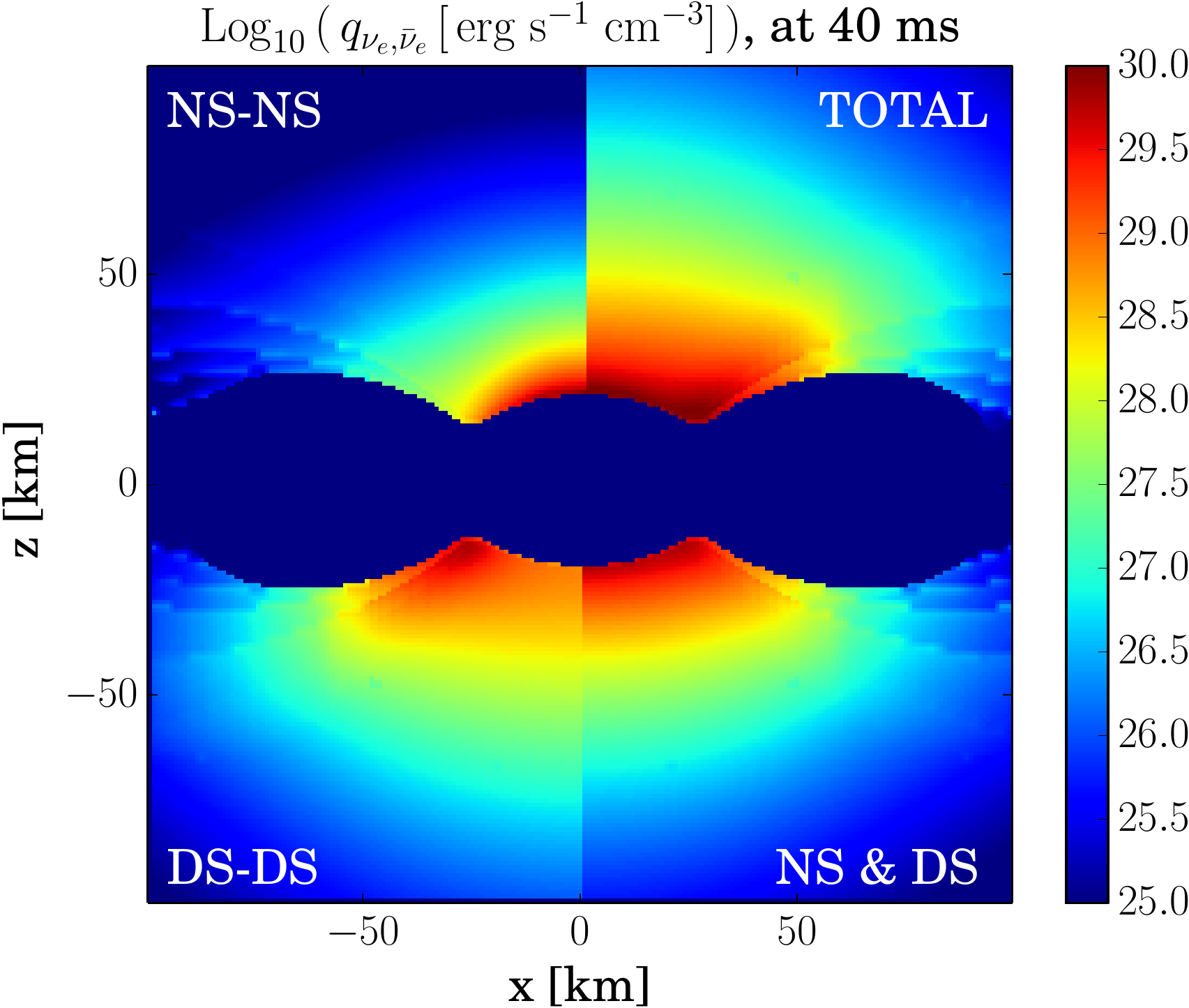}
 \hspace{0.1cm}
 \includegraphics[width=0.49 \linewidth]{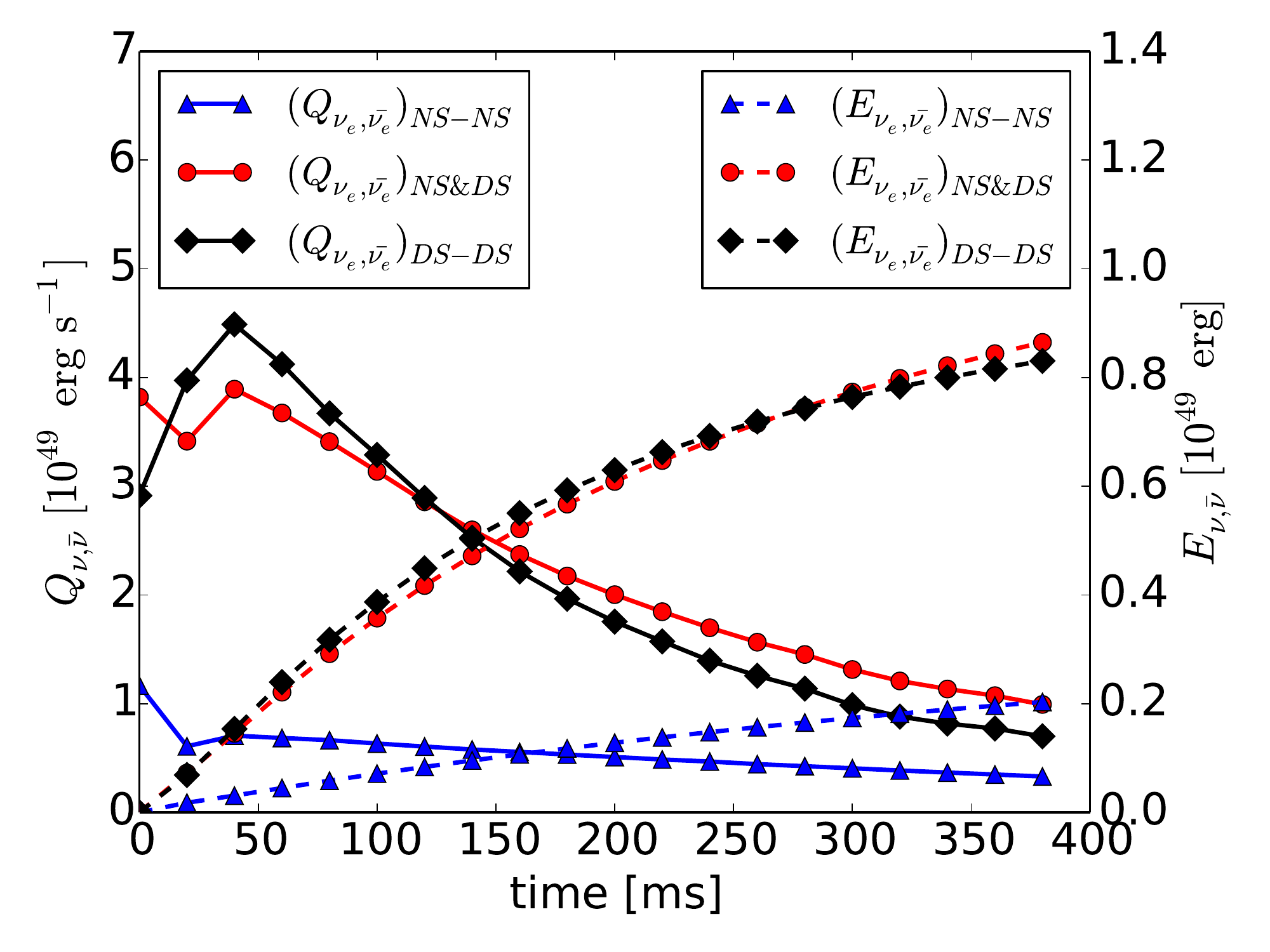} 
 \caption{Left: Energy deposition rate by electron neutrino pair annihilation at 40~ms inside our simulation.
 In the four quadrants, we plot the three different contributions obtained by tagging the emitted neutrinos based
 on their production site (neutron star (NS), or disk (DS)) together with their sum.
 Right: Contributions to the volume-integrated energy rate ($Q_{\nu,\bar{\nu}}$, solid lines) and cumulative deposited energy 
 ($E_{\nu,\bar{\nu}}$, dashed lines) for electron neutrino, as a function of the simulated time, coming from the NS-NS contribution
 (blue lines, triangles), the DS-DS contribution (black lines, diamonds), and the NS\&DS contribution 
 (red lines, circles). For both panels, we use our standard (Newtonian) results.}
 \label{fig: decomposition of the annihilation rate}
\end{figure}

The splitting of the local emissivity into a neutron star and a disk component allows also
the decomposition of the local radiation intensity,
$I_{\nu} = I_{\nu,{\rm NS}} + I_{\nu,{\rm DS}}$.
Since we deal with a pair process, the integrand in Eq.~(\ref{eq: dessart}) for the local energy deposition 
rate $q_{\nu,\bar{\nu}}$ can be expanded in four different contributions according to: 
\begin{eqnarray}
 q_{\nu,\bar{\nu}} = 
(q_{\nu,\bar{\nu}})_{\rm NS-NS} + (q_{\nu,\bar{\nu}})_{\rm DS-DS} +  \left[ (q_{\nu,\bar{\nu}})_{\rm NS-DS} + (q_{\nu,\bar{\nu}})_{\rm DS-NS} \right].
 \label{eq: q contributions} 
\end{eqnarray}
In the left panel of Figure \ref{fig: decomposition of the annihilation rate}, we show the 
different contributions for electron flavor neutrinos at 40~ms, alongside with their sum.
In the following, we will refer to the combined neutron star-disk contribution as 
$(q_{\nu,\bar{\nu}})_{\rm NS \& DS} \equiv \left[ (q_{\nu,\bar{\nu}})_{\rm NS-DS} + (q_{\nu,\bar{\nu}})_{\rm DS-NS} \right]$,
if not stated differently.
The different contributions deposit energy at different locations above the remnant.
The NS-NS contribution is more intense immediately above the surface of the MNS and it fades
rapidly for increasing radial distances. This reflects the quasi-spherical nature of the neutrino
surfaces. The annihilation is efficient only close to the emission surfaces, where the intensities are
larger and the cosine of the average propagation angle is $\langle \mu \rangle \sim 1/2$. For larger distances, the intensities 
decrease as $r^{-2}$ and enter the forward-peaked regimes, $\langle \mu \rangle \sim 1$ \cite{Janka:1991}.
The largest energy deposition rates for the DS-DS contribution are located above the region marking 
the transition between the MNS and the disk. The energy rate in the funnel above the MNS is still high, 
but it decreases while approaching the rotational axis. This behavior points to the fact that most of the 
DS-DS pair annihilations happen close to the neutrino emission surfaces 
(for radiation emitted from contiguous zones of the disk and annihilating at all angles) rather than in the funnel, 
confirming results from \citeasnoun{Dessart.etal:2009}.
The larger intensities of the former configuration compensate for the better collision angle of the latter.
Finally, for the NS\&DS contribution, neutrinos moving away
from the MNS annihilate with those emitted by the disk towards the funnel. The properties of this contribution
are a combination of the properties of the previous two. The results is an 
intense energy deposition both close to the MNS surface and above the transition region between the MNS and
the disk.

We integrate the energy deposition rates over the volume for the three different contributions
and the results are shown in the right panel of Figure \ref{fig: decomposition of the annihilation rate}.
The DS-DS and the NS\&DS contributions are comparable throughout the whole simulation time.
This result points out the potential relevance of a long-lived MNS, in comparison with a BH-torus system.
In fact, in the latter case, only the DS-DS contribution is expected to be present and the total 
energy deposited by neutrino pair annihilation can be reduced by a factor of (at least) two.

The steeper decrease of the DS-DS contribution is due to its quadratic dependence on the more rapidly
decreasing disk luminosity, $q_{\rm DS-DS} \propto L_{\nu,{\rm DS}} L_{\bar{\nu},{\rm DS}}$,
see Eq.~(\ref{eq: simple parametrization}),
while for the NS\&DS contribution we expect 
$q_{\rm NS\&DS} \propto \left( L_{\nu,{\rm NS}} L_{\bar{\nu},{\rm DS}} + L_{\nu,{\rm DS}} L_{\bar{\nu},{\rm NS}} \right)$. 
Despite the intense energy deposition associated with the NS-NS contribution
in the funnel above the MNS, the volume integrated rate is significantly sub-dominant, especially at
earlier times. This is a consequence of the smaller volume where the intense energy deposition occurs, 
see Eq.~(\ref{eq: volume element dependences}).

\begin{figure}[th]
 \includegraphics[width=0.49 \linewidth]{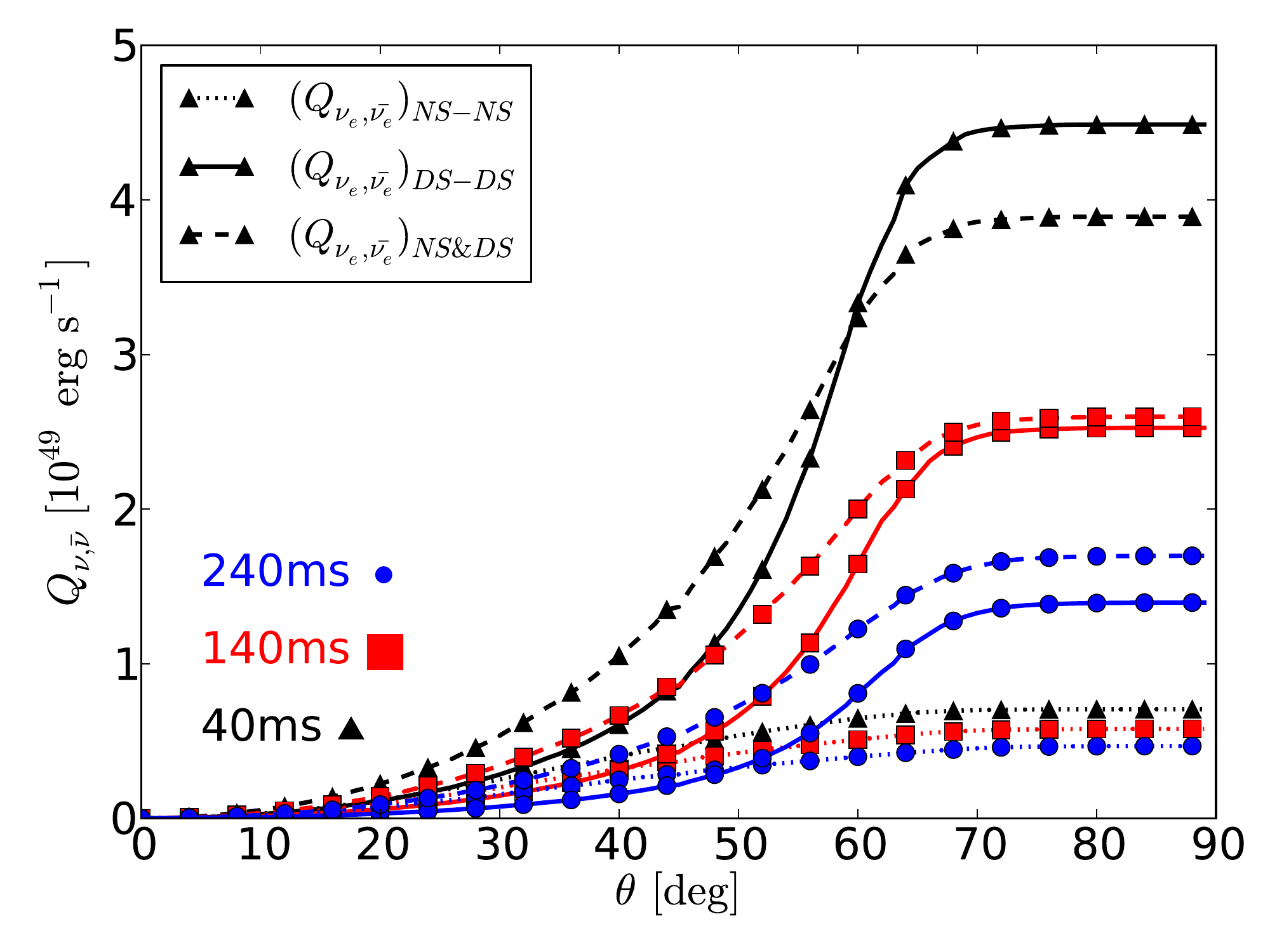} 
 \includegraphics[width=0.48 \linewidth]{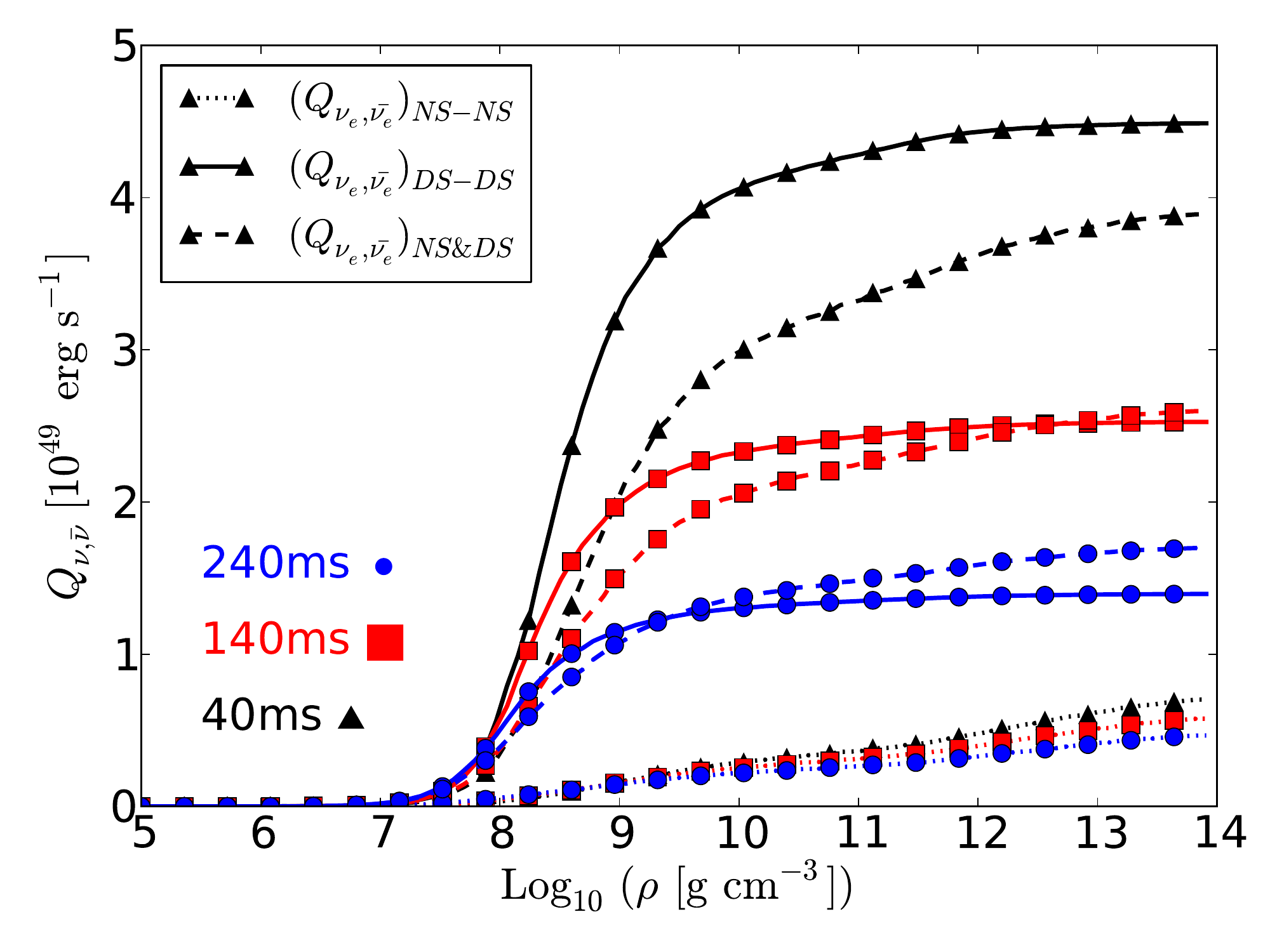}
 \caption{Energy deposition rate by electron neutrino pair annihilation at 40~ms (black triangles), 140~ms (red squares) and 240~ms 
 (blue circles) inside our simulation, as a function of the maximum polar angle (left) and of the maximum matter density where 
 the deposition occurs, for the NS-NS (dotted), DS-DS (solid) and NS\&DS (dashed) contributions.
 For both panels, we use our standard (Newtonian) results.}
 \label{fig: decomposition of the annihilation rate, angle and density cuts}
\end{figure}

To summarize the spatial dependence of the energy deposition, in Figure 
\ref{fig: decomposition of the annihilation rate, angle and density cuts} we present the volume integral of 
the energy deposited by $\nu_e,\bar{\nu}_e$ pair annihilation at three different times and in selected regions
of the domain. In particular, the integration
is performed only for regions where polar angles are smaller than $\theta$ or greater than $(\pi - \theta)$ (left panel), and 
densities are smaller than $\rho$ (right panel).
The energy deposited closer to the rotational axis and at lower densities is foreseen to contribute more to the formation of
a jet.
A comparison between the different terms reveals that the contributions involving the MNS
deposit a more significant fraction of their energy at high densities ($\rho > 10^{11} {\rm g \, cm^{-3}}$) compared with the DS-DS
contribution. However, the latter deposits energies also at larger angular distances from the rotational axis ($\theta \gtrsim 45^o$).

\begin{figure}
 \includegraphics[width=0.49 \linewidth]{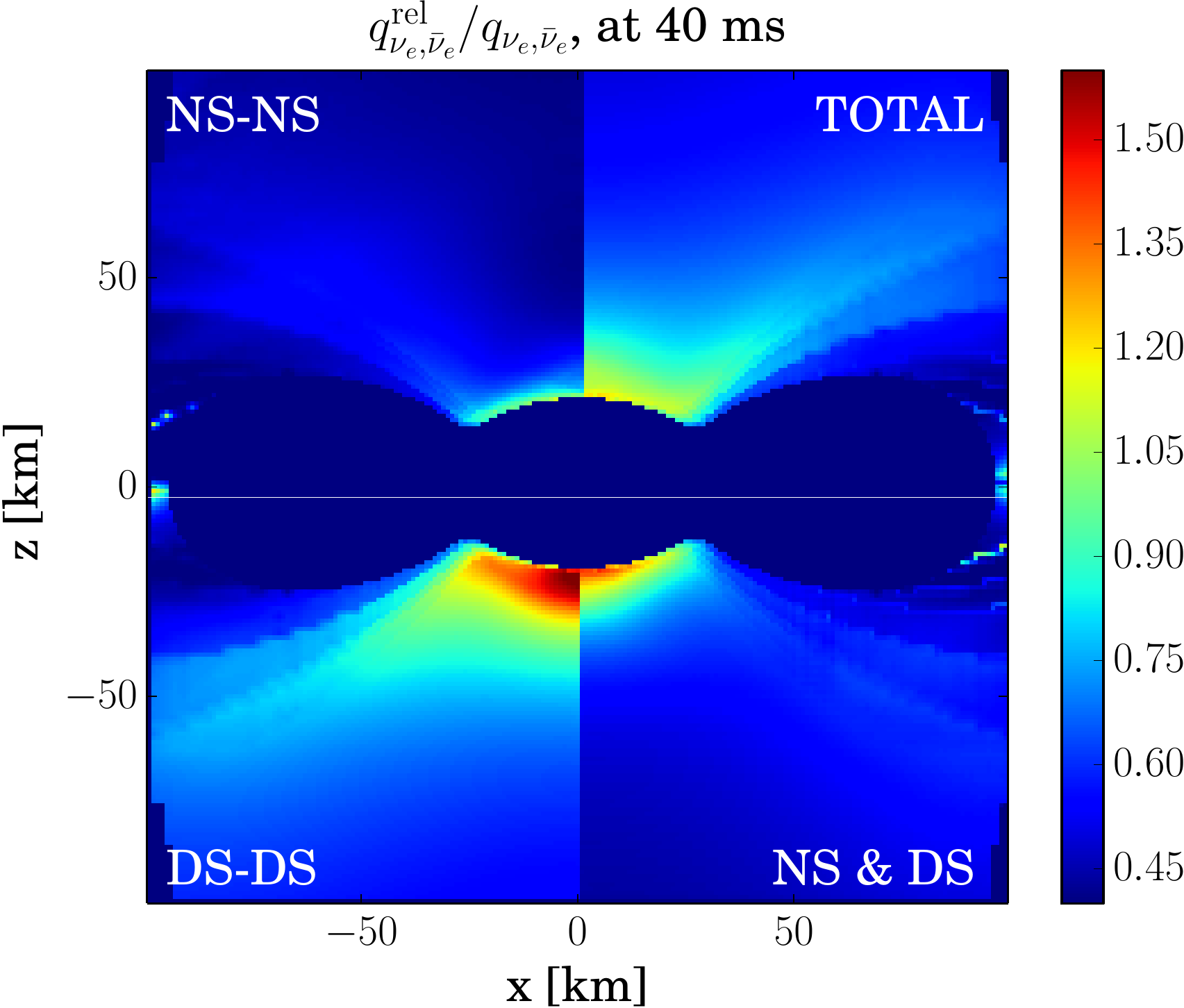}
 \includegraphics[width=0.49 \linewidth]{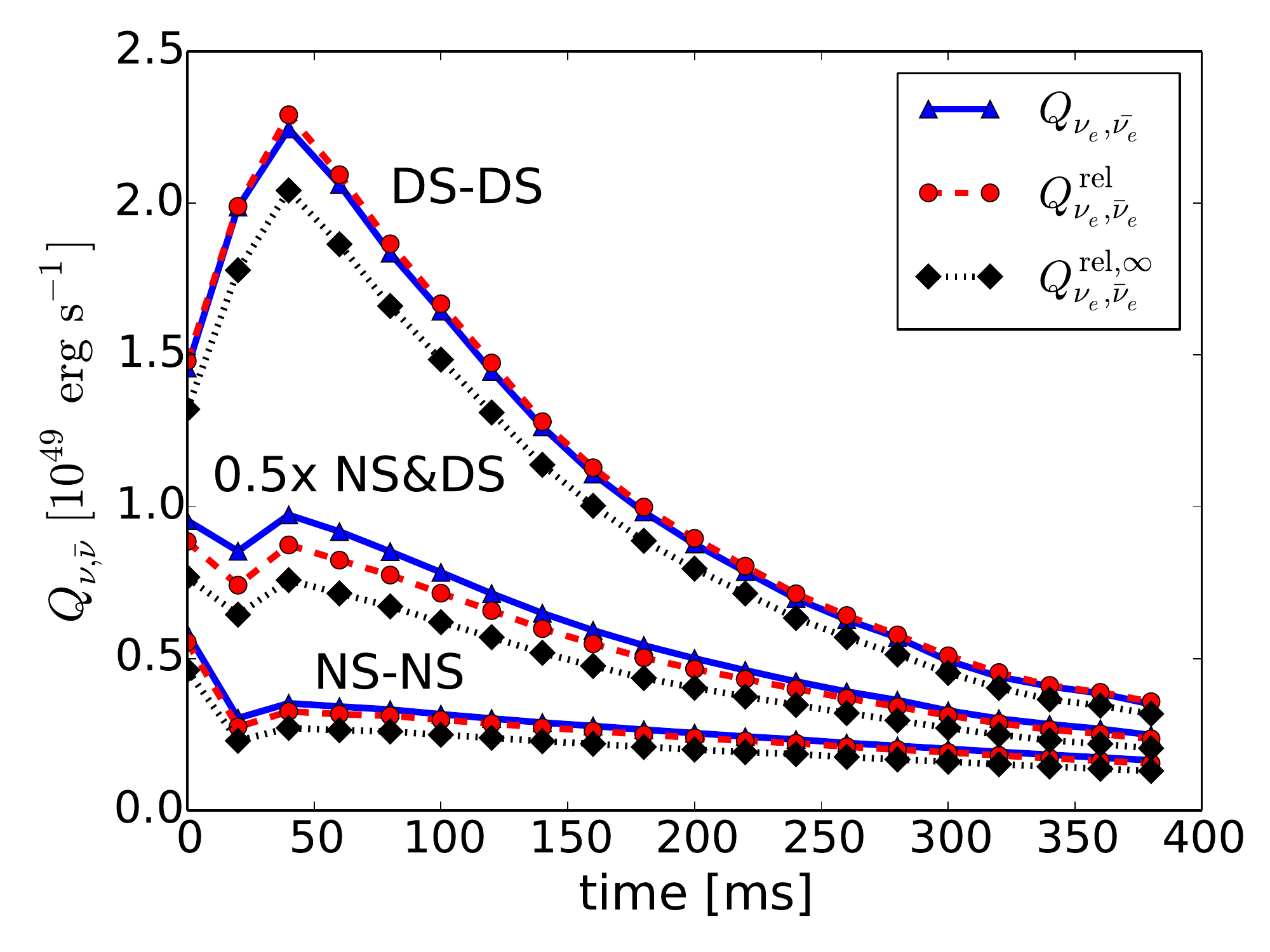}
 \caption{Left: Ratio between the energy deposition rates due to
 the annihilation of $\nu_e,\bar{\nu}_e$ pairs, with and without the inclusion of relativistic effects
 in the neutrino propagation, at 40~ms inside our simulation.
 The quantities $q_{\nu_e,\bar{\nu}_e}$ and $q^{\rm rel}_{\nu_e,\bar{\nu}_e}$ 
 are the Newtonian, the general relativistic energy deposition rate, respectively. 
 Each quadrant of the panel represents one of the components of the rate decomposition and their sum.
 Right: Volume-integrate rates for electron neutrinos, $Q_{\nu_e,\bar{\nu}_e}$ (blue lines, triangles), 
 $Q^{\rm rel}_{\nu_e,\bar{\nu}_e}$ (red lines, circles) and $Q^{\rm rel, \infty}_{\nu_e,\bar{\nu}_e}$ (black lines, diamonds), 
 for each of the three contributions (NS-NS, NS\&DS, DS-DS). For the NS\&DS contribution, we plot it halved for visualization purposes. }
 \label{fig: relativity effect on components}
\end{figure}

Finally, we show the impact of relativistic effects on the annihilation process.
From the left panel of Figure \ref{fig: relativity effect on components}, we can infer that
the DS-DS contribution is affected the most by relativistic effects. On the one hand, the beaming 
effect due to the matter motion in the disk, together with the Doppler effect and the gravitational redshift, 
reduces the annihilation rate in the funnel above the MNS. On the other hand, the gravitational blueshift 
and the light bending occurring close to the MNS and to the innermost neutrino surfaces increase the 
rates immediately above the MNS surface.
The same effects are visible, even if less pronounced, for the NS\&DS contributions. On the contrary, 
for the NS-NS contribution, the dominant gravitational redshift reduces the annihilation rates almost everywhere.
To quantify the global impact of the relativistic effects, we integrate the different contributions over the deposition volume, 
according to Eqs.~(\ref{eq: Q non rel}), (\ref{eqn: relativisitic local volume integrated energy rate}) 
and (\ref{eqn: relativisitic at infinity volume integrated energy rate}).
The results are shown in the right panel of Figure \ref{fig: relativity effect on components}. 
In the case of the DS-DS contribution, relativistic effects can even marginally increase the integrated 
energy deposition rate, as measured by local observers, leading to 
$(Q^{\rm rel}_{\nu,\bar{\nu}})_{\rm DS-DS} \approx (Q_{\nu,\bar{\nu}})_{\rm DS-DS}$.
In the case of the NS\&DS and NS-NS contributions, the larger volume and the more pronounced decrease 
in the energy deposition efficiency at large radii compensate the increase close to the remnant, such 
that $Q^{\rm rel}_{\nu,\bar{\nu}} \lesssim Q_{\nu,\bar{\nu}}$.
A more detailed analysis of the impact of special and general relativistic effects
on the energy deposition rates is presented in \ref{apx: analysis rel effects}.

\subsection{Dependence on the neutrino luminosity and conversion efficiency}

\begin{figure}
 \begin{center}
 \includegraphics[width=0.49 \linewidth]{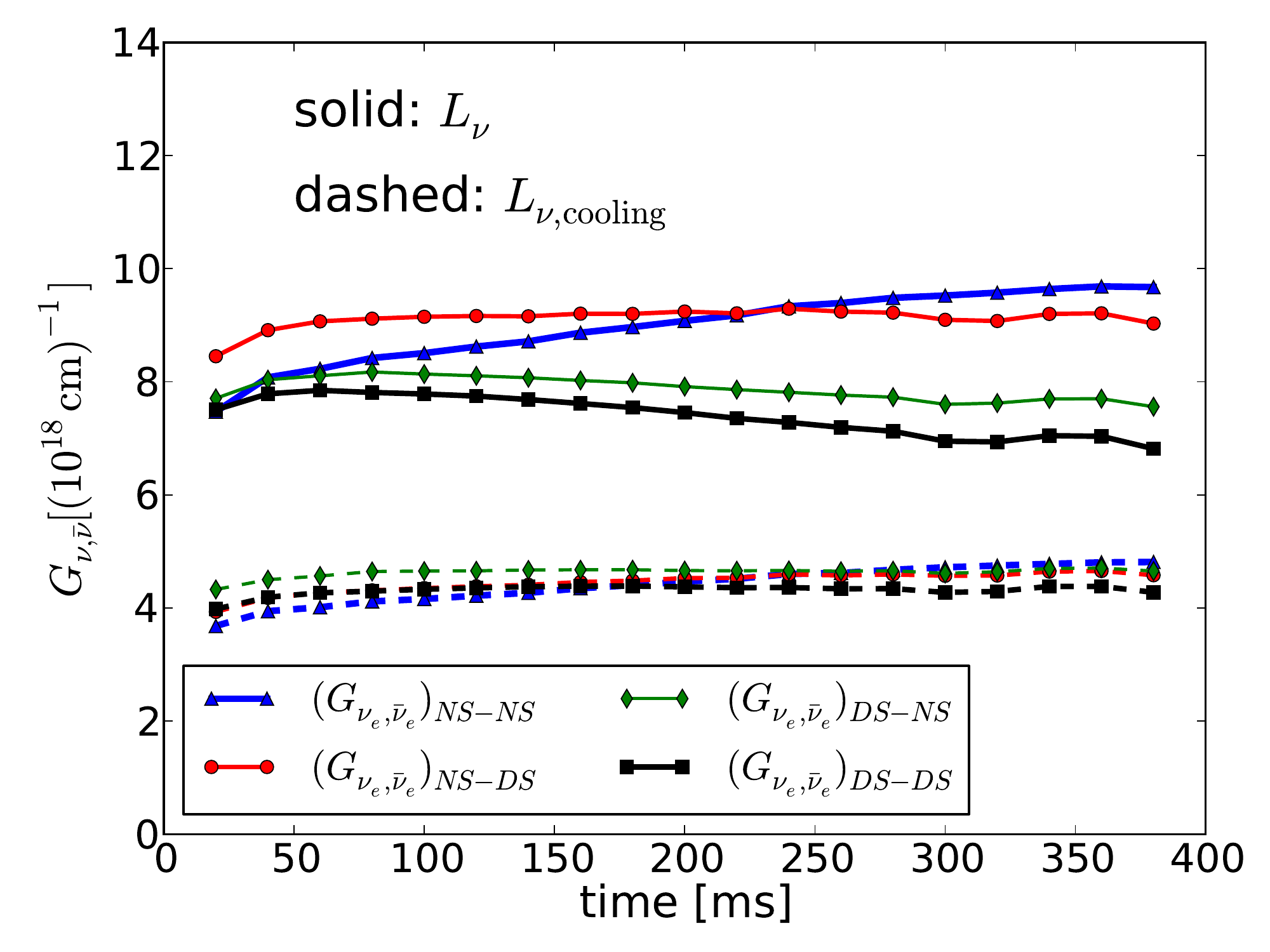}
 \includegraphics[width=0.49 \linewidth]{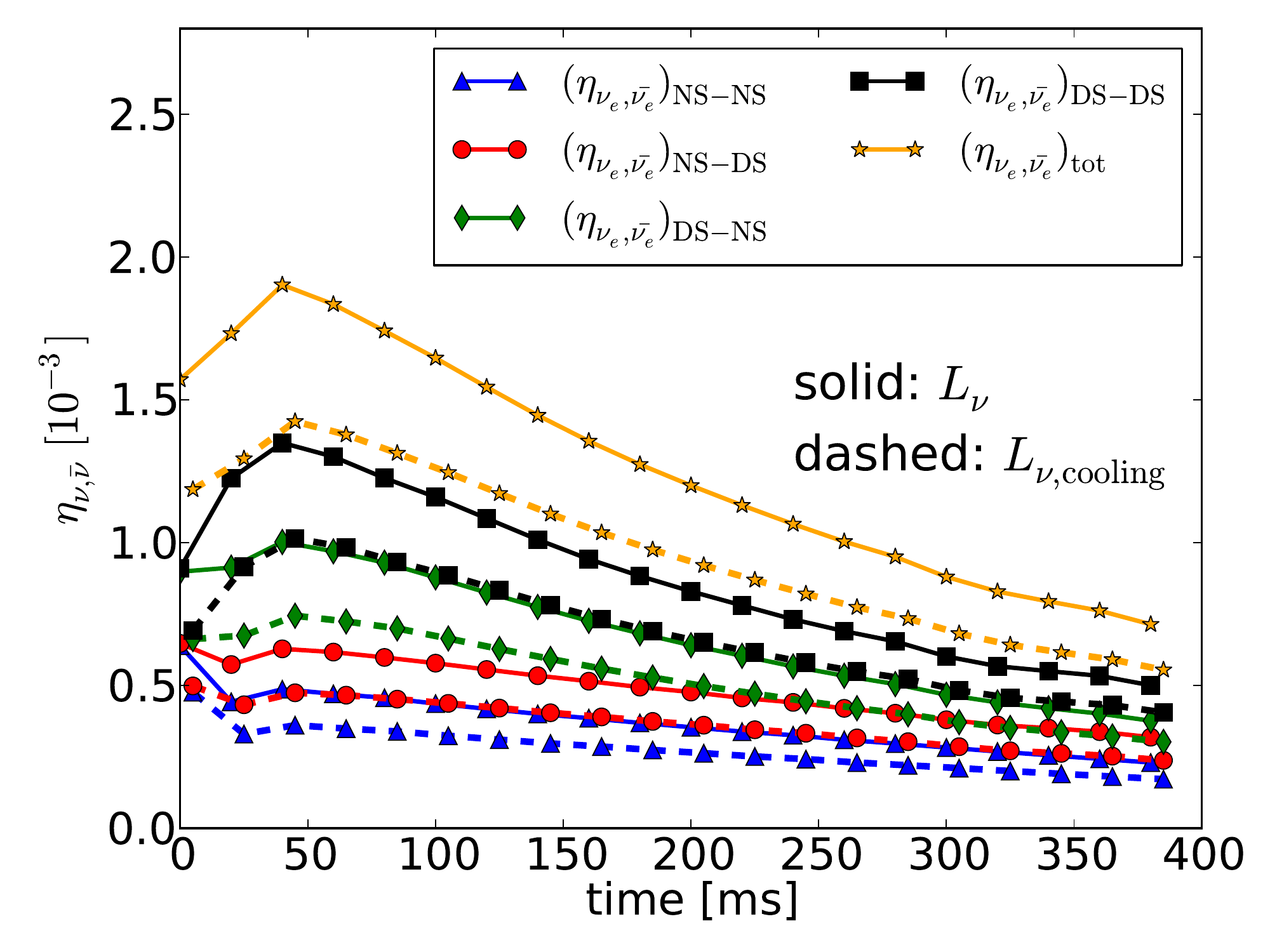}
 \end{center}
 \caption{Left: Geometrical factors $G_{\nu,\bar{\nu}}$ (left), defined in Eq.~(\ref{eq: simple parametrization decomposed}), 
 and conversion efficiency parameters $\eta_{\nu,\bar{\nu}} = Q_{\nu,\bar{\nu}}/(L_{\nu} + L_{\bar{\nu}})$ (right). 
 Both quantities are calculated for four distinct contributions to the energy deposition rate by $\nu_e,\bar{\nu}_e$ pairs, 
 using the net luminosities (solid lines) or the cooling luminosities (dashed lines). In the right panel, the dashed
 lines are shifted by $\Delta t = 5 \, {\rm ms}$ for visualization purposes.}
 \label{fig: G factor Lcool}
\end{figure}

In Eq.~(\ref{eq: simple parametrization}), we have introduced a possible parametrization of the 
volume-integrated energy deposition rates, $Q_{\nu,\bar{\nu}}$.
The advantage of this expression is to provide global information about the energy deposition rate 
based on basic properties of the neutrino emission ($L_{\nu}$, $L_{\bar{\nu}}$, neutrino mean energies) 
and on the geometry of the system ($G_{\nu,\bar{\nu}}$, $H_{\nu,\bar{\nu}}$).
In the following, we verify the validity of our expression and compute the unknown geometrical factors.
We take again advantage of the luminosity decomposition and we evaluate 
the geometrical factors separately for each of the contributions appearing 
in Eq.~(\ref{eq: q contributions}), i.e. $(G_{\nu,\bar{\nu}})_{ij}$ and $(H_{\nu,\bar{\nu}})_{ij}$ with
$i,j=\{{\rm NS,DS} \}$.
Since our decomposition distinguishes among contributions presenting different emission properties,  
we further assume that
$(G_{\nu,\bar{\nu}})_{ij} \approx (H_{\nu,\bar{\nu}})_{ij}$ in Eq.~(\ref{eq: simple parametrization}) and obtain
\begin{equation}
\left( Q_{\nu,\bar{\nu}} \right)_{i,j} \approx \frac{\sigma_0 (c_A^2 + c_V^2)_{\nu,\bar{\nu}}}{96 \pi^2 c (m_e c^2)^2}
L_{\nu,i} L_{\bar{\nu},j} \left( G_{\nu,\bar{\nu}} \right)_{i,j} 
\left[
\frac{\langle \epsilon^2_{\nu,i} \rangle}{\langle \epsilon_{\nu,i} \rangle} +
\frac{\langle \epsilon^2_{\bar{\nu},j} \rangle}{\langle \epsilon_{\bar{\nu},j} \rangle}
\right] \, ,
\label{eq: simple parametrization decomposed}
\end{equation}
where $\nu,\bar{\nu}$ can be both electron or heavy flavor neutrino pairs.
Differently from before, here we distinguish
between neutrinos coming from the MNS and antineutrinos coming from the 
disk, and vice versa, for the 
$\nu_e$,$\bar{\nu}_e$ contributions.
In the left panel of Figure~\ref{fig: G factor Lcool}, we present the temporal evolution of 
$G_{\nu_e,\bar{\nu}_e}$ for all the four different contributions to $Q_{\nu_e,\bar{\nu}_e}$. 
We tested both the usage of the luminosities at infinity (solid lines),
$L_{{\nu}_e}$, and the cooling luminosities (dashed line), $L_{\nu_e,{\rm cooling}}$ (see Section \ref{sec: simulation}) 
in Eq.~(\ref{eq: simple parametrization decomposed}).
We first notice that, despite the different geometrical properties of the MNS and disk emission,
the geometrical factors show similar values and similar trends with time, within 10\% of their average values.
This result corroborates the validity of parametrization (\ref{eq: simple parametrization decomposed}).
The geometrical factors obtained by the cooling luminosities exhibit a smaller spread, both among them and 
in time. This reflects the different impact of neutrino absorption outside the neutrino surfaces 
for the different contributions. In particular, the factor $\sim 2$ between the results computed using 
$L_{\nu}$ and $L_{\nu,{\rm cooling}}$ can be explained as 
$\left( \exp(-\tilde{\tau}_{\nu_e,\rm en}) \exp(-\tilde{\tau}_{\bar{\nu}_e,\rm en}) \right)^{-1}$,
where $\tilde{\tau}_{\nu_e,\rm en}$ and $\tilde{\tau}_{\bar{\nu}_e,\rm en}$ are the average values of
the energy optical depth at the last scattering surface. Since for $\nu_e$, the scattering and the energy
optical depth are close (see Figure (\ref{fig: nu_surfaces})), 
$\tilde{\tau}_{\nu_e,\rm en} \approx 0.5 \lesssim 2/3$, while for $\bar{\nu}_e$ the difference is 
larger, $\tilde{\tau}_{\bar{\nu}_e,\rm en} \approx 0.2$.

The situation is qualitatively different for heavy flavor neutrinos (not shown in the figure).
Their values are such that 
$ 4 \times 10^{-18}{\rm cm}^{-1} \lesssim (G_{\nu_x,\bar{\nu}_x})_{\rm NS-NS} \lesssim 5 \times 10^{-18}{\rm cm}^{-1}$,
$ (G_{\nu_x,\bar{\nu}_x})_{\rm NS-DS} \approx 6 \times 10^{-18}{\rm cm}^{-1}$, and
$ (G_{\nu_x,\bar{\nu}_x})_{\rm DS-DS} \approx 7 \times 10^{-18}{\rm cm}^{-1}$.
Since there is no significant $\nu_x$ absorption outside the neutrino surfaces, $L_{\nu_x} \approx L_{\nu_x,{\rm cooling}}$.
Thus, if we compare with the results obtained with $L_{\nu_e,{\rm cooling}}$ and $L_{\bar{\nu}_e,{\rm cooling}}$, we conclude
that the geometrical factors for $\nu_x$ are larger.
We speculate that the origin of these larger values depends on the smaller extension of the neutrino surfaces. On the one hand,
this provides a more isotropic emission, but on the other hand leads to less spatially distributed emission and 
larger neutrino densities. 

After having explored the dependency of the energy deposition rates on the neutrino luminosities, we can quantify 
the annihilation efficiency, as defined in Eq.~(\ref{eq: annihilation efficiency}).
In the right panel of Figure \ref{fig: G factor Lcool}, we plot $\eta_{\nu_e,\bar{\nu}_e}$ for the total rates and the 
four different contributions. In the latter cases, at the denominator, we select only the relevant luminosities, i.e. 
$(\eta_{\nu,\bar{\nu}})_{i,j} = (Q_{\nu,\bar{\nu}})_{i,j}/( L_{\nu,_i} + L_{\bar{\nu},j} ) $.
Also in this case, we compute $\eta_{\nu_e,\bar{\nu}_e}$ both using the total (solid lines) and the cooling luminosities
(dashed lines), and obviously the former are systematically larger than the latter. The annihilation efficiency of the single
components ranges between 0.03\% up to 0.15\%.
The behavior of the different components and their relative strength results from the non-trivial combinations of 
neutrino luminosities, geometrical factors, and mean energies. 
The most efficient energy conversion happens for the DS-DS contribution, while the less efficient is the
NS-NS contribution. We also compute the efficiency for the total rates (orange line) and find that it varies between
0.2\% (at the beginning, when the annihilation is dominated by the more efficient DS-DS and NS\&DS contributions)
and 0.07\% (at the end, when the NS-NS contribution becomes more relevant). We also notice that 
$(\eta_{\nu,\bar{\nu}})_{\rm tot}$ is larger than any other contribution. This is due to the fact that each component
of the luminosity powers two different energy deposition contributions. Thus, we can write
$(\eta_{\nu,\bar{\nu}})_{\rm tot} = \sum_{i,j} a_{i,j}(\eta_{\nu,\bar{\nu}})_{i,j} $ , where 
$\sum_{i,j} a_{i,j} = 2 $ .

\subsection{MNS collapse time scale}

\begin{table}
 \begin{tabular}{| c | c | c  c | c  c | c  c |}
 \hline
  flavor & contribution & \multicolumn{2}{c}{standard} & \multicolumn{2}{c}{relativistic (loc)} & \multicolumn{2}{c}{relativistic ($\infty$)} \\ 
         &              &  $\left( Q_{\nu,\bar{\nu}} \right)_0$  & $t_0$  & $\left( Q^{\rm rel}_{\nu,\bar{\nu}} \right)_0$  & $t_0$  & $\left(Q^{\rm rel,\infty}_{\nu,\bar{\nu}} \right)_0 $  & $t_0$  \\ \hline
  -      &  -           & $10^{49}~{\rm erg~s^{-1}}$           & s    & $10^{49}~{\rm erg~s^{-1}}$           & s   & $10^{49}~{\rm erg~s^{-1}}$ & s   \\ \hline \hline
  \multirow{4}{*}{$\nu_e,\bar{\nu}_e$} & NS-NS        &   3.96(-1)    & 0.445 & 3.70(-1) & 0.450  & 3.08(-1) & 0.454 \\
                                       & DS-NS        &   1.44        & 0.233 & 1.29     & 0.240  & 1.12     & 0.240 \\
                                       & NS-DS        &   8.86(-1)    & 0.253 & 7.97(-1) & 0.268  & 9.91(-1) & 0.268 \\
                                       & DS-DS        &   2.81        & 0.176 & 2.86     & 0.176  & 2.55     & 0.176 \\ \hline
  \multirow{4}{*}{$\nu_x,\bar{\nu}_x$} & NS-NS        &   8.44(-2)    & 0.514 & 7.64(-2) & 0.513  & 6.41(-2) & 0.513 \\
                                       & DS-NS/NS-DS  &   3.34(-3)    & 0.198 & 3.33(-3) & 0.205  & 2.88(-3) & 0.205 \\ 
                                       & DS-DS        &   1.09(-4)    & 0.126 & 1.27(-4) & 0.128  & 1.12(-4) & 0.128 \\ \hline
  \end{tabular}
 \caption{Table with the values obtained from the exponential interpolation of the different components of 
 $Q_{\nu,\bar{\nu}}$ (left), $Q^{\rm rel}_{\nu,\bar{\nu}}$ (center), and $Q^{\rm rel,\infty}_{\nu,\bar{\nu}}$ (right), 
 for $\nu_e,\bar{\nu}_e$ (top) and $\nu_x,\bar{\nu}_x$ (bottom), according to Eq.~(\ref{eq: interpolation}). 
 The integers in bracket correspond to the power of ten of the number.}
 \label{tab: exp fit}
\end{table}

At the end of our simulation, the annihilation rate has significantly decreased with
time, but the cumulative energy deposition has not saturated yet.
In addition, we have supposed that the MNS is stable against gravitational
collapse on a time scale  $> 380$~ms. 
In the following, we want to estimate the energy deposition at later time times (1 sec) and 
investigate the possible impact of BH formation on the total amount of deposited energy.

\begin{figure}
 \centering
 \includegraphics[width=0.60 \linewidth]{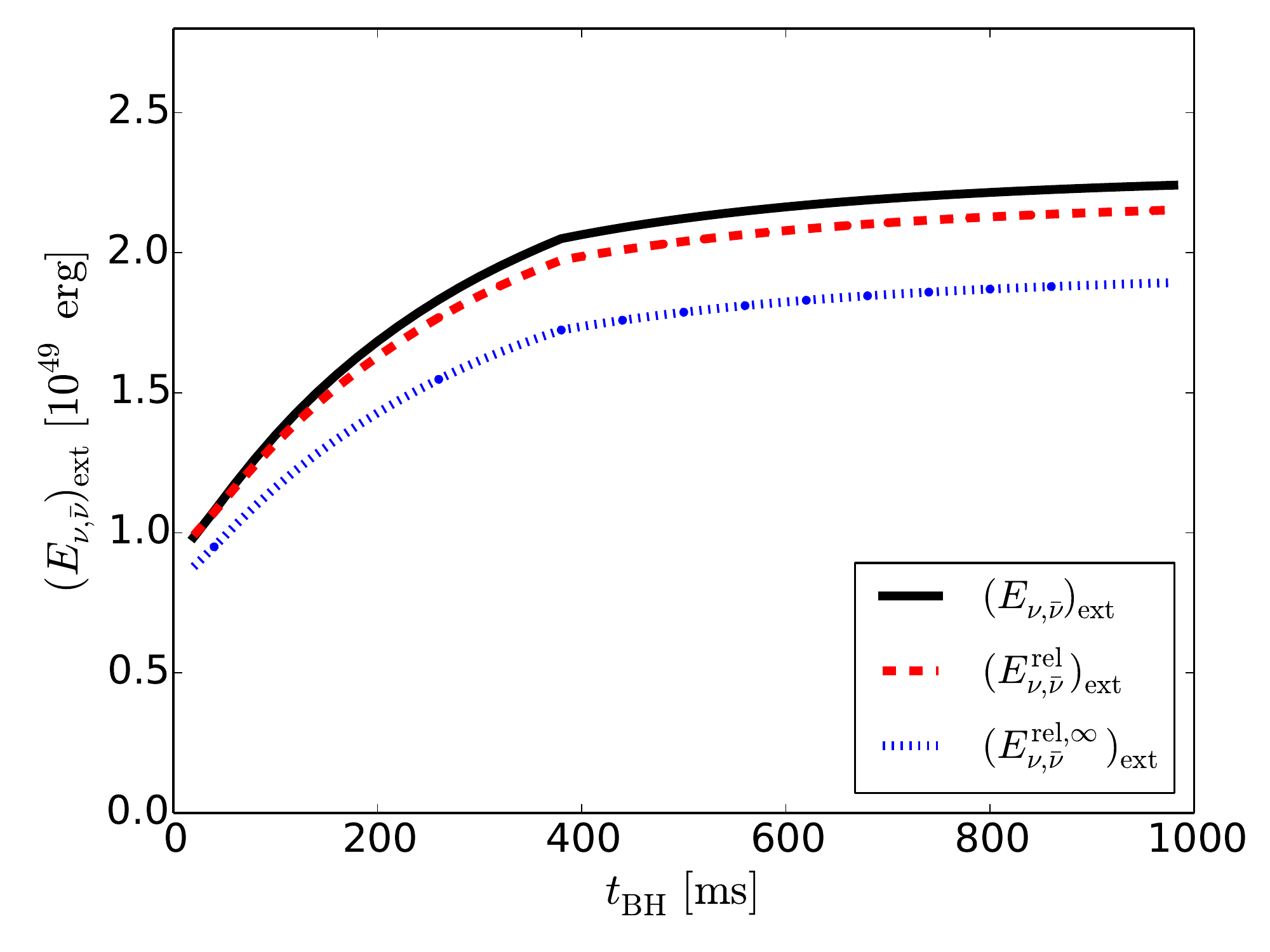}
 \caption{Cumulative energy deposition by all neutrino flavors at 1~sec,
 as a function of $t_{\rm BH}$, the collapse time of the MNS to a BH. 
 For $t>t_{\rm BH}$, contributions of $q_{\nu,\bar{\nu}} $involving NS neutrinos 
 (NS-NS and NS\&DS) are set to 0.  
 The black solid line represents
 the reference Newtonian calculations, while the red dashed and blue dotted lines refer
 to the integral of the local energy and to the energy at infinity, respectively, once
 relativistic effects in the neutrino propagation have been taken into account.}
 \label{fig: effect of BH formation}
\end{figure}

We interpolate the different contributions to $Q_{\nu,\bar{\nu}}$ assuming an exponential behavior,
\begin{equation}
 Q_{\nu,\bar{\nu}} \approx \left(Q_{\nu,\bar{\nu}} \right)_0 \exp{(-t/t_0)} \, .
 \label{eq: interpolation}
\end{equation}
In Table \ref{tab: exp fit}, we report the values obtained for the two neutrino species and for the four 
different contributions in Eq.~(\ref{eq: q contributions}), for our standard case and for calculations including 
relativistic effects in the neutrino propagation. To account for possible additional energy deposition, 
we extrapolate our fits for $t>380$~ms.
We define $t_{\rm BH}$ as the time when the BH 
forms and we assume that for $t>t_{\rm BH}$ all the contributions to $Q_{\nu,\bar{\nu}}$ involving the MNS
($(Q_{\nu,\bar{\nu}})_{\rm NS,NS}$ and $(Q_{\nu,\bar{\nu}})_{\rm NS\&DS}$) do not participate in the energy deposition.
In Figure \ref{fig: effect of BH formation}, we present $\left( E_{\nu,\bar{\nu}} \right)_{\rm ext}$, the cumulative energy 
deposition rate extrapolated up to one second, as a function of $t_{\rm BH}$. Since the DS-DS contribution represent $\sim 0.5$ 
of the total contribution, a quick MNS collapse leads to a reduction of $\lesssim$ 50\% of the final deposited energy.
Due to the decrease of the neutrino luminosities with time, a delayed collapse to a BH produces a smaller effect
on $E_{\nu,\bar{\nu}}$ as time increases. These results are qualitatively insensitive to the modelling of relativistic 
effects in the neutrino propagation.

\section{Uncertainties on the luminosities and comparison with short GRB energetics}
\label{sec: discussion}

\begin{table}
 \begin{tabular}{| c | c | c | c | c | c | c | c |}
 \hline
  label & GRB name & $E_{\gamma,{\rm iso}}$ & $E_{{\rm kin,iso}}$ & $\theta_{\rm jet,min}$ & $\theta_{\rm jet,max}$ & $E_{\rm true,min}$ & $E_{\rm true,max}$ \\ \hline
    -   &    -     & $10^{52}~{\rm erg}$    & $10^{52}~{\rm erg}$    & degree                 & degree                 & $10^{49}~{\rm erg}$    & $10^{49}~{\rm erg}$    \\ \hline \hline
  A & 051221A  & 1.3  & 0.16 & 5   & 8  & 5.55  & 14.21 \\
  B & 090426A  & 2.0  & 1.40 & 5   & 7  & 12.94 & 25.34 \\
  C & 111020A  & 0.17 & 0.48 & 3   & 8  & 0.891  & 6.326 \\
  D & 130603B  & 0.37 & 0.11 & 4   & 8  & 1.169  & 4.671 \\ \hline
  E & 050709   & 0.09 & 0.0026 & 15& $\sim$ 30 & 3.155  & 12.406 \\
  F & 050724A  & 0.24 & 0.18 & 25  & $\sim$ 30 & 39.35 & 56.27 \\
  G & 101219A  & 0.74 & 0.30 & 4   & $\sim$ 30 & 2.533  & 139.33 \\
  H & 111117A  & 0.55 & 0.06 & 3   & $\sim$ 30 & 0.8359 & 81.72 \\
  I & 120804A  & 3.4  & 1.10 & 13  & $\sim$ 30 & 115.3 & 602.88 \\
  J & 140903A  & 0.08 & 2.90 & 6   & $\sim$ 30 & 16.32 & 399.24 \\
  K & 1409030B & 0.4  & 0.28 & 9   & $\sim$ 30 & 8.372 & 91.10 \\ \hline
 \end{tabular}
 \caption{Table with measured short GRB energetics. The isotropized photon and kinetic energies 
 ($E_{\rm \gamma,iso}$ and $E_{\rm kin,iso}$) are taken from \citeasnoun{Fong.etal:2015}, for
 their fiducial models ($\epsilon_B = 0.1$ for all cases, but $\epsilon_B = 10^{-4}$ for F and
 but $\epsilon_B = 10^{-3}$ for J), as well as the jet opening angles. In case only a lower limit
 for $\theta_{\rm jet,min}$ is available, an upper limit of $\theta_{\rm jet,max}= 30^o$ is assumed. 
 The minimal and maximal true energies, $E_{\rm true,min}$ and $E_{\rm true,max}$ are computed 
 according to Eq.~(\ref{eq: E true}).}
 \label{tab: GRB observations}
\end{table}

\begin{figure}
 \centering
 \includegraphics[width=0.48 \linewidth]{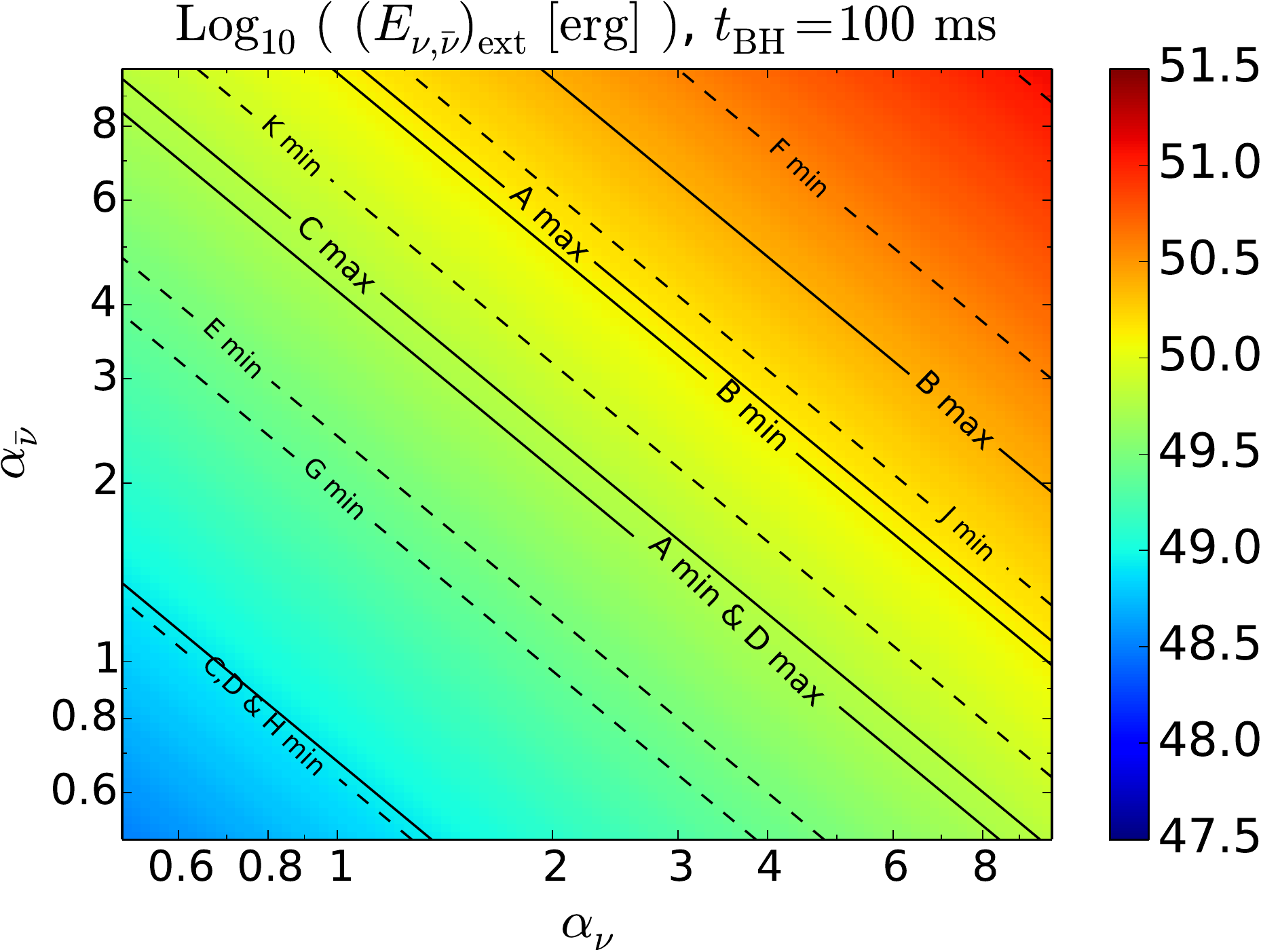}
 \hspace{0.1cm} 
 \includegraphics[width=0.48 \linewidth]{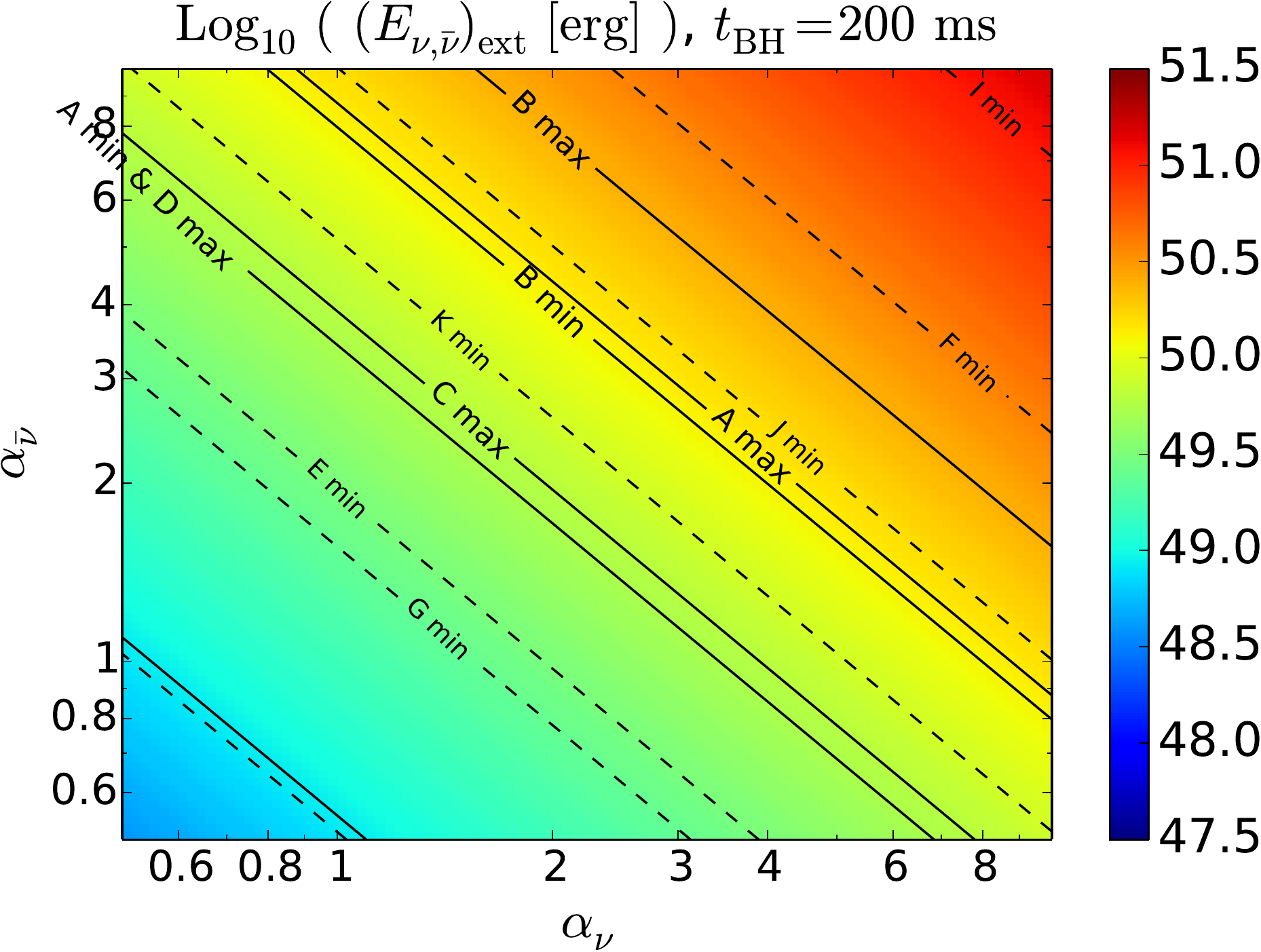}
 \includegraphics[width=0.48 \linewidth]{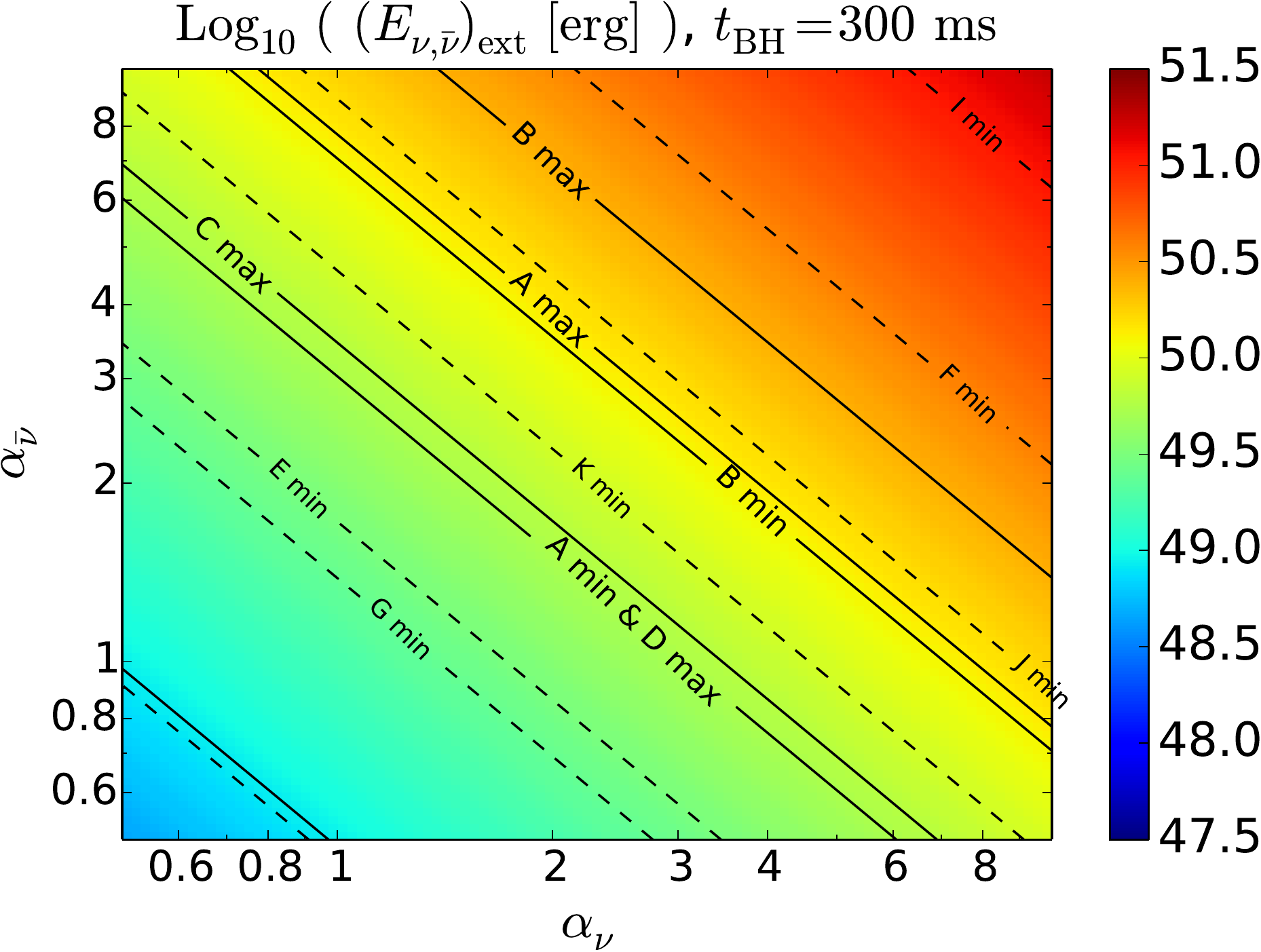}
 \hspace{0.1cm} 
 \includegraphics[width=0.48 \linewidth]{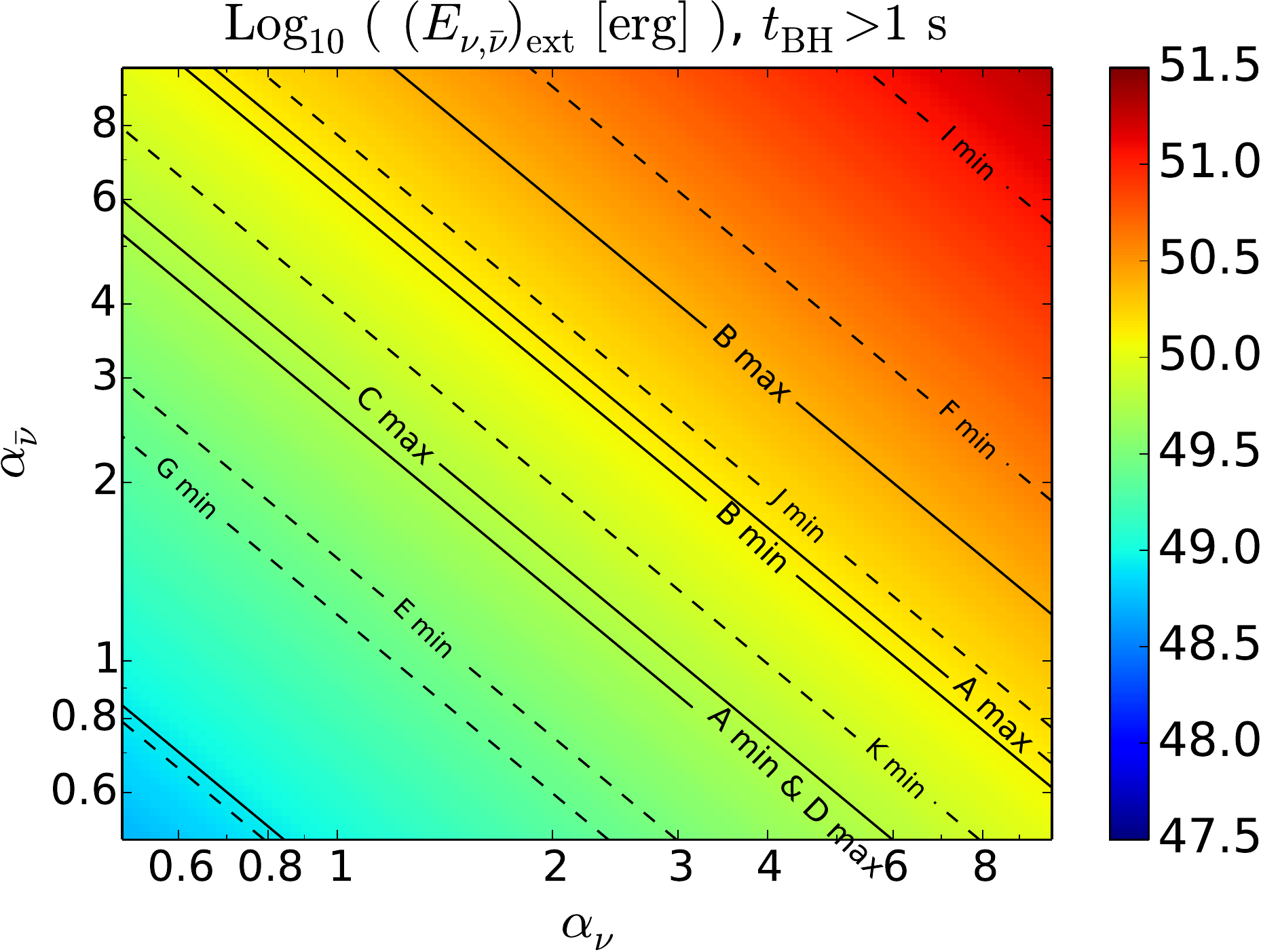}
 \caption{Cumulative energy deposition by $\nu_e$ and $\bar{\nu}_e$ at 1~s, obtained rescaling the neutrino and
 the antineutrino luminosities by independent constant factors $0.5 \leq \alpha_{\nu (\bar{\nu})} \leq 10$. The
 $\alpha_{\nu} = \alpha_{\bar{\nu}} = 1$ case the results of our simulation.
 The four panels refer to different time for the MNS collapse to a BH.
 Solid lines refer to the inferred minimum and maximum true energy associated with observed short 
 GRBs for which an estimate of the jet opening angle is available. Dashed lines refer to the minimum true 
 energy associated with observed short GRBs for which only lower limit estimates of the jet opening angle are 
 available. GRBs labels, energies and opening angles are detailed in Table \ref{tab: GRB observations}.}
 \label{fig: effect of luminosity rescaling, nu-antinu}
\end{figure}

\begin{figure}
 \centering
 \includegraphics[width=0.48 \linewidth]{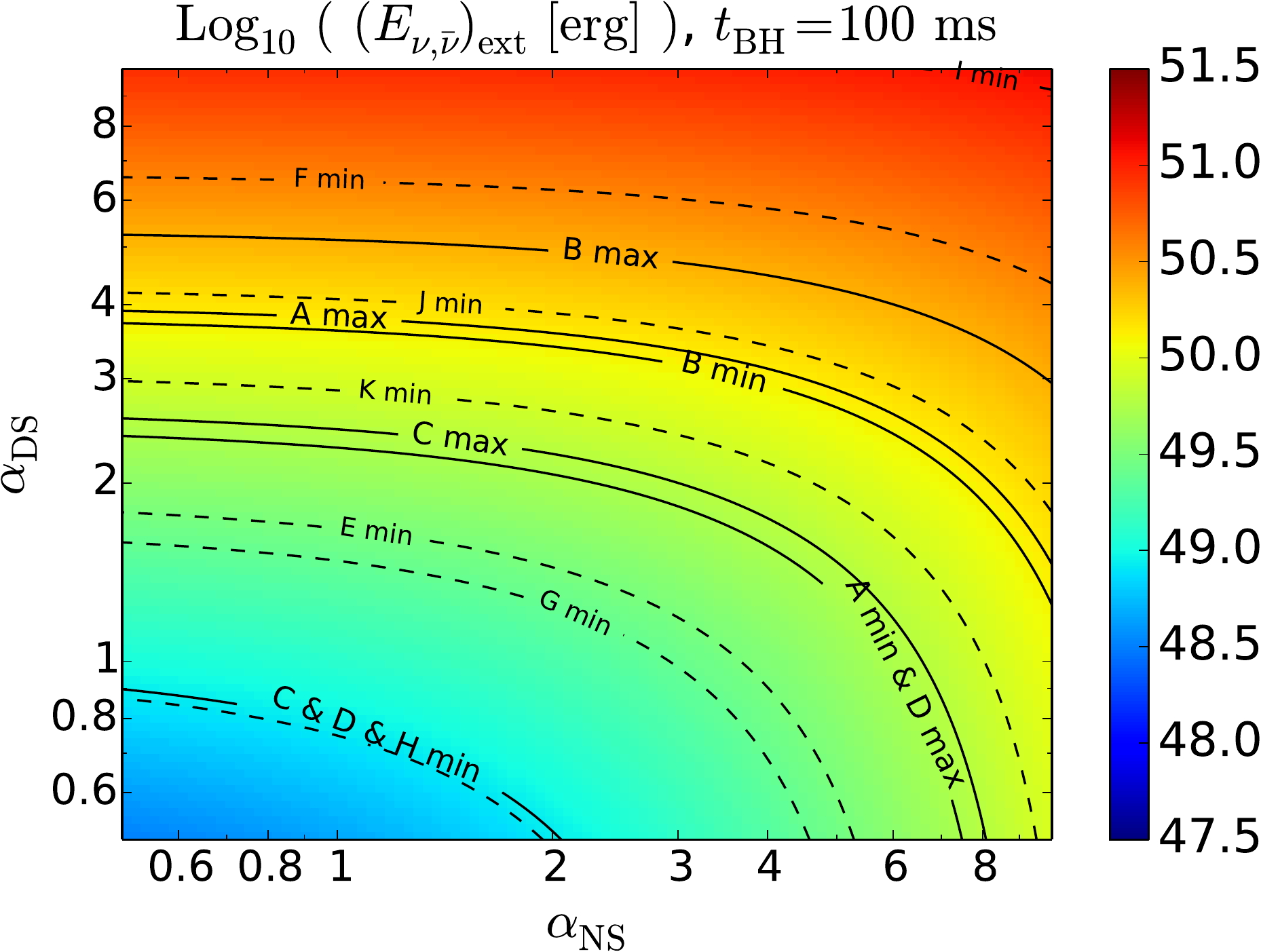}
 \hspace{0.1cm}
 \includegraphics[width=0.48 \linewidth]{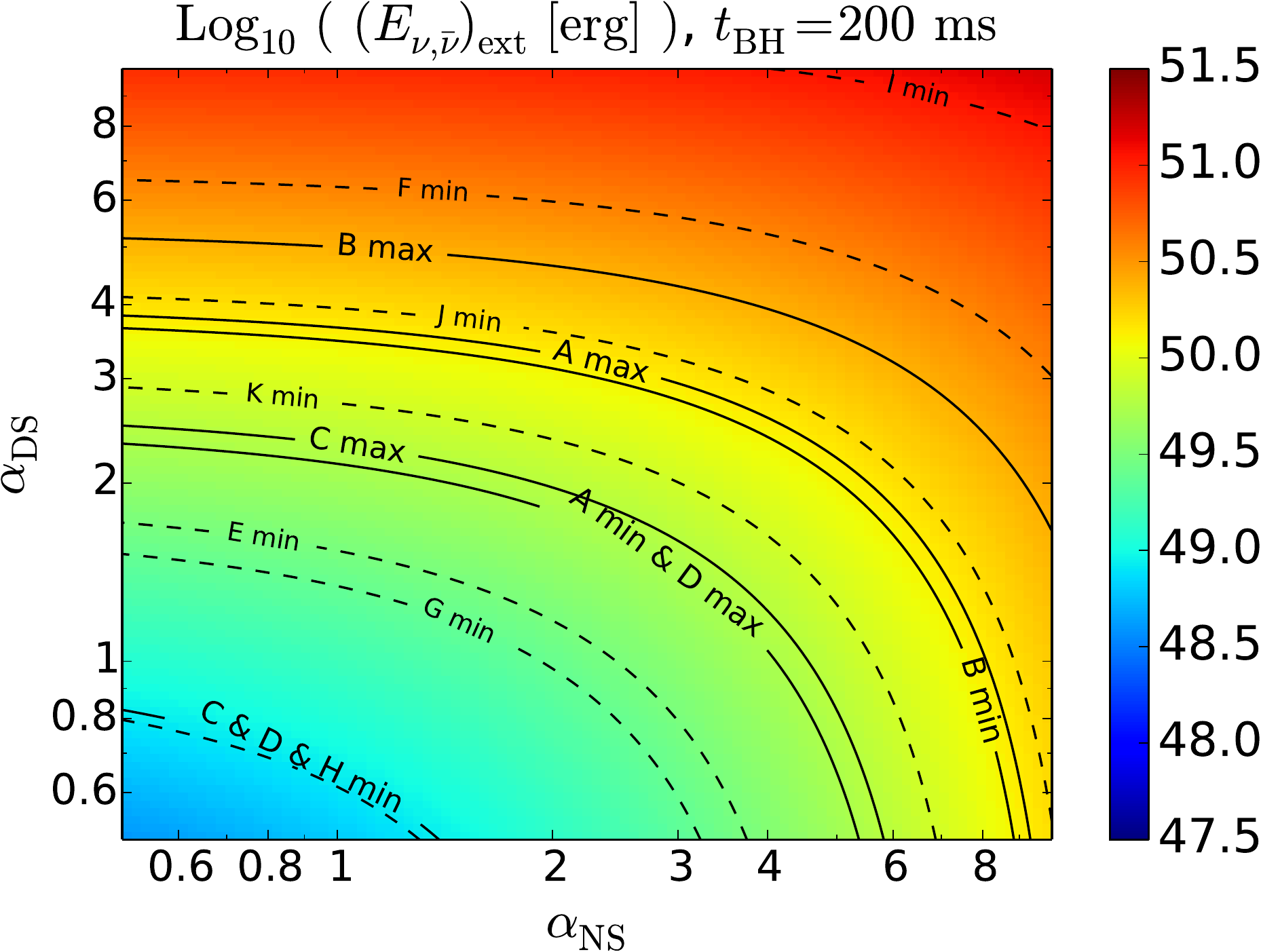}
 \includegraphics[width=0.48 \linewidth]{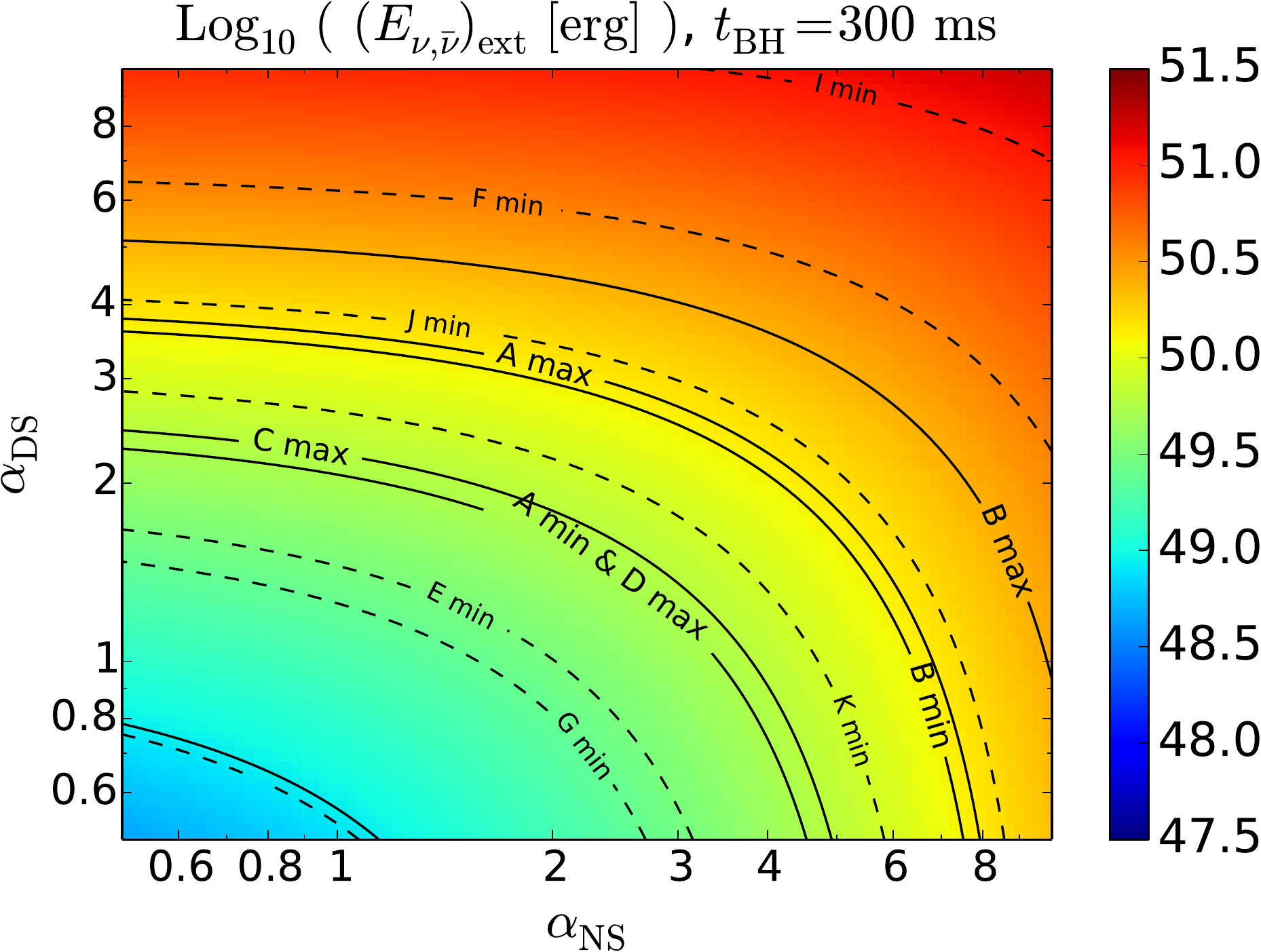}
 \hspace{0.1cm}
 \includegraphics[width=0.48 \linewidth]{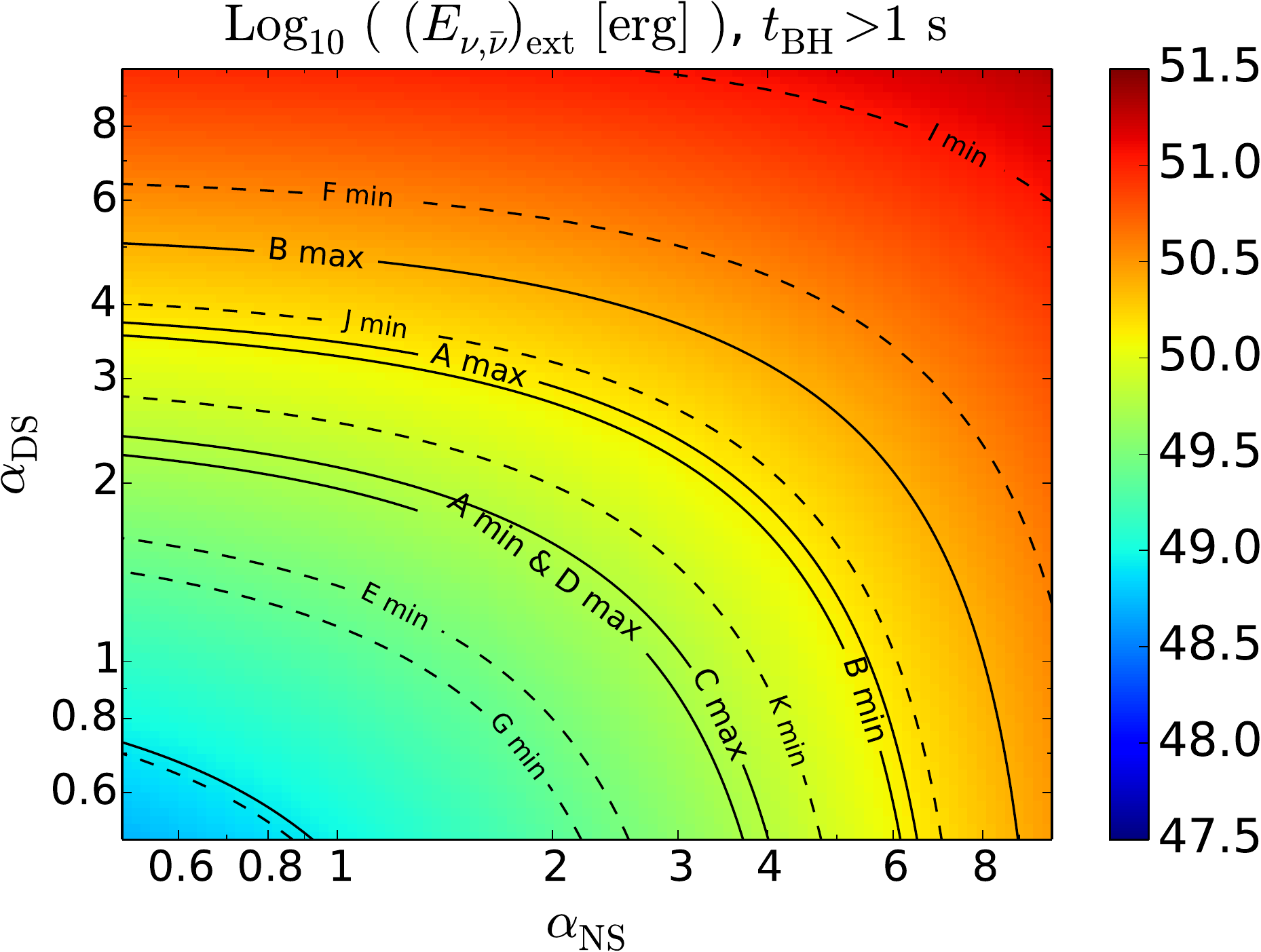}
 \caption{Same as in Figure \ref{fig: effect of luminosity rescaling, nu-antinu}, but rescaling the NS and
 the DS luminosities by independent constant factors $0.5 \leq \alpha_{\rm NS,DS} \leq 10$.}
 \label{fig: effect of luminosity rescaling, NS-DISK}
\end{figure}

The results obtained so far rely on calculations of the neutrino luminosities based on one single
hydrodynamical model. However, the large variety of initial conditions (NS masses, mass ratios and spins) are expected to translate
into a potentially large variety of neutrino luminosities. Moreover, the uncertainties
on the nuclear EOS, as well as the impact of relativistic dynamics and different neutrino treatments, 
can introduce noticeably differences in the intensity of the neutrino luminosities between different numerical models. 
Specifically, GR merger simulations tends to produce hotter remnant and larger neutrino luminosities 
\cite{Sekiguchi.etal:2015,Foucart.etal:2016,Radice.etal:2016}. A similar effect
is observed using softer nuclear EOS. Moreover, gray moment transport schemes seem to suggest 
lower $\nu_e$ luminosities, compared with gray leakage schemes \cite{Foucart.etal:2015a}.
To address the impact of diverse luminosities on the total deposited energy, we employ the proportionality
between $L_{\nu}L_{\bar{\nu}}$ and $Q_{\nu,\bar{\nu}}$ in Eq.~(\ref{eq: simple parametrization decomposed}), 
separately for each of the four distinct contributions appearing in Eq.~(\ref{eq: q contributions}).
We first assume to rescale separately $L_{\nu}$ and $L_{\bar{\nu}}$, i.e. $L_{\nu} \rightarrow \alpha_{\nu} L_{\nu}$
and $L_{\bar{\nu}} \rightarrow \alpha_{\bar{\nu}} L_{\bar{\nu}}$, by a constant factor 
$ 0.5 \leq \alpha_{\nu (\bar{\nu})} \leq 10.0$.
In Figure \ref{fig: effect of luminosity rescaling, nu-antinu}, we present the total cumulative deposited energy
as a function of $\alpha_{\nu}$ and $\alpha_{\bar{\nu}}$. The four different panels refer to four different BH formation
times and the $(\alpha_{\nu},\alpha_{\bar{\nu}})=(1,1)$ case corresponds to our standard simulation.
Luminosities that are twice as large as the computed ones lead to an energy deposition rate four times larger,
due to the linear dependence of $Q_{\nu,\bar{\nu}}$ on the product of $L_{\nu}$ and $L_{\bar{\nu}}$.
Assuming that neutrino pair annihilation is the only source of energy powering short GRBs and that the 
deposited energy is converted very efficiently in kinetic energy and photons inside the relativistic jet,
we can compare the GRBs inferred energy with the energy computed in our models. 
Thus, in the same figures, we show also the inferred total energy from the observation of eleven short GRBs, for which also
an estimate of the jet opening angle is available, see Table \ref{tab: GRB observations}, 
derived from \citeasnoun{Fong.etal:2015}.
In four cases (A-D), the temporal steepening of the afterglow decline rate is attributed to jet breaks, and this interpretation 
provides a measure of the jet opening angle. For the others, the absence of evidences of jet breaks
translates in a lower limit for $\theta_{\rm jet}$, depending on the time of the last observation. 
In these cases, an upper limit of $\theta_{\rm jet} \approx 30^o$ is assumed \cite{Rosswog.Ruiz:2002a}.
Following \citeasnoun{Piran:2004}, the true energy is computed from the isotropized total energy according to
\begin{equation}
 E_{\rm true} = \left( 1-\cos{\theta_{\rm jet}} \right) \left(  E_{\rm \gamma,iso} + E_{\rm kin,iso} \right) \, .
\label{eq: E true}
\end{equation}
In case the MNS collapses quickly to a BH ($t_{\rm BH}< 100~{\rm ms}$), the results obtained 
from our simulation are incompatible with all available observations. Also in the case of a long-lived MNS
($t_{\rm BH} > 1 \, {\rm s}$), our results are compatible only with the lower limits of the less
energetics GRBs. Significantly larger luminosities ($\alpha \gtrsim 4$) are needed to explain the energetics of a large fraction of the observed short GRBs. A very delayed collapse to BH ($t_{\rm bh}>1~{\rm s}$) decreases the required 
increase in luminosity by a factor $\lesssim 2$, compared with an early collapse ($t_{\rm bh}=100~{\rm ms}$).
Nevertheless, the most energetic short GRBs seem to require $\alpha \gtrsim 8$.

We repeat the calculations rescaling the NS and the DS contributions, without distinguishing $\nu_e$ from $\bar{\nu}_e$:
$L_{\nu, \rm NS} \rightarrow \alpha_{\rm NS} L_{\nu, \rm NS}$ and 
$L_{\rm DS} \rightarrow \alpha_{\rm DS} L_{\nu, \rm DS}$, and again $ 0.5 \leq \alpha_{\rm NS,DS} \leq 10.0$.
In Figure \ref{fig: effect of luminosity rescaling, NS-DISK}, we present the total cumulative deposited 
energy, extrapolated at one second, as a function of $\alpha_{\rm NS}$ and $\alpha_{\rm DS}$, for four 
different BH formation times.
Since the NS-NS contribution is smaller than any other contribution involving radiation 
coming from the disk, there is no more symmetry between $\alpha_{\rm NS}$ and $\alpha_{\rm DS}$, and 
variations along $\alpha_{\rm DS}$ have a significantly larger impact.
Thus, hotter disks and long-lived MNS provide larger energy deposition, and they 
are necessary ingredients to explain short GRBs energetics with neutrino pair
annihilation (at least for bursts of low and medium energy). 

\section{Discussion and conclusions}
\label{sec: conclusions}

In this work, we have investigated the energy and momentum deposition operated
by the annihilation of $\nu$-$\bar{\nu}$ pairs above the remnant of a BNS merger over a time scale
comparable with the expected disk lifetime ($\sim 400\,{\rm ms}$).
In particular, we have analyzed the implications of a long-lived MNS and studied
the impact of relativistic effects on the neutrino propagation and annihilation.
For our study, we have used results of the first long-term, three dimensional simulations 
of the aftermath of a BNS merger after the influence of neutrino cooling and heating \cite{Perego.etal:2014b}.
The neutrino emission was modeled via a spectral leakage scheme and neutrino pair annihilation rates have been
computed outside the neutrino surfaces using a detailed, spectral ray-tracing algorithm.

Our major findings are:
\begin{itemize}
 \item the presence of a MNS (instead of a BH) in the center increases the annihilation rate by a factor of $\sim 2$, due to the 
 interaction between neutrinos coming from the MNS with antineutrinos coming from the disk, and vice versa. Moreover, it
 increases the efficiency at which neutrinos can be emitted from the disk, due to the larger emission time and slower 
 accretion process on the MNS surface;
 \item energy and momentum depositions operated by neutrino pair annihilation are more intense 
  closer to the neutrino surfaces, i.e. 
 immediately above the MNS surface and above the region that marks the transition between the MNS and the disk
 ($q_{\nu,\bar{\nu}} \sim 10^{29}\,{\rm erg \, cm^{-3} \, s^{-1}}$, 
 $\left| \mathbf{p}_{\nu,\bar{\nu}} \right| \sim 10^{18}\,{\rm g \, cm^{2} \, s^{-2}}$). 
 Neutrino annihilation above the MNS occurs at lower polar angles, but larger matter densities, than neutrino annihilation above
 the innermost part of the disk ($\lesssim 50 \, {\rm km}$). The location of the annihilation rate and the level of baryonic pollution are 
 expected to influence the dynamical effect of the energy and momentum deposition;
 \item the net momentum provided by neutrino annihilation is mostly pointing outwards: upwards above the MNS and
 the densest part of the disk, more radially at large distances from the center. The efficiency at which net momentum is deposited, 
 with respect to the energy deposition, is low close to the remnant ($\lesssim 0.5$), while efficient ($\gtrsim 0.75$) far from it;
 \item the energy and momentum depositions operated by heavy flavor neutrinos are more than one order of magnitude ($\sim 1/30$) 
 smaller than the ones by electron flavor neutrinos;
 \item volume-integrated energy deposition rates change from $Q_{\nu,\bar{\nu}} \approx 9 \times 10^{49} {\rm erg \, s^{-1}}$ at
 the beginning of our simulation to $Q_{\nu,\bar{\nu}} \approx 2 \times 10^{49} {\rm erg \, s^{-1}}$ at the end (400~ms).
 They are, in good approximation, proportional to the product 
 of the neutrino luminosities, $L_{\nu} L_{\bar{\nu}}$. Interestingly, the proportionality term that contains all the geometrical 
 dependences of the process ($G_{\nu,\bar{\nu}}$) is fairly similar for all the different contributions of $Q_{\nu,\bar{\nu}}$ and
 rather constant in time;
 \item the efficiency at which the emitted neutrinos annihilate above the remnant is rather low ($\eta_{\nu,\bar{\nu}} \lesssim 0.2\%$),
 especially in comparison with the efficiency at which neutrinos are emitted from the disk ($\eta_{\rm acc} \gtrsim 5 \%$);
 \item the inclusion of relativistic effects in the neutrino propagation does not change the results we have obtained
 with the Newtonian calculations qualitatively. From a more detailed quantitative analysis, we observe that the beaming of the neutrino radiation 
 emitted from the rotating disk reduces the annihilation rate above the disk and in the funnel; the gravitational redshift further
 reduces the amount of energy deposited at large distances  (a few tens of kilometers) from the remnant, while 
 close to the compact object neutrinos emitted from the disk experience gravitational blueshift and light bending, 
 which increase their energy deposition efficiency, but reduce the momentum deposition one;
 \item we computed a cumulative energy deposition of $\approx 2.0 \times 10^{49} \, {\rm erg}$ at 400~ms and estimated
 $\approx 2.2 \times 10^{49} \, {\rm erg}$ at 1~s. An early collapse ($t_{\rm BH} \lesssim 100 \, {\rm ms}$) of the MNS
 to a BH decreases the total energy deposition by $\sim 1/2$, due to the progressive decrease 
 of the annihilation rate. A late collapse 
 ($t_{\rm BH} \gtrsim 300 \, {\rm ms}$) is significantly less relevant.
 \end{itemize}

In the following, we compare our results with some recent calculations of neutrino pair annihilation
above compact binary mergers in the literature.
The initial conditions used in our BNS merger aftermath simulation (Section \ref{sec: simulation}) are very close 
to the ones used by \citeasnoun{Dessart.etal:2009}. Thus, this work allows a direct comparison, even if
our calculations span a much longer time. 
In their work, \citeasnoun{Dessart.etal:2009} computed the annihilation rates during the first 100~ms after the BNS merger 
using two different approaches: first, a neutrino gray leakage scheme 
and an annihilation rate formalism based on \citeasnoun{Ruffert.etal:1996} and \citeasnoun{Ruffert.etal:1997}; second, a $S_n$ neutrino 
transport scheme \cite{Livne.etal:2004,Ott.etal:2008} and a moment formalism for the annihilation rate calculations. 
Comparing with our results, on the one hand our spatial distribution of the energy deposition rate,
Figure \ref{fig: anni rate different timesteps}, is very similar to the one obtained by \citeasnoun{Dessart.etal:2009} using their 
second, more accurate approach.
On the other hand, significant qualitative differences are visible when comparing with results obtained with their gray 
leakage scheme. In particular, despite the usage of a leakage scheme, we do not observe a difference of five orders of magnitude
between electron and heavy flavor neutrino annihilation rates, rather of a factor 10-100, more similar to the $S_n$ results.
This comparison reveals that more accuracy can be reached using spectral approaches. Moreover, it confirms that taking into 
account that neutrinos diffusing from optically thick regions are ultimately emitted from the last scattering neutrino surfaces
is key to describe the neutrino intensities and their angular distribution in optically thin conditions (rather than being emitted 
isotropically from their production site).
This effect is more evident for heavy flavor neutrinos, since they decouple much deeper inside the remnant.
A more gentle decrease of the total deposited energy with time is observed in our calculations, compared with the 
steeper decrease found by \citeasnoun{Setiawan.etal:2006} and \citeasnoun{Dessart.etal:2009}.
Consequently, if we integrate the $S_n$ results obtained by \citeasnoun{Dessart.etal:2009} during the first 100 ms, 
we obtain $\approx 5 \times 10^{48} \, {\rm erg}$, smaller than our result
$E_{\nu,\bar{\nu}}(t=100 \, {\rm ms}) \approx 8 \times 10^{48} \, {\rm erg} $.
This is due to the different temporal evolution of the neutrino luminosities between \citeasnoun{Dessart.etal:2009} and \citeasnoun{Perego.etal:2014b}.
This difference depends on the diverse neutrino treatments 
and dimensionality of the two models, as already discussed in \citeasnoun{Perego.etal:2014b}. 
The larger accretion rate obtained in our simulation powers more intense neutrino luminosities, which decrease more 
gradually than in \citeasnoun{Dessart.etal:2009}.

\noindent \citeasnoun{Richers.etal:2015} computed neutrino annihilation rates in models of compact binary merger remnants
using a Monte Carlo radiative transfer code for neutrinos, and comparing with a gray leakage scheme \cite{Metzger.Fernandez:2014}.
If we restrict the volume integrals of the energy deposition rate only to the two cones of 45$^o$ around the rotational axis, the energy 
deposited at 400~ms is $E_{\nu,\bar{\nu}}(t = 400 \, {\rm ms},45^o) \approx 1.3 \times 10^{49} \, {\rm erg}$, more than six times larger 
than their results at 3~s assuming a long-lived MNS ($1.9 \times 10^{48} {\rm erg} $). 
Since we do not expect the energy deposition to be significantly large for $t > 0.5 \, {\rm s}$, we explain this discrepancy 
with differences in the two models.
In particular, the differences that \citeasnoun{Richers.etal:2015} discussed in relation with \citeasnoun{Dessart.etal:2009} apply also to our case.
Their MNS is larger, while their disk is significantly smaller (by a factor of 6) and less dense than ours. As a consequence, their optically
thin disk produces lower luminosities, and a more spherical and dilute radiation distribution. 
The different role of the disk in the two models is confirmed by the BH-torus results. In their calculations, the cumulative energy 
in the 45$^o$ cones at 300~ms is $2.8 \times 10^{46} \, {\rm erg}$, much lower than the corresponding disk-disk contribution we have 
computed in our model $ \left( E_{\nu,\bar{\nu}} \right)_{\rm DS-DS}(t = 300 \, {\rm ms},45^o) \approx 6.5 \times 10^{48} \, {\rm erg}$.

\noindent Our disk-disk contribution can be also compared with the results of \citeasnoun{Just.etal:2016}, where the annihilation of neutrinos above
BH-torus systems was computed. Our results at 400~ms, $ \left( E_{\nu,\bar{\nu}} \right)_{\rm DS-DS}(t = 400 \, {\rm ms}) \approx 8.5 \times 10^{48} \, {\rm erg}$,
differ only by $\sim 30 \%$ with the results of their equal mass (1.45 $M_{\odot}$-1.45 $M_{\odot}$) 
binary merger remnant calculation, despite the usage of different NS masses, EOSs, hydrodynamics and neutrino treatments. 
In particular, the smaller and faster decreasing neutrino luminosities are compensated by a larger global 
annihilation efficiency, $\bar{\eta}_{\nu,\bar{\nu}} \approx 0.23 \%$, compared with our results ($\approx 0.13\%$ at peak).

\noindent Finally, we notice that the effects of the inclusion of relativistic effects in the neutrino propagation are qualitatively compatible with results presented in \citeasnoun{Birkl.etal:2007} and \citeasnoun{Zalamea.Beloborodov:2011}, for BH-torus calculations. In the latter case,
the lack of an extended compact object in the center increases the positive effects close to the BH horizon and 
the annihilation rate, compared with Newtonian calculations. Since our hot MNS is rather extended ($R_{\rm MNS} \approx 20 \, {\rm km}$), 
these effects are less relevant. For a softer nuclear EOS, we expect a large impact and an increase of the energy deposition rates.

The comparison of the deposited energy we have computed with the inferred energy 
of observed short GRBs does not support the annihilation of neutrino pairs as a sufficient mechanism to power short GRB jets.
Neutrino pair annihilation is still a possible central engine only if the luminosities are significantly larger than the ones we have
computed. A factor of 2-3 is usually necessary to explain low energy short GRBs, while a much larger increase ($\gtrsim 8$) 
is necessary for the most energetic ones. Smaller jet opening angles ($\theta_{\rm jet} \lesssim 10^o$) reduce the 
intrinsic GRB energetics and require smaller luminosity magnification factors.
Hotter merger remnants, emerging from general relativistic simulations, possibly employing softer nuclear EOSs, are expected to power
significantly larger neutrino luminosities (up to a few times $10^{53}{\rm erg \, s^{-1}}$ in the first ms after the merger). However, 
it is unclear if such higher luminosities can be sustained for a long enough time scale. Moreover, general relativistic models predict 
less massive disks, with presumably smaller lifetimes \citeaffixed{Giacomazzo.etal:2013}{e.g.,}.

The deposition of an amount of energy compatible with short GRB energetics is not enough to explain the formation of a relativistic jet. 
The presence of a relatively baryon-free region, where the deposition occurs, is also required to accelerate matter to relativistic speeds. 
Recently, \citeasnoun{Just.etal:2016} simulated the formation and expansion of jets powered by neutrino pair annihilation in BH-torus systems. 
They pointed out the possible role of the dynamic ejecta as an obstacle for the formation of a relativistic jet, especially in the case of 
BNS mergers. On the one hand, a long-lived MNS can potentially pollute more heavily the regions above the remnant 
via baryon-rich ejecta, in the form of $\nu$-driven \cite{Perego.etal:2014b} or magnetically-driven \cite{Siegel.etal:2014} winds.
On the other hand, the presence of a large scale $\nu$-driven wind, mainly generated from the disk surface, can collimate the jet 
within a cone of small opening angle \citeaffixed{Murguia.etal:2016}{$\theta_{\rm jet} \lesssim 20^o$,}. Moreover, our results suggest 
that a long-lived MNS increases the amount of deposited energy and net momentum significantly. 
Nevertheless, the required high luminosities and the difficulties in keeping the funnel above the remnant baryon-free suggest
that it is unlikely that neutrino pair annihilation can power the formation of relativistic jets alone. 
If they still provide the bulk of the jet energy, another mechanism is necessary to keep the funnel free from
intense baryon contaminations. Otherwise, an alternative engine is needed and it can work together with neutrino annihilation. 
If this requires the gravitational collapse of the MNS, the propagation of the jet inside non-relativistic baryon-rich ejecta
imposes constraints to the duration of the MNS phase, $t_{\rm BH} \lesssim 0.4 \, {\rm s}$, \cite{Murguia.etal:2014,Murguia.etal:2016}.

\ack
We thank Raffaella Margutti, Giancarlo Ghirlanda and Gabriele Ghisellini for useful discussions. 
We acknowledge support from the Helmholtz-University Investigator grant No. VH-NG-825, and from
the European Research Council Grant No. 677912 EUROPIUM. This work was also supported by a grant 
from the Swiss National Supercomputing Centre (CSCS) under project ID 667.

\appendix

\section{Implementation of relativisitc effects}
\label{apx: implementation rel effects}

In this appendix, we illustrate in detail the implementation of relativistic effects in our 
annihilation rate calculations.

We model the spacetime outside the MNS as a stationary spherically symmetric spacetime
generated by the mass $M_{\rm NS}$. 
In the global coordinates $(ct,r,\theta,\phi)$ the metric $g_{\mu \nu}$ is
\begin{equation}
 {\rm d}s^2 = g_{\mu \nu} {\rm d}x^{\mu}{\rm d}x^{\nu} = - \Gamma^2 \, (c {\rm d}t)^2 + \Gamma^{-2} \, \left( {\rm d}r \right)^2 +
 r^2 \, \left( {\rm d}\theta \right)^2 + r^2 \sin^2{\theta} \left( {\rm d}\phi \right)^2 \, ,
 \label{eqn: metric}
\end{equation}
where $\Gamma = \left( 1 - 2GM_{\rm NS}/(r \, c^2) \right)^{1/2} $.
Since $M_{\rm disk}\ll M_{\rm NS}$, we neglect the effects of the accretion disk mass.
We also ignore the effect of the MNS rotation on the spacetime.
The spin parameter of the MNS is $\hat{a}_{\rm NS} \equiv c J_{\rm NS} / (GM^2_{\rm NS}) \approx 0.5$, 
where $J_{\rm NS}$ is the MNS angular momentum extracted from our simulation.
Although the MNS spin is not negligible, $\hat{a}$ is still significantly lower than 1 and $R_{\rm NS} \gtrsim 2 R_{g}$.
Thus, we assume that rotational effects on the neutrino propagation do not change the 
results obtained with the Schwarzschild metric qualitatively.

We compute the energy ($q^{\rm rel}_{\nu,\bar{\nu}}$) and momentum ($\mathbf{p}^{\rm rel}_{\nu,\bar{\nu}}$)
deposition rates according to Eqs.~(\ref{eq: dessart})
and (\ref{eq: dessart momentum}), valid in the local stationary 
frame of each annihilation point.
In this locally inertial frame, we can introduce a fixed tetrad 
$\left\{ \mathbf{e}_\mu \right\}_{\mu=0 \dots 3}$, i.e. an orthornormal tetrad fixed with respect 
to the spatial global coordinates. This basis is defined such that $e_{\mu} \cdot e_{\nu} = \eta_{\mu \nu}$,
where $\eta_{\mu \nu}$ is the Minkoswki metric. For the fiducial observer associated with
this frame, all the relevant physical quantities are defined as in the flat spacetime.
Let's consider neutrinos of energy $\epsilon_{A}$ emitted around a direction $\mathbf{n}_{A}$ 
from a point $A$ with emissivity $\eta_{A}(\epsilon_{A},\mathbf{n}_{A})$,  
as measured by the local stationary observer.
These neutrinos travel to $B$ (the annihilation point) where they arrive with 
an energy $\epsilon_{B}$ and a direction $\mathbf{n}_{B}$, as measured by the local stationary observer.
The energy at the two points are related by the gravitational redshift formula:
\begin{equation}
 \epsilon_{B} = \left( \frac{\Gamma_A}{\Gamma_B} \right) \epsilon_{A} .
 \label{eqn: energy shift}
\end{equation}
The direction $\mathbf{n}_{B}$ is obtained by solving the geodetic motion of a null 
particle in the Schwarzschild metric traveling from $A$ to $B$. 
In Section (\ref{subsec: geodetic motion}), we detail our method to compute the geodetic motion.
Now we want to compute the neutrino intensity at B based on the emissivity at $A$: 
the observer at $B$ measures a number of neutrinos equal to the number of particles emitted at 
$A$ towards $B$, diminished by the particles absorbed along the propagation path.
We estimate the latter quantity similarly to what we have done 
in the non-relativistic case, Eq.~(\ref{eq: energy conservation between A and B}):
\begin{equation}
 {\rm d}N_{A} (1-\exp{(-\Delta \tau_{\rm en})}) = {\rm d}N_{B} \, .
 \label{eq: relation between A and B neutrinos in GR}
\end{equation}
Expressing Eq.~(\ref{eq: relation between A and B neutrinos in GR}) in terms of the local intensity at $B$ and emissivity at $A$, we obtain
\begin{equation}
 \frac{\eta_{A} {\rm d}V_{A} {\rm d}\Omega_{A} {\rm d}\epsilon_{A} {\rm d}t_{A} }{\epsilon_{A}} = 
 \frac{I_{B} {\rm d}A_{B} {\rm d}\Omega_{B} {\rm d}E_{B} {\rm d}t_{B}  }{E_{B}} \, ,
\end{equation}
and from that
\begin{equation}
\frac{I_{B} {\rm d}\Omega_{B} {\rm d}E_{B}}{E_{B}} =
\left( \frac{\Gamma_A}{\Gamma_B} \right) \left( \frac{{\rm d}\Omega_{A}}{{\rm d}A_{B}} \right)
\frac{\eta_{A} {\rm d}V_{A} {\rm d}\epsilon_{A}}{\epsilon_{A}} \, .
\label{eqn: intensity in arrival point}
\end{equation}
The calculation of $({\rm d} \Omega_A /{\rm d} A_B )$ is also reported in Section (\ref{subsec: geodetic motion}).

At the point $A$, the radiation is emitted by matter moving with a velocity 
$\mathbf{v}_{\rm fl} = \boldsymbol{\beta}~c$, as seen by the stationary observer.
The frame comoving with the fluid at $A$ is related 
to the local stationary frame by a Lorentz boost $\Lambda(\boldsymbol{\beta})$.
If we define $\gamma \equiv \left( 1 - \beta^2 \right)^{-1/2}$, 
$\cos{\psi} = ( \mathbf{p}_{A} \cdot \boldsymbol{\beta} ) 
/ ( p_{A}~\beta )$,
$\beta = \left| \boldsymbol{\beta} \right|$, and 
$p_{A} = \left| \mathbf{p}_{A} \right|$, then
the energy measured by the local static observer and the emission energy in the comoving frame, 
$\epsilon_0$, are related by the Doppler shift formula:
\begin{equation}
 \epsilon_{A} = \frac{\epsilon_{A,0}}{\gamma \left( 1 - \beta \cos{\psi} \right)} \, .
 \label{eq: energy comoving frame}
\end{equation}
The emissivity transforms between the local and the comoving frames according to
\begin{eqnarray}
 \eta_A(\epsilon_{A},\mathbf{n}_{A}) & = \eta_A(\epsilon_{A,0},\mathbf{n}_{A,0}) 
 \frac{1}{\gamma^2 \left( 1 - \beta \cos{\psi} \right)^3} \, .
 \label{eqn: emissivity transformation law}
\end{eqnarray}
We notice here that, instead of the usual transformation law for the emissivity, 
stating that $\eta/\epsilon^2$ is a relativistic invariant, we adopted the receiver point 
of view \citeaffixed{Rybicki.Lightman:1979}{e.g.,}, because we express
the received intensity at $B$ based on the emissivity at $A$, Eq.~(\ref{eq: energy comoving frame}). 
In this case, the time interval does not transform as ${\rm d}t = \gamma {\rm d}t_0$
(emitter's point of view), but as 
${\rm d}t = \gamma \left( 1 - \beta \cos{\psi}  \right) {\rm d}t_0$, due to the Doppler effect.
In the comoving frame, the emission is assumed to originate from a volume 
${\rm d}V_{A,0} = \gamma{\rm d}V_{A}$ and to be isotropic in the half plane 
defined by a direction $\langle {\mathbf{n}}_{A,0} \rangle$, according to 
Eq.~(\ref{eq: isotropic emissivity}). 
To relate our relativistic calculations with the input data coming from our Newtonian simulation, we
make the following assumptions. First, we assume that Eq.~(\ref{eq: eta and R_nu relationship})
is valid in the comoving reference frame.
Second, we assume that the favored emission direction $\mathbf{n}_{\tau}$ of our Newtonian calculations
corresponds to $\langle \mathbf{n}_{A} \rangle$, the average emission direction as seen by the local 
stationary observer. The sign of the scalar product 
$ \langle \mathbf{n}_{A,0} \rangle \cdot \mathbf{n}_{A,0} $,
required in Eq.~(\ref{eq: isotropic emissivity}) for Eq.~(\ref{eqn: emissivity transformation law}), is computed 
using $\mathbf{k}$ and $\mathbf{w}$, two vectors parallel to $ \langle \mathbf{n}_{A,0} \rangle $ 
and $ \mathbf{n}_{A,0} $, respectively, and obtained from the Lorentz transformation of the 
corresponding transformed propagation directions $ \langle \mathbf{n}_{A} \rangle$ and 
$ \mathbf{n}_{A} $:
\begin{equation}
\mathbf{k} =  \frac{1}{1-\boldsymbol{\beta} \cdot \mathbf{n_{\tau}}}
\left[
\frac{1}{\gamma} \mathbf{n_{\tau}} - \left( 1 - \frac{\gamma}{\gamma + 1} \boldsymbol{\beta} \cdot \mathbf{n_{\tau}} \right) \boldsymbol{\beta}
\right]
\end{equation}
and
\begin{equation}
\mathbf{w} =  \frac{1}{1-\boldsymbol{\beta} \cdot {\mathbf{n}}_{A}}
\left[
\frac{1}{\gamma} \mathbf{n}_{A} - \left( 1 - \frac{\gamma}{\gamma + 1} \boldsymbol{\beta} \cdot \mathbf{n}_{A} \right) \boldsymbol{\beta}
\right] \, .
\end{equation}
Thus,
\begin{eqnarray}
 & {\rm sign}\left( \langle \mathbf{n}_{A,0} \rangle \cdot \mathbf{n}_{A,0} \right) = 
 {\rm sign}\left( \mathbf{k} \cdot \mathbf{w} \right) = {\rm sign}\left( 
 \frac{1}{ \left( 1-\boldsymbol{\beta}\cdot \mathbf{n}_{\tau} \right) 
           \left(1-\boldsymbol{\beta}\cdot \mathbf{n}_{A} \right)} \right. \nonumber \\
 &         \left.  \left[ 1 + \left( \boldsymbol{\beta}\cdot \mathbf{n}_{\tau} \right)\left( \boldsymbol{\beta}\cdot \mathbf{n}_{A} \right) - \frac{1}{\gamma^2}\left(1- \mathbf{n}_{\tau} \cdot \mathbf{n}_{A} \right) - \boldsymbol{\beta}\cdot \mathbf{n}_{\tau} - \boldsymbol{\beta}\cdot \mathbf{n}_{A} \right]
 \right).
 \label{eq: sign of inner product}
\end{eqnarray}

In a first step, the spectral neutrino fluxes at the annihilation point are computed from the 
simulation data according to the expressions Eqs.~(\ref{eqn: intensity in arrival point}), 
(\ref{eqn: emissivity transformation law}) and (\ref{eq: sign of inner product}),
on an energy grid shifted with respect to the 
grid used in our hydrodynamics simulation, according to Eqs.~(\ref{eqn: energy shift}) and 
(\ref{eq: energy comoving frame}).
In a second step, the spectral intensities are interpolated back on the original energy grid.
The interpolation is performed in such a way that the energy-integrated neutrino flux
is preserved.

\subsection{Geodetic motion}
\label{subsec: geodetic motion}

Here, we clarify the method we have adopted to compute the geodetic motion of a null particle between $A$ and $B$
in the metric (\ref{eqn: metric}).
We first find the rotation $\Psi$ that transforms the global coordinate system to a new coordinate system
$(ct',r',\theta',\phi')$ such that the origin, 
and the points $A$ and $B$ lie on the plane $\theta'=\pi/2$. 
Due to angular momentum conservation, the geodetic motion of neutrinos is also confined on this plane.
We define $\phi'_{A}$ and $\phi'_{B}$ as the polar angles of $A$ and $B$ on the $\theta'=\pi/2$ plane, and
$\phi'_{B} = \phi'_{A} - \phi_{AB} $, where $\phi_{AB}$ is the angle between the position vectors of 
$A$ and $B$ in the original coordinate system. Due to the axisymmetry of the problem around the 
$z'$ axis, the choice of $\phi'_A$ is arbitrary.
Since $\Psi$ is a spatial rotation, $r'=r$ and $t'=t$.
The solution of the elliptic integrals that describe exactly the geodetic motion would be computationally 
too expensive for our annihilation rate calculations. Thus, we use a method derived from the approximated 
analytic expression provided in \citeasnoun{Beloborodov:2002}.
The trajectory of a null particle, expressed in polar coordinates $(r',\phi')$ on the
$\theta'=\pi/2$ plane, can be approximated by
 \begin{eqnarray}
  r'(\phi') \approx & \left[ \frac{R_g^2 (1-\cos{(\phi'- \Delta)})^2}{4(1+
  \cos{(\phi'-\Delta)})^2} + \frac{b^2}{\sin^2{(\phi'-\Delta))}} \right]^{1/2} \nonumber \\
  & - \frac{R_g (1-\cos{(\phi'-\Delta)})}{2(1+\cos{(\phi'-\Delta)})} \, .
  \label{eq: second trajectory equation}
 \end{eqnarray}
In the previous expression, $b$ is the impact parameter, one of the trajectory integrals, and 
at any point of the trajectory, it can be related to the 
propagation angle $\alpha$ by
  \begin{equation}
   \sin{\alpha} = \frac{b}{r'} \sqrt{1-\frac{R_g}{r'}} \, .
  \end{equation}
The propagation angle $\alpha$ is defined 
as the angle between the local propagation direction of the neutrino and the local radial 
direction, measured in clockwise direction. If we denote $\alpha_0$ as
the initial propagation angle, the deflection angle $\psi_0$ is equal to 
\begin{equation}
  \cos{\psi_0}=1- \left( 1-\cos{\alpha_0} \right) \left( 1-\frac{R_g}{r_A} \right)^{-1}.
\end{equation}
Concerning $\Delta$, if $0 \leq \alpha_0 < \pi$, then $\Delta = \phi_A-\psi_0$. In this case, 
$\phi'$ takes values in the interval $[\phi_A -\psi_0, \phi_A]$, such that 
$r'(\phi_A) = r_A$ and $r'(\phi_A-\psi_0) = + \infty$.
On the other hand, if $\pi \leq \alpha_0 < 2\pi$, then $\Delta = \phi_A+\psi_0-2\pi$. 
In this other case, $\phi'$ takes values in the interval $[\phi_A, \phi_A + \psi_0]$, and 
$r'(\phi_A) = r_A$ and $r'(\phi_A+\psi_0) = + \infty$.
The approximated formula replaces the exact elliptic integrals with high accuracy for $r \gtrsim 2 R_g$.
This approximation is justified because in our case $R_g \approx 8 \, {\rm km}$ and $R_{\rm NS} \approx 20 \,{\rm km}$. 

Let's assume first that $r_A < r_B$.
Our method consists on the numerical solution of the equation $f(\alpha_0) = 0$, where
\begin{eqnarray}
 f(\alpha_0) = & r_B - \left[ \frac{R_g^2 (1-\cos{(\phi_B- \Delta)})^2}{4(1+
 \cos{(\phi_B-\Delta)})^2} + \frac{b^2}{\sin^2{(\phi_B-\Delta))}} \right]^{1/2} \nonumber \\
 & + \frac{R_g (1-\cos{(\phi_B-\Delta)})}{2(1+\cos{(\phi_B-\Delta)})} \, ,
 \label{eqn: NewRap equation}
\end{eqnarray}
obtained by imposing the passage of the trajectory from $A$ {\it and} $B$.
If a solution of Eq.~(\ref{eqn: NewRap equation}) is found, it allows the calculation 
of the propagation directions $\mathbf{n}'_{A}$ and $\mathbf{n}'_{B}$ 
in the rotated local reference frames.
In the cases where $r_B < r_A$, we use the time invariance of the geodetic motion. We solve for
the trajectory of a neutrino traveling from $B$ to $A$ and then invert the propagation directions
$ \mathbf{n}_{\rm inv} \rightarrow \mathbf{n}_{\rm dir} = - \mathbf{n}_{\rm inv}$.
Finally, we apply the inverse rotation $\Psi^{-1}$ to these vectors to find the components of
$\mathbf{n}_{A}$ and $\mathbf{n}_{B}$ for the original local stationary observers.

The quantity $({\rm d \Omega_{A}}/{\rm d}A_{B})$, required in Eq.~(\ref{eqn: intensity in arrival point}), 
is computed based on our approximated expressions of the trajectory of the null geodesics.
In particular, we consider radiation emitted at $A$ with an initial small spread ${\rm d}\alpha_0$ 
in the propagation angle around the direction $\mathbf{n}_i(\alpha_0)$, 
such that ${\rm d}\Omega_{A} \approx \pi {\rm d}\alpha_0^2$.
This radiation reaches $B$ with a central direction $\mathbf{n}_f(\alpha_f)$,
within a surface ${\rm d}A_{B}=\pi D^2$. The quantity $D$ is the length
of the perpendicular displacement ${\rm d}\mathbf{l}_B$, defined such as
\begin{equation}
 {\rm d}\mathbf{l}_B =  - D \sin{\alpha_f} {\mathbf{\hat{r}'_B}} + D \cos{\alpha_f} {\boldsymbol{\hat{\phi}'_B}} \, ,
\label{eqn: expression for displacement 1} 
\end{equation}
assuming to have expressed the final direction as $\mathbf{n}_f = \cos{\alpha_f} \mathbf{\hat{r}'_B} + 
\sin{\alpha_f} {\boldsymbol{\hat{\phi}'_B}}$. 
The displacement can also be expressed in terms of the variation of the polar coordinates
\begin{equation}
{\rm d}\mathbf{l}_B = {\rm d}r'_B \mathbf{\hat{r}'_B} + 
r'_B {\rm d}\phi'_B {\boldsymbol{\hat{\phi}'_B}} \, .
\label{eqn: expression for displacement 2}
\end{equation}
If we consider the expression of the trajectory, Eq.~ (\ref{eqn: NewRap equation}), 
as a function of both $\phi$ and $\alpha_0$, we compute the total differential as:
\begin{equation}
 {\rm d}r' = \frac{\partial r'}{\partial \alpha_0} \, {\rm d}\alpha_0 +
             \frac{\partial r'}{\partial \phi'} \, {\rm d}\phi' .
 \label{eqn: trajectory differentials}            
\end{equation}
Comparing Eqs.~(\ref{eqn: expression for displacement 1}) and 
(\ref{eqn: expression for displacement 2}) and using 
(\ref{eqn: trajectory differentials}), we obtain
\begin{equation}
 \left( \frac{{\rm d}\Omega_{A}}{{\rm d}A_{B}} \right) \approx 
 \left[ \left( \left. - \frac{\partial r'}{\partial \phi'} \right|_B \frac{\cos{\alpha_f}}{r'_f} - \sin{\alpha_f} \right) 
 \left( \left. \frac{\partial r'}{\partial \alpha_0} \right|_B \right)^{-1} \right]^2 \, .
 \label{eq: ratio domega dA}
\end{equation}

\section{Analysis of relativistic effects on the energy deposition rates}
\label{apx: analysis rel effects}

\begin{figure}
 \includegraphics[width=0.49 \linewidth]{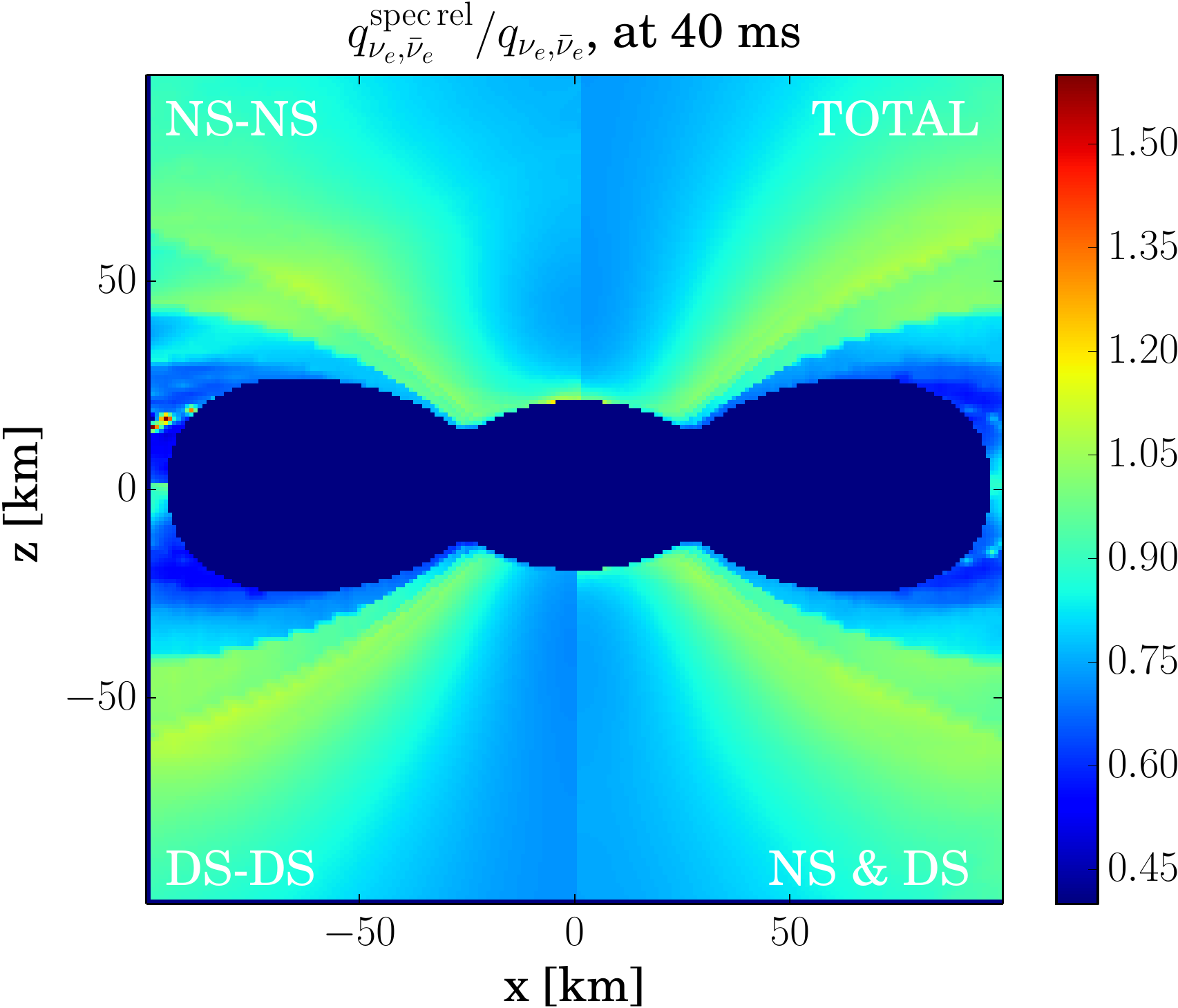}
 \includegraphics[width=0.49 \linewidth]{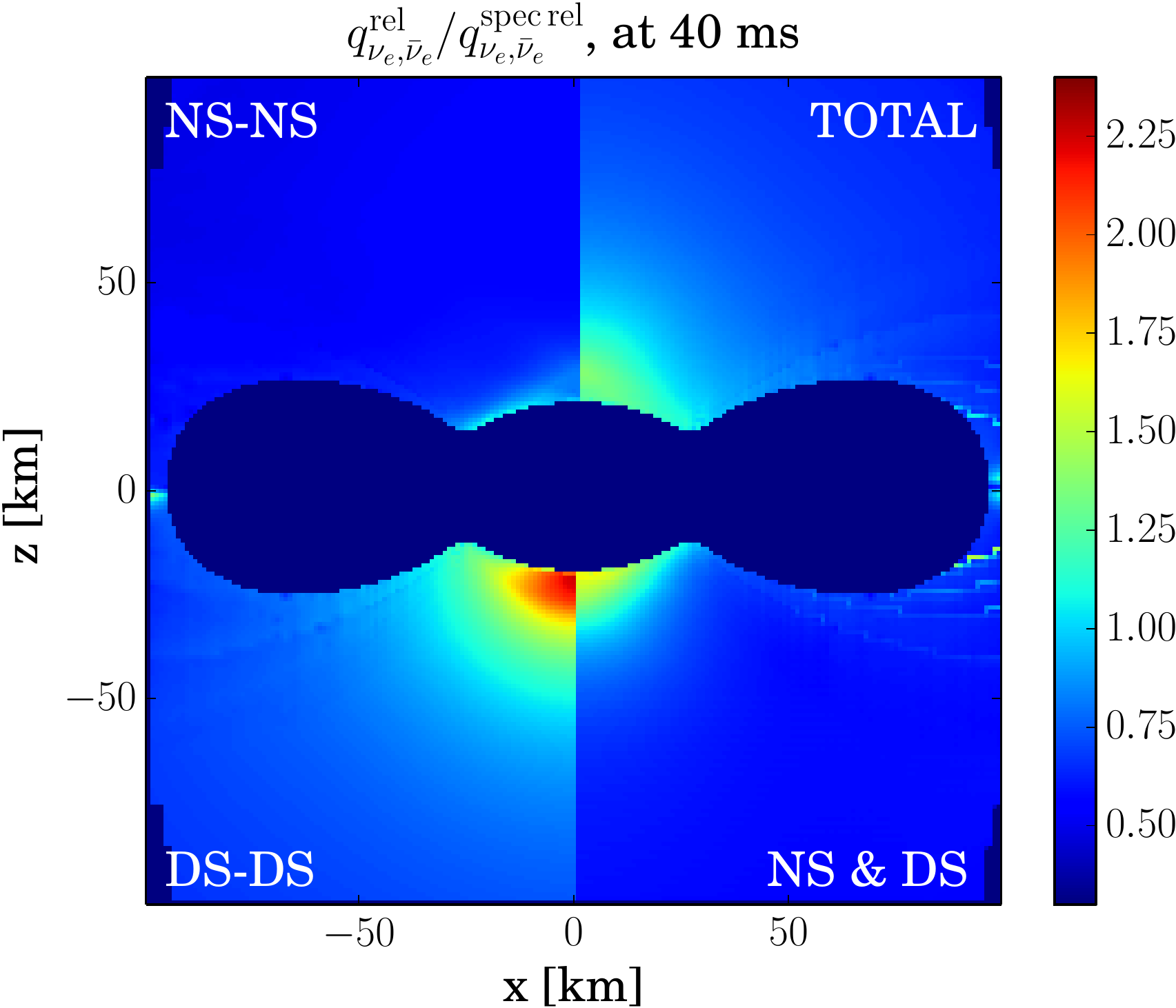}
 \caption{Same as in the left panel of Figure \ref{fig: relativity effect on components}, but for
  $q^{\rm spec \, rel}_{\nu_e,\bar{\nu}_e}/q_{\nu_e,\bar{\nu}_e}$ (left) and 
  $q^{\rm rel}_{\nu_e,\bar{\nu}_e}/q^{\rm spec \, rel}_{\nu_e,\bar{\nu}_e}$ (right).}
 \label{fig: detailed relativity effect on components}
\end{figure}

Special and general relativistic corrections have potentially different effects on the
four different contributions to the energy deposition rate presented in Eq.~(\ref{eq: q contributions}),
due to the different origin and emission geometry.
In the color coded panels of Figure \ref{fig: detailed relativity effect on components},
we show the ratios between $q_{\nu_e,\bar{\nu}_e}$ computed with different levels of approximations
for the propagation of the neutrinos outside the neutrino surfaces. The rates refer to 40~ms in our simulation. 
We consider the standard Newtonian rates, $q_{\nu,\bar{\nu}}$, (Section \ref{sec: method Newtonian}), 
as well as the relativistic ones, $q^{\rm rel}_{\nu,\bar{\nu}}$ (Section \ref{sec: method rel} 
and \ref{apx: implementation rel effects}). As intermediate step, we consider 
also $q^{\rm spec \, rel}_{\nu,\bar{\nu}}$ , the energy deposition 
rate obtained by taking into account only special relativistic effects due to the fast motion 
of matter inside the remnant (relativistic Doppler and beaming), but neglecting all the 
relativistic effects due to the curved spacetime (i.e., we assume $\Gamma=1$ in the spacetime metric). 
The different quadrants correspond to the different contributions to $q_{\nu_e,\bar{\nu}_e}$. 
The ratio $q^{\rm spec \, rel}_{\nu,\bar{\nu}}/q_{\nu,\bar{\nu}}$ (left panel) shows
that the beaming effect and the Doppler effect reduce the efficiency of the annihilation process significantly 
in the funnel and, more in general, far from the disk plane, for all the different contributions. 
At the same time, they increase the energy deposition efficiency close to the remnant. 
Since neutrinos coming from the disk are more subject to
these effects, the largest relative increase is verified for the DS-DS contribution.
General relativistic effects modify special relativistic deposition rates mainly in the proximity 
of the MNS, where the curvature is more pronounced.
In particular, the NS-NS contribution benefits only of
the radiation bending immediately above the MNS, while in the rest of the volume the redshift decreases the 
energy deposition rate.
For the DS-DS contribution, the better collision angle due to radiation bending and the gravitational blueshift 
of neutrinos moving towards smaller radii increase significantly the deposition energy efficiency.
A similar, although weaker, effect is observed also in the case of the NS\&DS contribution. In this case, neutrinos
coming from the MNS are more subject to the gravitational redshift and this explains the faster decrease of
the $q_{\nu,\bar{\nu}}$ ratio at larger radii, compared with the DS-DS case.

%


\end{document}